\documentclass[fleqn,usenatbib]{mnras}

\usepackage[T1]{fontenc}
\usepackage{ae,aecompl}

\usepackage[figurename=Figure.,
                  justification=RaggedRight, 
                  labelfont={bf, footnotesize}, 
                  textfont={footnotesize},position=top]{caption}

\usepackage{graphicx}	
\usepackage{amsmath}	
\usepackage{amssymb}	
\usepackage{float}
\usepackage{dblfloatfix}
\usepackage{blindtext}
\usepackage{booktabs}
\usepackage{rotating}
\usepackage{natbib}
\usepackage{fmtcount}
\usepackage{threeparttablex}
\usepackage{multirow}
\usepackage{morefloats}
\usepackage{dblfloatfix}
\usepackage{dcolumn}
\usepackage{amsmath}
\usepackage{latexsym}
\usepackage{longtable}
\usepackage{lipsum} 
\usepackage{placeins}
\usepackage{textcomp}
\usepackage{gensymb}
\usepackage{mathtools}
\usepackage{afterpage}
\usepackage{caption}
\usepackage{hyperref}

\usepackage{amsmath}
\usepackage{longtable}
\usepackage{rotating}
\usepackage{tikz}
\usepackage{float}
\usepackage{lipsum}
\usepackage{isotope}

\title[SN~2012au: Photometry, polarimetry, and spectroscopy]{Photometric, polarimetric, and spectroscopic studies of the luminous, slow-decaying Type Ib SN~2012au}

\author[S. B. Pandey et al.]{S. B. Pandey$^{1}$\thanks{E-mail: shashi@aries.res.in},
Amit Kumar$^{1,2}$ \thanks{E-mail: amitkundu515@gmail.com}, Brajesh Kumar$^{1}$ \thanks{E-mail: brajesh@aries.res.in}, G. C. Anupama$^{3}$, S. Srivastav$^{3,4}$, \newauthor D. K. Sahu$^{3}$, J. Vinko$^{5,6,7}$, A. Aryan$^{1,8}$, A. Pastorello$^{9}$, S. Benetti$^{9}$, L. Tomasella$^{9}$, \newauthor Avinash Singh$^{1}$, A. S. Moskvitin$^{10}$, V. V. Sokolov$^{10}$, R. Gupta$^{1,8}$, K. Misra$^{1}$, P. Ochner$^{9}$ \newauthor and S. Valenti$^{11}$ \\ \\ 
$^{1}$Aryabhatta Research Institute of Observational Sciences, Manora Peak, Nainital - 263001, India\\
$^{2}$School of Studies in Physics and Astrophysics, Pandit Ravishankar Shukla University, Chattisgarh 492010, India\\
$^{3}$Indian Institute of Astrophysics, II Block, Koramangala, Bengaluru 560034, India\\
$^{4}$Astrophysics Research Centre, School of Mathematics and Physics, Queen's University Belfast, BT7 1NN\\
$^{5}$Konkoly Observatory, Research Center for Astronomy and Earth Sciences, Konkoly Thege M. ut 15-17, Budapest 1121, Hungary\\
$^{6}$Department of Optics and Quantum Electronics, University of Szeged, Dom ter 9, Szeged 6720, Hungary\\
$^{7}$Department of Astronomy, University of Texas, Austin, TX 79712, USA\\
$^{8}$Department of Physics, Deen Dayal Upadhyaya Gorakhpur University, Gorakhpur 273009, India\\
$^{9}$INAF - Osservatorio Astronomico di Padova, Vicolo dell'Osservatorio 5, 35122, Padova, Italy\\
$^{10}$Special Astrophysical Observatory of the Russian Academy of Sciences (SAO RAS), Nizhnij Arkhyz 369167, Russia\\
$^{11}$Department of Physics, University of California,1 Shields Avenue, Davis, CA 95616-5270, USA
}
\date{Accepted XXX. Received 2021; in original form ZZZ}
\date{Accepted --------, Received 2021; in original form ------}
\pubyear{}
\begin{document}
\label{firstpage}
\pagerange{\pageref{firstpage}--\pageref{lastpage}}
\maketitle

\begin{abstract}
Optical, near-infrared (NIR) photometric and spectroscopic studies, along with the optical imaging polarimetric results for SN 2012au, are presented in this article to constrain the nature of the progenitor and other properties. Well-calibrated multiband optical photometric data (from --0.2 to +413 d since $B$-band maximum) were used to compute the bolometric light curve and to perform semi-analytical light-curve modelling using the {\tt MINIM} code. A spin-down millisecond magnetar-powered model explains the observed photometric evolution of SN~2012au reasonably. Early-time imaging polarimetric follow-up observations (--2 to +31 d) and comparison with other similar cases indicate signatures of asphericity in the ejecta. Good spectral coverage of SN~2012au (from --5 to +391 d) allows us to trace the evolution of layers of SN ejecta in detail. SN~2012au exhibits higher line velocities in comparison with other SNe~Ib. Late nebular phase spectra of SN~2012au indicate a Wolf--Rayet star as the possible progenitor for SN~2012au, with oxygen, He-core, and main-sequence masses of $\sim$1.62 $\pm$ 0.15 M$_\odot$, $\sim$4--8 M$_\odot$, and $\sim$17--25 M$_\odot$, respectively. There is a clear absence of a first overtone of carbon monoxide (CO) features up to +319 d in the $K$-band region of the NIR spectra. Overall analysis suggests that SN~2012au is one of the most luminous slow-decaying Type Ib SNe, having comparatively higher ejecta mass ($\sim$\,4.7 -- 8.3 M$_\odot$) and kinetic energy ($\sim$\,[4.8 -- 5.4] $\times$ 10$^{51}$ erg). Detailed modelling using {\tt MESA} and the results obtained through {\tt STELLA} and {\tt SNEC} explosions also strongly support spin-down of a magnetar with mass of around 20 M$_\odot$ and metallicity Z = 0.04 as a possible powering source of SN 2012au.
\end{abstract}


\section{Introduction} \label{sec:int}
Type Ib supernovae (SNe) are a subclass of stripped-envelope core-collapse SNe (SESNe), as the outer hydrogen (H) envelopes of their progenitors are partially or completely removed because of the higher mass-loss rate before the explosion \citep{Wheeler1987, Filippenko1997, Gal-Yam2017, Prentice2017, Shivvers2017, Modjaz2019}. Weak signatures of H found in a few SNe~Ib may be attributed to the thin layer of H and its continuous stripping from the progenitor \citep{Branch2006, Elmhamdi2006, Valenti2011, Hachinger2012}. SNe~Ib also serve as important events to probe deeper towards the proposed transition of H-poor SNe~IIb to He-free SNe~Ic \citep{Filippenko1993, Dessart2015, Yoon2015}. H-deficient Wolf--Rayet (WR) stars ($\gtrsim$\,20-25\,M$_\odot$) are believed to be possible progenitors for Type IIb/Ib SNe, as they have lost their H envelope partially/completely due to stellar winds \citep{Heger2003, Georgy2009, Yoon2015}. On the other hand, low-mass progenitors ($\gtrsim$\,11\,M$_\odot$) in binary systems are also thought to be progenitors of these SESNe, where the primary star loses its H envelope through mass-transfer to a companion star \citep[][]{Podsiadlowski1992, Nomoto1995, Smartt2009, Yoon2010, Smith2011}. Envelope stripping is crucial in aiding understanding of the evolution of massive stars, post-explosion interaction, and the properties of the resulting SNe \citep{Yoon2015, Gilkis2019}. SNe~Ib seem to have a higher/lower degree of envelope stripping and more/less massive progenitors in comparison to SNe~IIb/Ic \citep{Fang2019}. More massive progenitors, in comparison with those for SNe of Type IIb/Ib/Ic, are responsible for producing H-deficient superluminous SNe \citep[SLSNe~I:][]{Moriya2018, Gal-Yam2019, Inserra2019}.

Probing pre-explosion images of nearby SNe sites is the best way to identify and determine physical properties like masses of progenitor stars \citep{Gal-Yam2007, Van2017}. So far, because of the limitation of achievable spatial resolution, only a handful of SESNe are known, including iPTF13bvn \citep[Ib:][]{Cao2013, Folatelli2016}, SN~2017ein \citep[Ic:][]{Kilpatrick2018, Van2018, Xiang2019}, and the recent SN~2019yvr \citep[Ib:][]{Kilpatrick2021}. Using pre-explosion images, a binary system for iPTF13bvn \citep{Fremling2014, Kuncarayakti2015, Folatelli2016} and a single WR star for SN~2019yvr were found as potential progenitor candidates \citep{Kilpatrick2021}. However, since the site hosting SN~2017ein is very crowded, possibilities of the progenitor being a WR star, a binary system with a high-mass candidate, and an unresolved young compact star cluster exist together \citep{Xiang2019}.

The most acceptable explosion mechanism of such SESNe is radioactive decay of $\rm ^{56}Ni$ (RD); however, the light curves of some SESNe (e.g., Type Ib/c SN~2005bf: \citealt{Maeda2007}; Type Ic SN~2019cad: \citealt{Gutierrez2021}; and many SLSNe~I; \citealt{Inserra2013, Nicholl2017, Kumar2021}) are explained by the existence of a spin-down millisecond magnetar (MAG) as a powering source \citep{Maeda2007, Kasen2010, Woosley2010}. Near-peak photometric data are important in performing analytical light-curve modelling, which helps in obtaining the power mechanism and estimating various physical parameters such as the ejecta mass ($M_{ej}$), the kinetic energy of the explosion ($E_K$), and the $\rm ^{56}Ni$ mass \citep{Chatzopoulos2012, Wheeler2015}. By contrast, late-phase spectral observations help in investigating the nature of the explosion mechanism/progenitor and in constraining the composition of the inner layers of the expanding ejecta, which allows us to estimate the amount of oxygen mass (M$_{O}$), He-core mass ($M_{He}$), and zero-age main-sequence mass (M$_{ZAMS}$); see \cite{Uomoto1986}, \cite{Thielemann1996}, and \cite{Fang2019}.

More useful information about such SNe can be probed using near-infrared (NIR) spectroscopy, which is important to understand molecular emission features not observed in the optical or UV regions (e.g., the carbon monoxide (CO) features in the $K$-band region) and is crucial to investigate dust formation \citep{Morgan2003}. Due to the large number of collisionally excitable energy levels of CO emission, it acts as a coolant and can shed light on the dust production rate in SN ejecta \citep{Gearhart1999, Morgan2003, Liljegren2020}. The NIR spectra of SNe~Ib exhibit strong features of He\,{\sc i} $\lambda$10 800 and C\,{\sc i}, which give an idea about the extent of envelope stripping in these objects. He\,{\sc i} features are found to be weaker/absent in SNe~Ic (e.g. SN~2007gr: \citealt{Hunter2009}); however, in addition to this, broad-line SNe Ic exhibit weak signatures of C\,{\sc i} (SN~2020bvc: \citealt{Rho2021}), which indicate large envelope mass stripping in such events. Additionally, SNe~Ib present stronger He\,{\sc i} $\lambda$20 580 in comparison with that displayed in the NIR spectra of SNe~Ic.

Asphericity is another generic property of SESNe, as half of these well-known events exhibit aspherical features owing to various possible effects like dust and the clumpy nature of ejecta \citep{Leonard2006, Maeda2008, Wang2008, Taubenberger2009, Tanaka2017}. The level of asphericity in the ejecta could be estimated using the observed degree of polarization \citep{Shapiro1982, McCall1984, Hoflich1991}. The asphericity of the explosion can also be established independently by investigating the emission-line profiles (e.g., [O\,{\sc i}] and [Ca\,{\sc ii}]) in optical nebular spectra, as suggested by \cite{Mazzali2001,Mazzali2005, Maeda2002,Maeda2006, Taubenberger2009}. Among SESNe, observational signatures indicate a higher level of asphericity in SNe connected to gamma-ray bursts (GRB-SNe) than normal ones \citep{Maeda2008}. Therefore, understanding explosion/ejecta geometry is vital in probing the nature of possible progenitor and explosion mechanisms for these energetic events.

In the literature, based on early photometric and spectroscopic optical studies up to $\sim$+140 d, \cite{Takaki2013} claimed SN~2012au had properties similar to hypernovae, whereas, based on the comparatively higher absolute magnitude of SN~2012au, \cite{Milisavljevic2013} also suggested it as a possible golden link between SLSNe~I and low-luminosity counterparts. Based on radio and $X$-ray observations \cite{Kamble2014} suggested a smooth circumstellar environment around SN~2012au. However, using a very late-time spectrum (after 6.2 years of explosion), \cite{Milisavljevic2018} did not observe any signature of CSMI in the very late-time spectrum (at +2270 d) of SN~2012au and claimed it as a pulsar wind nebula remnant. In general, SLSNe~I are $\sim$2-3 magnitudes brighter and have roughly three times broader light-curve peaks than classical SNe \citep{Gal-Yam2019, Inserra2019}. However, investigations of a larger photometric sample suggest that SNe~Ib/c, Ic-BL, and SLSNe-I display a continuum in luminosity distribution \citep{DeCia2018}. The near-maximum spectra of SLSNe~I have unique dominant O\,{\sc ii} features, but late-phase spectra exhibit a resemblance with SNe~Ib/c \citep{Pastorello2010, Quimby2018}. For this reason, SLSNe~I and SESNe are expected to share similar powering mechanisms, although with diverse physical parameters for the former, such as comparatively higher $M_{ej}$ or re-shocking of the ejecta by a central engine \citep{Nicholl2015}. The observational features of SN~2012au further prompted re-investigation using new sets of optical imaging photometric/polarimetric and the spectroscopic data and modelling resources used in the present study (e.g. {\tt MINIM}: \citealt{Chatzopoulos2013}; {\tt MESA}: \citealt{Paxton2018}; {\tt STELLA}: \citealt{Blinnikov2006}; {\tt SNEC}: \citealt{Morozova2015}; {\tt SYNAPPS}: \citealt{Thomas2011}).

\begin{figure*}
\includegraphics[scale=0.32]{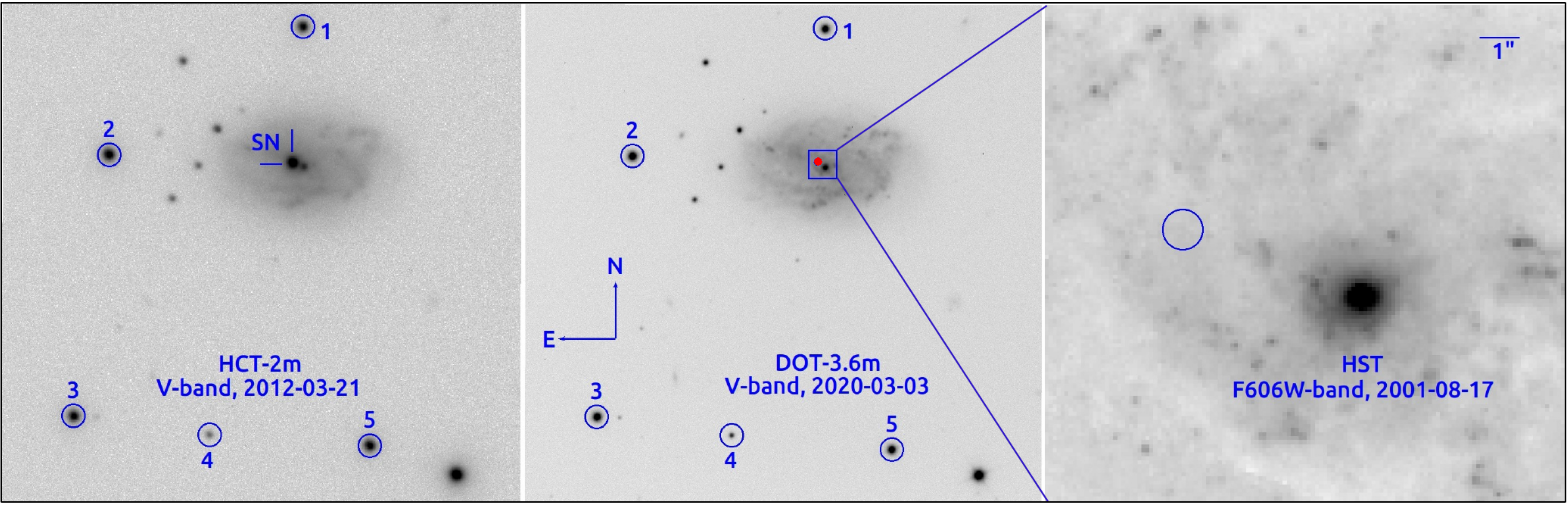}
\caption{Identification chart of SN~2012au and local secondary stars (IDs 1--5). The $V$-band images obtained using HCT-2m and DOT-3.6m taken on 2012 March 21 and 2020 March 3 are shown in the left and middle panels, respectively. The field of view is roughly 3.5 $\times$ 3.5 arcmin$^2$, and both images are astrometrically matched. The right panel shows a pre-explosion HST image observed on 2001 August 17, zoomed-in near the central region of the host galaxy NGC~4790. The SN~2012au location is marked in the left panel, and the same location is also indicated in the middle panel with a red dot after the SN has faded. The same location is shown with a circle (0.5-arcsec radius) in the pre-explosion HST image. North is up, and east is to the left.}
\label{fig:find-chart}
\end{figure*}

The presentation of the analysis is structured as follows. Multiband optical light curves of SN~2012au are presented in Section~\ref{sec:Multiband}. Section~\ref{sec:MINIM} discusses semi-analytic model fitting to the bolometric light curve of SN~2012au. A polarimetric study is presented in Section~\ref{sec:polarimetry}. In Section~\ref{sec:spectra}, we present the optical and NIR spectroscopic evolution and estimation of basic parameters of SN~2012au using the nebular phase spectrum and also compare the line velocities of SN~2012au with those of other SNe~Ib. Spectral comparisons of SN~2012au with other SESNe are discussed in Section~\ref{sec:comwithother}. The {\tt MESA} modelling of the progenitor is presented in Section~\ref{sec:MESA}. We discuss our results in Section~\ref{sec:Descussion} and conclude in Section~\ref{sec:CONclusion}. Throughout the work, the phase is computed with reference to the $B$-band maximum light, and magnitudes are expressed using the Vega scale.

\section{Light Curves of SN~2012au} \label{sec:Multiband}

\begin{figure*}
\includegraphics[angle=0,scale=0.9]{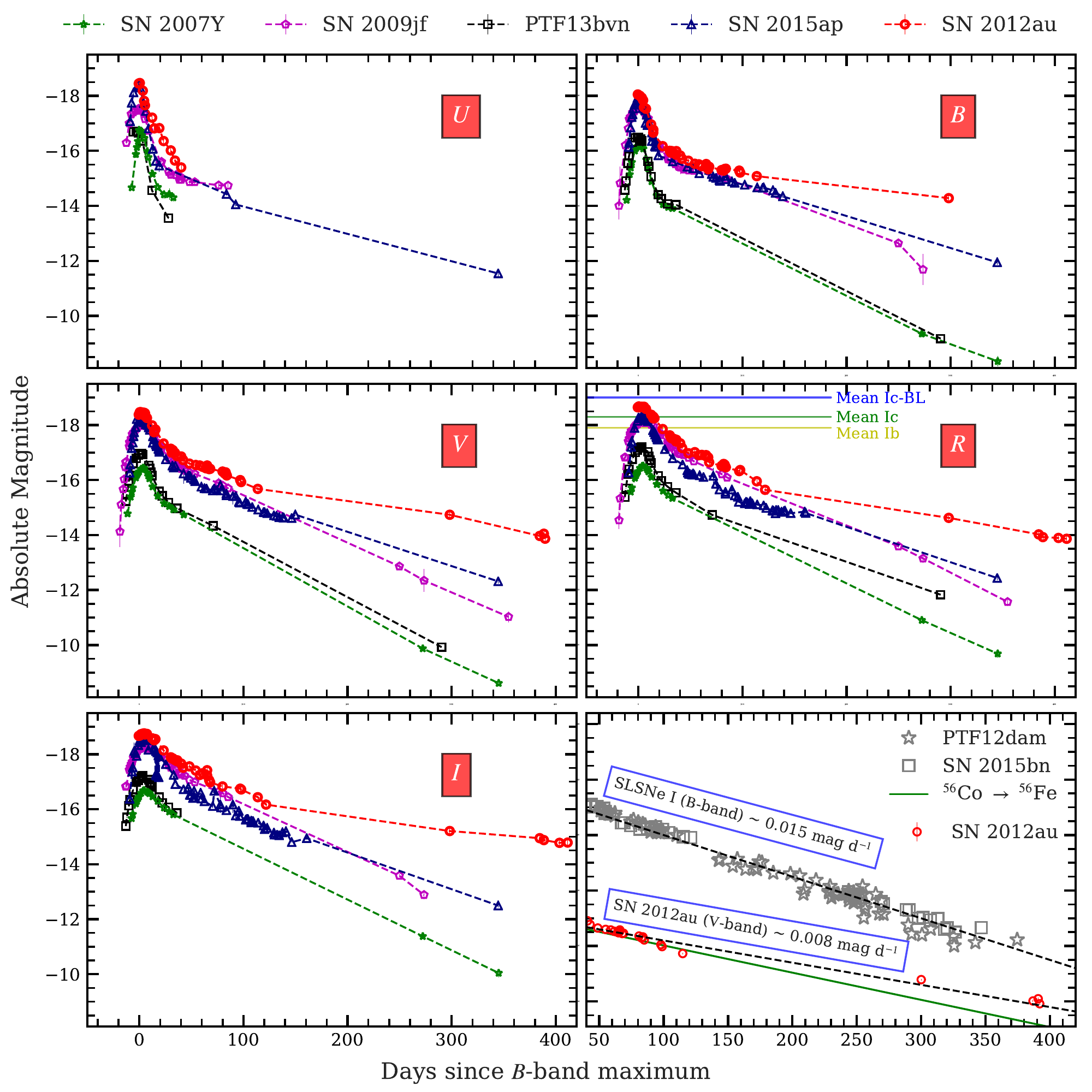}
\caption{Multiband light-curve evolution of SN~2012au and other SNe~Ib. The SN~2012au light curve peaks earlier in the bluer bands, as observed in other similar SNe \citep{Taddia2018}. SN~2012au is found to be one of the highly luminous SNe~Ib, with a shallower post-peak decay rate in comparison with other SNe~Ib. In the lower-right panel, the \textit{V}-band post-peak decay rate of SN~2012au is compared with \textit{B}-band late-time light curves of two low-redshift slow-decaying SLSNe~I (PTF12dam and SN~2015bn) and the \isotope[56]{Co} $\rightarrow$ \isotope[56]{Fe} decay curve.}
\label{fig:multiband}
\end{figure*}

The photometric monitoring of SN~2012au (J2000: $\alpha$ = $12^{\rm h}54^{\rm m}52\fs18$ and $\delta$ = $-10\degr 14\arcmin 50\farcs2$) was carried out using several facilities, including the Sampurnanand Telescope (ST-1.04m\footnote{\label{note1}\url{https://aries.res.in/index.php/facilities/astronomical-telescopes}}), Devasthal Fast Optical Telescope (DFOT-1.3m$^{\ref{note1}}$), Himalayan Chandra Telescope (HCT-2m\footnote{\url{https://www.iiap.res.in/iao/cycle.html}}), 2.2m telescope on Calar Alto (CAHA-2.2m\footnote{\url{http://www.caha.es/observing-mainmenu-148/telescopes-aamp-instruments-mainmenu-155}}), and Big Telescope Alt-azimuth (BTA-6m\footnote{\url{https://www.sao.ru/tb/tcs/}}). 

Our $U$-,$B$-,$V$-,$R$-, and $I$-band observations at 44 epochs spanned $\sim$\,414 d. Data reduction was performed through standard procedures described in \cite{Kumar2021}. Photometric calibrations were made using the secondary standards in the SN~field, through independent observations of the Landolt standard fields with HCT-2m and with the recently commissioned Devasthal Optical Telescope (DOT-3.6m\footnote{\label{note1}\url{https://aries.res.in/facilities/astronomical-telescopes/36m-telescope}}) operated by ARIES, Nainital, India \citep{Kumar2018}. The PG~0918 and SA~110 fields were observed with HCT-2m on 2012 March 21 and 2012 April 21. Similarly, on 2020 March 3, observations of the PG~1323 field were performed with the 4K $\times$ 4K CCD Imager mounted at DOT-3.6m \citep{Pandey2018} to fine-tune the photometric calibrations. Using the Landolt standards, transformations to the standard system were derived by applying average colour terms and photometric zero-points. Average atmospheric extinction values in different bands for the Hanle and Devasthal sites were adopted from \cite{Stalin2008} and \cite{Mohan1999}, respectively. Five secondary standard stars were used to calibrate the SN magnitudes. A finding chart indicating these stars (IDs 1--5) and the location of SN~2012au is shown in the left and middle panels of Fig.~\ref{fig:find-chart}, and their magnitudes are listed in Table~\ref{Ap:table1}. A zoomed view of the SN~2012au field (about 0.7$\times$0.7 arcmin$^2$) as observed on 2001 August 17 in the F606W filter using the Hubble Space Telescope ($HST$) is also shown in the right panel of Fig.~\ref{fig:find-chart}. The SN~location is near the bright nucleus of the host galaxy. Consequently, the host contamination must be removed to infer the uncontaminated SN~flux. In this study, template images of the field (without the supernova) were subtracted to obtained the final calibrated magnitudes. For this purpose, template images were acquired on 2014 February 26 with HCT-2m. All the images were first aligned with the GEOMAP and GEOTRAN tasks in {\sc IRAF}. Then, {\sc IRAF}-based scripts were used to perform template subtraction and intensity matching of the template images to the frames with the SN. It is also worth mentioning that the photometry published by \cite{Takaki2013} was published without applying image subtraction. The marginal differences among photometry of \textit{BVRI} filters can be explained in terms of possible contamination by the host, based on our present careful analysis. Finally, aperture photometry was performed on the subtracted images, and the calibrated SN~magnitudes are listed in Table~\ref{Ap:table2}.

The light curves of SN~2012au (in red) thus derived in the \textit{U}, \textit{B}, \textit{V}, \textit{R}, and \textit{I} bands are presented in Fig.~\ref{fig:multiband}. We correct the data of SN~2012au for a Galactic extinction of $E(B -V)$ = 0.043 mag \citep{Schlafly2011} and a host galaxy extinction of $E(B -V)$ = 0.02 $\pm$ 0.01 mag. The host galaxy extinction is adopted from \cite{Milisavljevic2013}, calculated using the equivalent width of Na\,{\sc i} D absorption; however, \cite{Takaki2013} assumed negligible host extinction. In the present analysis, the data have also been corrected for cosmological expansion. We derived absolute magnitudes from the extinction-corrected apparent magnitudes using

\begin{equation}\label{eq:mag}
M = m - 5\,\log(d_L/10~{\rm pc}) + 2.5\,\log(1+z)
\end{equation}

Here, $M$ corresponds to the absolute magnitude, $m$ represents the apparent magnitude, and $d_L$ is the luminosity distance \citep{Hogg2002}. We adopt a distance to the host NGC~4790 of $\approx$\,23.5 $\pm$ 0.5 Mpc, as given by \cite{Milisavljevic2013}. 

\begin{table*}
\scriptsize
  \begin{center}
    \caption{List of four SNe~Ib and two SLSNe~I along with SN~2012au, and the following parameters: MJD$_{B,{\rm peak}}$, $M_{B,{\rm peak}}$, $\delta$m$_{40-400}^*$, M$_{ej}$, E$_k$, M$_{O}$, and M$_{ZAMS}$.}
    \label{tab:tablecomp}
    \addtolength{\tabcolsep}{-4pt}
    \begin{tabular}{cccccccccccccc} 
    \hline \hline
       Object &  Redshift & Distance & $E(B-V)$ & MJD$_{B,{\rm peak}}$ & $M_{B,{\rm peak}}$& $\delta$m$_{40-400}$ & M$_{ej}$ & E$_k$ & M$_{O}$  & M$_{ZAMS}$ & Source \\

      $ $ & ($z$) & (Mpc) & (mag; total)& & (mag) & $(\frac{mag}{10 d})$ & (M$_\odot$) & (10$^{51}$ erg) & (M$_\odot$) & (M$_\odot$) &  \\
      \hline

       SN~2007Y & 0.0046 & 19.3 & 0.11 & 54161.4 & $-$16.22 $\pm$ 0.02 & 0.18 & 0.4 & 0.1  & 0.2 &3.3 & \cite{Stritzinger2009}\\

       SN~2009jf & 0.0079 & 34.2 & 0.11 & 55119.2 & $-$17.57 $\pm$ 0.02 & 0.15 & 4$-$9 & 3$-$8  & 1.3 & 20$-$25 &\cite{Sahu2011} \\

      &   &    &   &   &   &  & &   &   &   & \cite{Valenti2011} \\

      iPTF13bvn& 0.0045 & 22.5 & 0.04 & 56474.2 & $-$16.50 $\pm$ 0.02 & 0.19 & 1.9 & 0.9  & 0.7 & 15$-$17 &\cite{Folatelli2016}\\

        &    &   &   &   &  & &   &   &   & & \cite{Fremling2016} \\

      SN~2015ap & 0.0114 & 45.8 & 0.04 & 57283.4 & $-$17.89 $\pm$ 0.08 & 0.11 & 3.8 & 5.0  & 0.9 & 12$-$20 & \cite{Prentice2019}\\

      &   &    &   &   &   &   & &     &   & & \cite{Gangopadhyay2020} \\
      
      PTF12dam & 0.1070 & 482.2 & 0.03 & 56093.7 & $-$21.68 $\pm$ 0.07 & 0.15 & 13 & 3.0  & --- & $>$60 & \cite{Nicholl2013}\\      
     
      SN~2015bn & 0.1140 & 529.0 & 0.32 & 57103.4 & $-$23.21 $\pm$ 0.10 & 0.15 & 15 & 3.4 & 9 & $>$40 & \cite{Nicholl2016}\\      

    SN~2012au & 0.0045 & 23.5 &  0.06 & 56006.4 & $-$18.06 $\pm$ 0.12 & 0.08 & 4.7$-$8.3 & 4.8$-$5.4 & 1.6 & 17$-$25 & This work\\
      \hline
    \end{tabular}
    \begin{tablenotes}[para,flushleft]
    $^*$Decay rate in mag per 10 d has been measured by linear fitting to the light curve from $\sim$+40 to +400 d.
    \end{tablenotes}
  \end{center}
\end{table*}

A third-order spline function was fitted around the approximate peak of the $B$-band light curve to measure the date of \textit{B}-band maximum (MJD$_{B,peak}$ = 56006.4 $\pm$ 0.5) and the corresponding absolute peak magnitude ($M_{B,{\rm peak}} = -18.06 \pm 0.12$ mag). In the \textit{R} band, SN~2012au reached a peak brightness ($m_{R,peak}$ = 13.18 $\pm$ 0.02 mag) at MJD$_{R,peak}$ = 56009.2 $\pm$ 0.5, with an absolute peak magnitude M$_{R,peak}$ = $-$18.67 $\pm$ 0.11 mag. The M$_{R,peak}$ of SN~2012au appears to be brighter in comparison with the mean M$_{R,peak}$ calculated for SNe~Ib and Ic, whereas it is lower in comparison with SNe~Ic-BL \citep[][see the middle right panel of Fig.~\ref{fig:multiband}]{Drout2011}. It indicates that SN~2012au is one of the brightest SESNe. SN~2012au has been also observed in $J$, $H$, and $K$ bands using Son OF ISAAC (SOFI) on the New Technology Telescope (NTT-3.58m\footnote{\label{note1}\url{https://www.eso.org/public/teles-instr/lasilla/ntt/}}) at 56001.4 ($\sim -$5 d) and the apparent magnitudes are 13.40 $\pm$ 0.06, 13.33 $\pm$ 0.10, and 12.69 $\pm$ 0.20 mag, respectively. 

SN~2012au seems to have comparatively shallower post-peak decay rate (see Fig.~\ref{fig:multiband}) in all observed passbands in comparison with SNe~Ib. From the peak to +40 d, the $U$-band light curve decays by $\sim$0.076 mag d$^{-1}$, whereas the evolution is flatter in the redder bands ($\sim$0.029 mag d$^{-1}$ in the \textit{I} band). At late phases (+40 to 400 d), the \textit{BVRI} light curves decay at $\sim$0.008 mag d$^{-1}$, marginally slower than the $\rm ^{56}Co$ $\rightarrow$ $\rm ^{56}Fe$ decay rate (0.0098 mag d$^{-1}$), see the bottom right panel of Fig.~\ref{fig:multiband}. 

The \textit{U}, \textit{B}, \textit{V}, \textit{R}, and \textit{I} light curves of SN~2012au are compared with those of SN~2007Y \citep{Stritzinger2009}, SN~2009jf \citep{Sahu2011, Valenti2011}, iPTF13bvn \citep{Folatelli2016, Fremling2016}, and SN~2015ap \citep{Gangopadhyay2020, Prentice2019}. For comparison, the four SNe~Ib are chosen as they have peak coverage and late-time photometric data ($\gtrsim$+250 d) in at least five bands (\textit{UBVRI}); see Fig.~\ref{fig:multiband}. SN~2007Y has data in the \textit{u$'$g$'$BVr$'$i$'$} bands, and these were transformed to the \textit{UBVRI} bands using the equations given by \cite{Jordi2006}. The MJD$_{B,{\rm peak}}$ values for the four SNe~Ib are estimated independently (see Table~\ref{tab:tablecomp}), and the light curves (in the unit of absolute magnitudes) are obtained using Equation~\ref{eq:mag}. SN~2012au exhibits the brightest $M_{B,{\rm peak}}$ in comparison with other events. At later epochs ($\sim$+40 to +400 d), SN~2012au light curves decay more slowly than the sampled SNe~Ib, see Table~\ref{tab:tablecomp}. In the bottom right panel of Fig.~\ref{fig:multiband}, the \textit{V}-band post-peak ($\sim$+40 to +400 d) decay of SN~2012au is compared with the $B$-band post-peak decay rates of two well-studied slow-decaying SLSNe~I: PTF12dm \citep{Nicholl2013} and SN~2015bn \citep{Nicholl2016}, which have late-time photometric observations. During +40 to +400 d, SN~2012au presents an even slower decay rate (0.008 mag d$^{-1}$) than PTF12dam, SN~2015bn (0.015 mag d$^{-1}$), and a theoretical decay rate of \isotope[56]{Co} $\rightarrow$ \isotope[56]{Fe}. In summary, SN~2012au is one of the brightest SESNe~Ib with slower post-peak decay.

\begin{figure}
\includegraphics[width=\columnwidth]{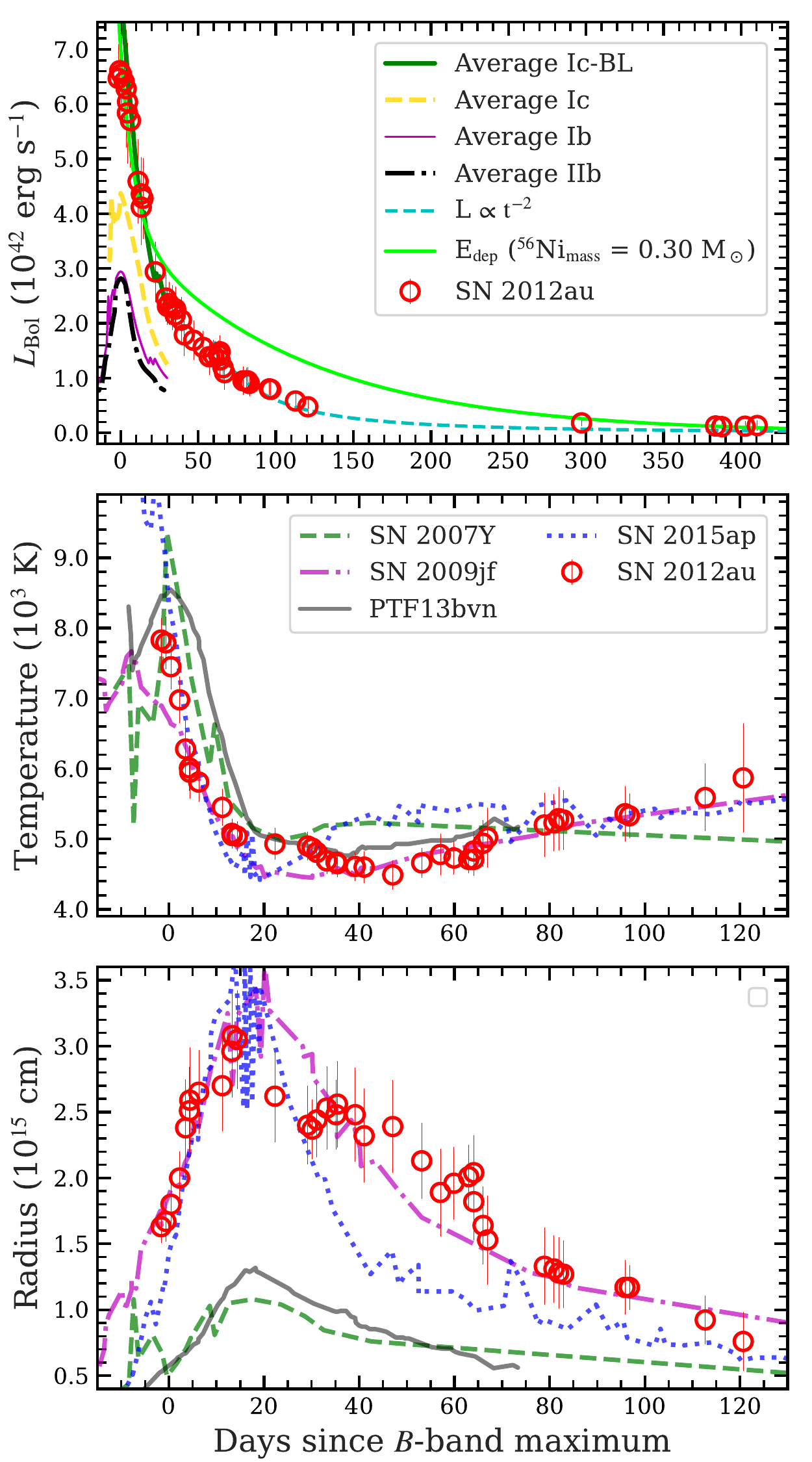}
\caption{Top panel: bolometric light curve of SN~2012au, shown along with the average bolometric light curves of SNe~IIb, Ib, Ic, and Ic-BL \citep{Lyman2016}. The full deposition rate of $\rm ^{56}Ni$ $\rightarrow$ $\rm ^{56}Co$ $\rightarrow$ $\rm ^{56}Fe$ decay chain, assuming $M_{Ni} = 0.30 M_\odot$, is also shown in a lime colour. A standard equation of magnetic dipole ($L~\varpropto$ $t^{-2}$) is also overplotted on the late-time bolometric light curve of SN~2012au (in cyan).} Middle and bottom panels: $T_{BB}$ and $R_{BB}$ evolution of SN~2012au, obtained through a BB fit to the photometric data. $T_{BB}$ and $R_{BB}$ evolution of SN~2012au is close to that inferred for SN~2009jf, another bright SN~Ib \citep{Sahu2011}.
\label{fig:bolotempradifig}
\end{figure}

\subsection{Bolometric light curve of SN~2012au} \label{sec:bolometric}

The quasi-bolometric (\textit{W1W2M2UBVRI}) light curve of SN~2012au was generated using the PYTHON-based code {\tt Superbol} \citep{Nicholl2018a}. {\it Swift}-UVOT data of SN~2012au in the \textit{W1W2M2} bands (from $-$4 to +32 d) were adopted from \cite{Milisavljevic2013}. Most of the data points were available in the $R$ band, while data points at similar epochs in the other bands were obtained through interpolation or extrapolation, assuming a constant colour. To add the NIR flux contribution, we extrapolated the blackbody (BB) spectral energy distribution (SED) by integrating the observed UV$-$optical fluxes. The data points have also been dereddened using $E(B-V) = 0.063$ mag, whereas the flux and wavelength values in individual bands were reported relative to the rest frame. In this way, we obtained the bolometric light curve (UV to NIR) from $-$0.2 to +413 d, with a peak luminosity of $\sim (6.56 \pm 0.70) \times 10^{42}$ erg s$^{-1}$ at MJD $\approx$ 56005.8 $\pm$ 1.0, which is higher in comparison with the estimated mean peak luminosity for SNe~IIb, Ib, and Ic, however lower than SNe Ic-BL \citep{Lyman2016}; see the upper panel of Fig.~\ref{fig:bolotempradifig}. The peak bolometric luminosity of SN~2012au is also higher than all SNe~IIb, Ib, and Ic (except SN~2018cbz) reported by \cite{Prentice2019}.

As discussed above, the late-time decay rate of the SN~2012au light-curve is even shallower than the theoretical decay curve of \isotope[56]{Co} $\rightarrow$ \isotope[56]{Fe} (see Fig.~\ref{fig:multiband}). It is known that the luminosity of a magnetar-powered engine decays as $L~\varpropto$ $t^{-\alpha}$ with $\alpha$ = 2, a standard equation of a magnetic dipole \citep{Kasen2010, Woosley2010}, see also \citealt{Nicholl2018b}. The late-time bolometric light curve of SN~2012au seems to trace the pattern well (in cyan colour, see the upper panel of Fig.~\ref{fig:bolotempradifig}) and indicate a magnetar origin.

We estimated the value of synthesized $\rm ^{56}Ni$ mass ($M_{\rm Ni}$) of $\sim 0.27 \pm$ 0.07~M$_\odot$ using the relation between $M_{peak}$ and $M_{Ni}$ given by \cite{Lyman2016}. We also calculated $M_{Ni}$ $\sim0.30~M_\odot$, through a comparison with the synthetic light curve of total energy production assuming a radioactive engine powered (in lime colour, see upper panel of Fig.~\ref{fig:bolotempradifig}) by the decay of the $\rm ^{56}Ni$ $\rightarrow$ $\rm ^{56}Co$ $\rightarrow$ $\rm ^{56}Fe$ decay chain \citep{Nadyozhin1994}, closer to that estimated above. The estimated $M_{Ni}$ for SN~2012au in this study ($\sim$\,0.30 M$_\odot$) is in good agreement with that suggested by \citet[][$\sim$0.3 $M_\odot$]{Takaki2013}.

Using the bolometric light curve of SN~2012au, we estimated $M_{ej}$, $E_{k}$, and the characteristic time-scale $(T_0)$ using equations 1, 3, and 5, respectively \citep{Wheeler2015}. We adopted an electron-scattering opacity ($\kappa$) = 0.05 cm$^2$ g$^{-1}$ and opacity to gamma rays ($\kappa_\gamma$) = 0.03 cm$^2$ g$^{-1}$ for the present analysis. To calculate the parameters discussed, a rise time ($t_r$) = 16.3 $\pm$ 1.0 d was estimated from the date of explosion \citep{Milisavljevic2013} and the peak bolometric luminosity, while the photospheric velocity at peak luminosity ($v_{ph,peak}$ $\sim$\,12 500 $\pm$ 500 km s$^{-1}$) was adopted from the spectral analysis (see Section~\ref{sec:velocomp}). Using the above values of $t_r$ and $v_{ph,peak}$, we obtained $M_{ej}$, $E_{k}$, and $T_0$ of $\sim$5.1 $\pm$ 0.7 $M_\odot$, $\sim$\,4.8 $\pm$ 0.6 $\times$ 10$^{51}$ erg, and $\sim$66.0 $\pm$ 9.4 d, respectively. The values of $M_{ej}$ and $E_{k}$ estimated for SN~2012au appear to be close to those found for SN~2009jf, whereas they are larger than those inferred for other SNe~Ib listed in Table~\ref{tab:tablecomp}, indicating SN~2012au as one of the most highly energetic SN~Ib. The $M_{\rm ej}$ value for SN~2012au is higher than those of SNe~IIb, Ib, and Ic discussed by \cite{Wheeler2015} and also higher than the mean values estimated for different samples of SNe~IIb, Ib, Ic, and Ic-BL by \cite{Drout2011}, \cite{Lyman2016}, and \cite{Prentice2019}, whereas it is nearly equal to those estimated for a sample of SNe~Ic and SNe~Ic-BL by \cite{Taddia2015}. The evaluated $E_{k}$ of SN~2012au is higher than the mean values calculated for a sample of SNe~Ib ($\sim$1.2 $\times$ 10$^{51}$ erg) and Ic ($\sim$1.0 $\times$ 10$^{51}$ erg), but lower in comparison with Ic-BL ($\sim$11 $\times$ 10$^{51}$ erg) and engine-driven SNe ($\sim$9 $\times$ 10$^{51}$ erg), reported by \cite{Drout2011}.

The evolution of BB temperature ($T_{BB}$) and the radius ($R_{BB}$) of SN~2012au are presented in the middle and lower panels of Fig.~\ref{fig:bolotempradifig}, respectively. The $T_{BB}$ and $R_{BB}$ parameters were calculated by modelling the SED at some epochs by fitting a BB function using the {\tt Superbol} code \citep{Nicholl2018a}. The $T_{BB}$ and $R_{BB}$ evolution are shown only up to +114 d, because the BB approximations are poorer at later epochs. From peak to $\sim$+12 d, the $T_{BB}$ of SN~2012au decreased from $\sim 8000$ to 4700 K, whereas from $\sim$+12 to +60 d, it remains nearly constant at around 4600 K. At later epochs (after $\sim$+60 d), $T_{BB}$ increases slowly from $\sim 4600$ to 6000 K. The $T_{BB}$ evolution of SN~2012au is in good agreement with that of the SN~Ib sample of \cite{Prentice2019}. On the other hand, from the peak to $\sim+12$ d, $R_{BB}$ of SN~2012au increases from $\sim1.4 \times 10^{15}$ to $3.0 \times 10^{15}$ cm, to decrease again after +114 d to $\sim 0.5 \times 10^{15}$ cm. The $T_{BB}$ and $R_{BB}$ evolution of SN~2012au is also compared with those estimated for SN~2007Y, SN~2009jf, iPTF13bvn, and SN~2015ap (see middle and lower panels of Fig.~\ref{fig:bolotempradifig}). The $T_{BB}$ and $R_{BB}$ values for iPTF13bvn and SN~2015ap are taken from \cite{Fremling2014} and \cite{Aryan2021}, respectively, whereas they are calculated independently for SN~2007Y and SN~2009jf using the {\tt Superbol} code \citep{Nicholl2018a}. Near the peak, SN~2012au exhibits lower temperature in comparison with SN~2007Y, iPTF13bvn, and SN~2015ap and higher than SN~2009jf, although, after $\gtrsim$+20 d all the SNe presented manifest a nearly similar trend of temperature evolution (see middle panel of Fig.~\ref{fig:bolotempradifig}). On the other hand, SN~2012au shares similar $R_{BB}$ evolution to SN~2009jf and is also closer to SN~2015ap, although throughout the evolution SN~2012au presents higher values of $R_{BB}$ in comparison with those estimated for SN~2007Y and iPTF13bvn (see lower panel of Fig.~\ref{fig:bolotempradifig}).

\subsubsection{Ejecta mass and kinetic energy} 
\label{sec:MejEk}

\begin{figure}
\begin{centering}
\includegraphics[width=\columnwidth]{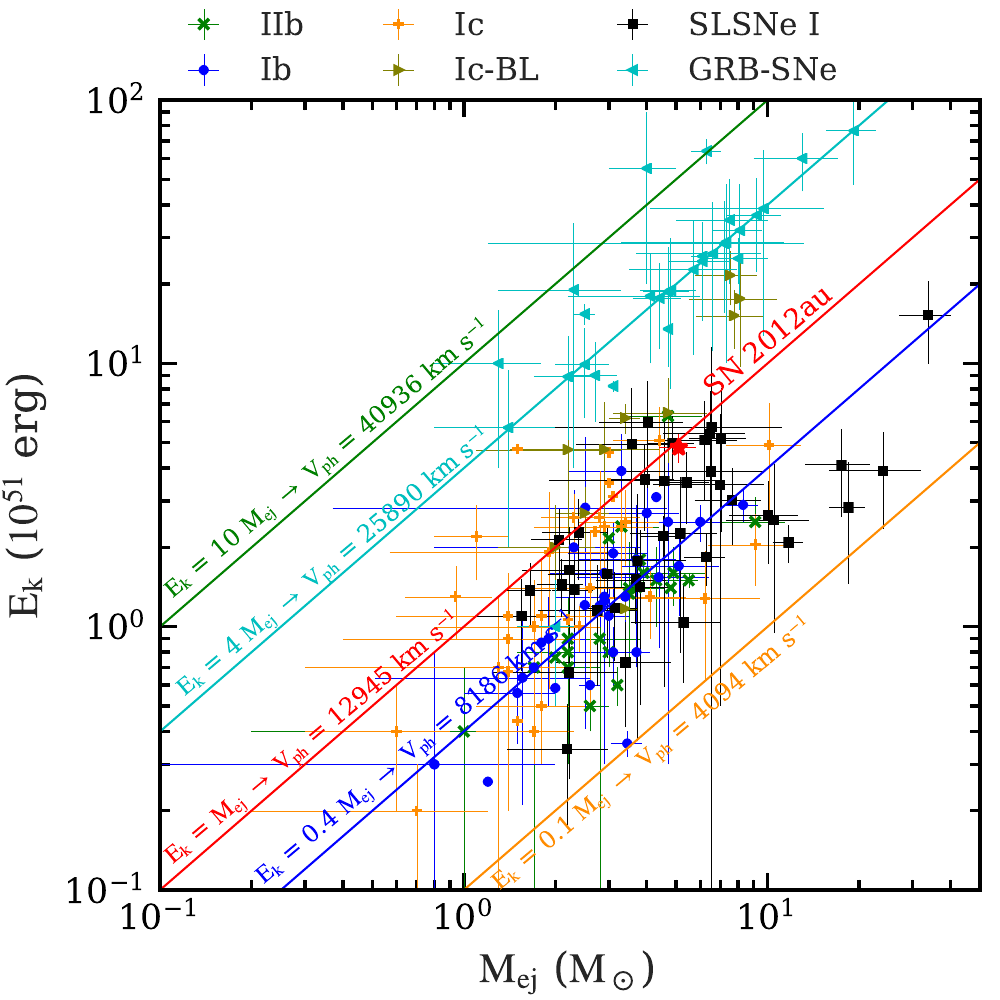}
\caption{$M_{ej}$ versus $E_{K}$ diagram for SN~2012au and a sample of SNe~IIb, Ib, Ic, Ic-BL, GRB-SNe, and SLSNe~I as discussed in Section~\ref{sec:MejEk}. SN~2012au appears to have higher $E_{K}$ and $M_{ej}$ in comparison with most SNe~IIb (in green), Ib (in blue), and Ic (in orange), however lower than the values for Ic-BL and GRB-SNe.}
\label{fig:Mej_Ek} 
\end{centering}
\end{figure}

The $M_{ej}$ and $E_{K}$ values of SN~2012au estimated using the bolometric light-curve analysis are compared with those of classical SNe~IIb (23), Ib (33), Ic (33), Ic-BL (10), GRB-SNe (30), and SLSNe~I (40) in Fig.~\ref{fig:Mej_Ek}. The parameters for SNe~IIb, Ib, Ic, and Ic-BL were taken from \cite{Taddia2015, Wheeler2015, Lyman2016, Taddia2018, Prentice2019} and those of GRB-SNe from \cite{Cano2017}, while those of SLSNe~I were from \cite{Nicholl2015,Nicholl2017}. The $E_K$ values were not quoted by \cite{Nicholl2015,Nicholl2017} and \cite{Prentice2019}, so we used equation 2 of \cite{Wheeler2015} to infer $E_K$ from $M_{ej}$ and photospheric velocity ($v_{ph}$). Different types of SNe are plotted with different legends and colours, while SN~2012au is plotted with a red star.

From Fig.~\ref{fig:Mej_Ek}, it is clear than SN~2012au has the highest $M_{ej}$ and $E_K$ values in comparison with most of the Type IIb (in green), Ib (in blue), and Ic (in orange). It also indicates that SN~2012au is one of the most energetic SNe~Ib. SNe~IIb, Ib, and Ic have a similar range of values for $E_K$ and $M_{ej}$, while SNe~Ic-BL seem to have comparatively high $E_K$ values, and GRB-SNe exhibit the highest $E_K$ values. The $E_K$ estimates for SLSNe~I are similar to those for SNe~IIb, Ib, and Ic, but they have a larger range of ejecta masses. In Fig.~\ref{fig:Mej_Ek}, ratios of $E_K$/$M_{ej}$ (from 0.1 to 10) for corresponding $v_{ph}$ values (from $\sim$4000 to 41 000 km s$^{-1}$) are also shown in different colours. Most GRB-SNe converge to around $v_{ph}$ $\sim$ 25 000 km s$^{-1}$, whereas SNe~IIb, Ib, and Ic have a lower $v_{ph}$ range (6000$-$13 000 km s$^{-1}$). Most SNe~IIb, Ib, and Ic lie below the $v_{ph}$ curve of $\sim$13 000 km $s^{-1}$ (red line), which shows comparatively higher $v_{ph}$ for SN 2012au.

\begin{figure*}
\includegraphics[angle=0,scale=1.1]{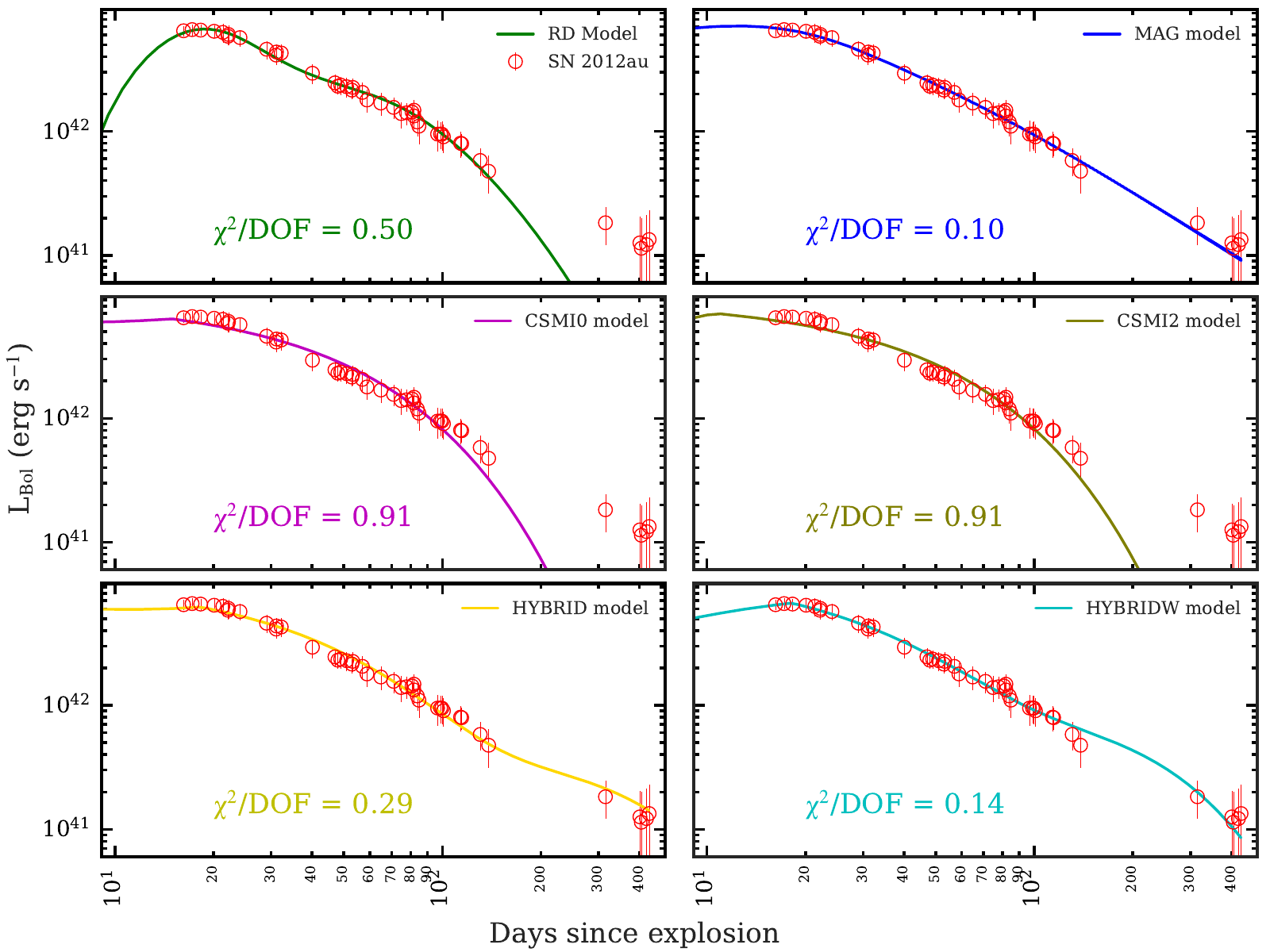}
\caption{Semi-analytic light curve models (RD, MAG, CSMI0, CSMI2, HYBRID, and HYBRIDW) fitted to the bolometric light curve of SN~2012au using the {\tt MINIM} code \citep{Chatzopoulos2013}, shown in six different panels. The photometric data points of SN~2012au are shown by hollow red circles, whereas the six modelled light curves are colour-coded with the best-fitting modelled light curves.}
\label{fig:minim}
\end{figure*}

\begin{table*}
\scriptsize
 \begin{center}
  \begin{threeparttable}
    \caption{Best-fitting parameters for the RD, MAG, CSMI0, CSMI2, HYBRID, and HYBRIDW models using the {\tt MINIM} code as described in \citet{Chatzopoulos2013}. The best fit parameters obtained from the MAG model (the most suitable one for SN~2012au) are shown in bold.}
    \label{tab:minim}
    \addtolength{\tabcolsep}{-2pt}
    \begin{tabular}{c c c c c c c c c c c}

    \hline \hline
          &  &  &  & \textbf{RD model}& & & & & \\

         $A_{\gamma}$ & $M_{\rm Ni}$ & $t_d$ &  & & & & $M_{\rm ej}$ & $\chi^2$/DOF &\\

         & ($M_\odot$) & (d) &  &  & & & ($M_\odot$) & & \\
      \\

     99.39 $\pm$ 17.96 & 0.24 $\pm$ 0.05 & 9.88 $\pm$ 3.40 & &  &  & &  1.14 $\pm$ 0.78 & 0.50 & \\

    \hline \hline
    
             &  & & & \textbf{MAG model}& & & & & \\

    $R_0$ & $E_p$ & $t_d$ & $t_p$ & $v_{\rm exp}$ & $P_i$ & $B$ &  $M_{\rm ej}$ & $\chi^2$/DOF & \\

    ($10^{13}$ cm) & ($10^{51}$ erg) & (d) & (d) & ($10^3$ km s$^{-1}$) & (ms) & ($10^{14}$ G) &  ($M_\odot$) & &\\
      \\

    \textbf{0.36} $\pm$ \textbf{0.13} &  \textbf{0.06} $\pm$ \textbf{0.01} & \textbf{20.83} $\pm$ \textbf{2.28} & \textbf{24.49} $\pm$ \textbf{2.12} & \textbf{11.66} $\pm$ \textbf{0.58}  &  \textbf{18.26} $\pm$ \textbf{0.01} & \textbf{8.05} $\pm$ \textbf{0.15} & \textbf{4.72} $\pm$ \textbf{1.03} & \textbf{0.10} &\\

    \hline \hline
    
               &  & & & \textbf{CSMI0 model}& & & & & \\
    
    $R_p$ & $M_{\rm ej}$ & $M_{\rm csm}$ & $\dot{M}$ & $M_{\rm Ni}$  & $v_{\rm exp}$ &  & & $\chi^2$/DOF &\\
    ($10^{13}$ cm) & ($M_\odot$) & ($M_\odot$) & ($M_\odot$ yr$^{-1}$) & ($M_\odot$) & ($10^3$ km s$^{-1}$) & & &\\
      \\

     0.10 $\pm$ 0.01 & 7.44  $\pm$ 0.31 & 1.57 $\pm$ 0.09 & 0.00012 $\pm$ 0.00001 & 0.0  $\pm$ 0.0 & 8.47 $\pm$ 0.16 & & & 0.91 &\\

    \\
              &  & & & \textbf{CSMI2 model}& & & & & \\    

     0.06 $\pm$ 0.03 & 6.77 $\pm$ 1.33 & 1.87 $\pm$ 0.24 & 0.47 $\pm$ 0.01 & 0.0 $\pm$ 0.0 & 8.22 $\pm$ 0.52 & & & 0.91 &\\
       
    \\
              &  & & & \textbf{HYBRID model}& & & & & \\

     0.11 $\pm$  0.01 & 8.00  $\pm$ 0.16 & 1.38 $\pm$ 0.03 & 0.0001  $\pm$ 0.00001  & 0.07 $\pm$ 0.01 & 7.96 $\pm$ 0.02 & & & 0.29 &\\

   \\
              &  & & & \textbf{HYBRIDW model}& & & & & \\    

     0.02 $\pm$  0.05 & 4.35  $\pm$ 0.90 & 1.64 $\pm$ 0.05 & 0.28  $\pm$ 0.02  & 0.09 $\pm$ 0.01 & 7.40 $\pm$ 0.19 & & & 0.14 &\\
    
    \hline 
    \end{tabular}
    \begin{tablenotes}[para,flushleft]
    \end{tablenotes}
  \end{threeparttable}
  \end{center}
\end{table*}

\section{Semi-analytical light curve modelling using {\tt MINIM}} 
\label{sec:MINIM}

The analytical model fitting to the bolometric light curve of SN~2012au was attempted using the {\tt MINIM} code \citep{Chatzopoulos2013}. MINIM is a semi-analytical light curve modelling technique used to fit the light curves of SESNe. It uses the Price algorithm to look for the global minimum of the $\chi^{2}$ hypersurface within the allowed parameter volume. Corresponding to the minimum  $\chi^{2}$, it provides a set of supernova explosion parameters including $M_{\rm Ni}$, $M_{\rm ej}$, ejecta opacity, etc. The working process of the {\tt MINIM} code is described in detail by \cite{Chatzopoulos2013}, see also \cite{Wheeler2017}. Using {\tt MINIM}, we attempted to fit the bolometric light curve of SN~2012au using various models like RD, MAG, constant density CSMI (CSMI0), wind-like CSMI (CSMI2), HYBRID (CSMI0 + RD), and HYBRIDW (CSMI2 + RD): see Fig.~\ref{fig:minim}, briefly described below.

RD model: the radioactive decay model was originally developed by \cite{Arnett1982} and it works on the assumptions discussed by \cite{Arnett1982}, \cite{Valenti2008}, \cite{Chatzopoulos2009}, and \cite{Chatzopoulos2012, Chatzopoulos2013}. Under this model, $\gamma$-rays produced by the \isotope[56]{Ni} $\rightarrow$ \isotope[56]{Co} $\rightarrow$ \isotope[56]{Fe} decay chain get trapped and thermalize the ejecta from centre to surface, powering the SN light curve. See equation 9 in \cite{Chatzopoulos2012} for the form of the resulting light curve based on $M_{Ni}$, initial time ($t_{ini}$), effective diffusion time-scale ($t_d$), and optical depth for $\gamma$-rays ($A_\gamma$) as fitting parameters. The $M_{ej}$ value is calculated using equation 10 of \cite{Chatzopoulos2012} assuming integration constant ($\beta$) = 13.8, $\kappa$ = 0.05 cm$^2$ g$^{-1}$ \citep{Drout2011, Milisavljevic2013}, and $v_{ph,peak}$ of = 12 500 $\pm$ 500 km s$^{-1}$, which was adopted from the spectral analysis (see Section~\ref{sec:velocomp}).

MAG model: under this model, a newly formed rapidly rotating millisecond magnetar (magnetized neutron star) is generated in the centre of the SN \citep{Maeda2007, Kasen2010, Woosley2010}. In this scenario, conversion of magneto-rotational energy into radiation causes the newly-born “magnetar” to spin down while simultaneously heating the SN ejecta. Refer to equation 13 in \cite{Chatzopoulos2012} for the form of the resulting light curve which is mainly based on the magnetic dipole spin-down formula. The fitting parameters of the MAG model are $t_{ini}$, progenitor radius ($R_0$), magnetar rotational energy ($E_p$), $t_d$, magnetar spin-down time-scale ($t_p$), and ejecta expansion velocity $v_{\rm exp}$. Similarly to the RD model, in the MAG model $M_{ej}$ is calculated using equation 10 of \cite{Chatzopoulos2012} assuming $\beta$ = 13.8, $\kappa$ = 0.05 cm$^2$ g$^{-1}$, and $v_{ph,peak}$ = 12 500 $\pm$ 500 km s$^{-1}$. The initial period of the magnetar ($P_i$) is given by $P_i = 10 \times (2 \times 10^{50}$erg s$^{-1}$ /$E_{p})^{0.5}$ ms and the magnetic field of the magnetar ($B$) is calculated using $B = 10^{14} \times (1.3 P_{10}^{2}/t_{p,yr})^{0.5}$ G; here, $t_{p,yr}$ is the magnetar spin-down time-scale in years \citep{Chatzopoulos2013}.

CSMI model: as we know, the circumstellar environment is not always a vacuum and the ejecta may be surrounded by a thick CSM shell generated by mass loss from the massive star. SN ejecta can then collide and interact violently with the CSM. Eventually, shock heating can cause a very bright light curve, governed by the diffusion of thermalized photons to the photosphere that is fixed at the outer radius of the shell \citep{Chevalier1994}. Here, CSMI0 and CSMI2 models are designated to represent constant density and wind-like CSM, respectively. See equations 14 and 20 in \cite{Chatzopoulos2012} for the form of the resulting light curves. The fitting parameters in the case of the CSMI model are progenitor radius before the explosion ($R_p$), $M_{ej}$, CSM mass ($M_{\rm csm}$), mass-loss rate ($\dot{M}$), and $v_{\rm exp}$.

HYBRID model: nowadays, many of the complex light curves of peculiar SNe are not easy to explain using the single powering sources discussed above. Hence more than one powering source is proposed, these are termed as ``HYBRID'' models \citep{Chatzopoulos2012, Chatzopoulos2013, Moriya2018, Chatzopoulos2019, Wang2019}. In the present study, we also tried to re-generate the bolometric light curve of SN~2012au using combinations of CSMI and RD models. Refer to equation 21 in \cite{Chatzopoulos2012} for the form of the resulting light curve of the CSM+RD model. The fitting parameters used in the HYBRID model are the same as those discussed in the CSMI model plus $M_{Ni}$. Details of all the models discussed above are given in \cite{Chatzopoulos2012,Chatzopoulos2013}.\\

All six models (RD, MAG, CSMI0, CSMI2, HYBRID, and HYBRIDW) discussed above could reproduce the bolometric light-curve of SN~2012au with $\chi^2$/DOF $\lesssim$1 (where DOF stands for degrees of freedom); see Fig.~\ref{fig:minim}. Here, we caution that $\chi^2$/DOF is used as an indicator for selecting the model parameters that fit the data best and not as a statistical probe for judging the significance of the models. In this situation, the best $\chi^2$ value alone cannot be used as a criteria to declare the most suitable model. Therefore, we considered various physical criterion to select different fitting parameters, as recently suggested in \cite{Kumar2021}. The best-fitting parameters thus obtained from the {\tt MINIM} modelling using the six models discussed are tabulated in Table~\ref{tab:minim}.

The RD model fits the light curve of SN~2012au nicely, but the $M_{\rm ej}$ is comparatively smaller than the value inferred from photometric analysis in Section~\ref{sec:bolometric}. For this reason, the RD model is excluded from the possible powering mechanisms for SN~2012au. In the case of CSMI0, CSMI2, HYBRID, and HYBRIDW model fittings, $v_{\rm exp}$ is significantly lower than that obtained from the spectral analysis (see Section~\ref{sec:spectrophpara}). In addition, in the case of CSMI2 and HYBRIDW models the $\dot{M}$ values are unreasonably high (0.47 and 0.28 $M_\odot$ yr$^{-1}$), so we exclude the possibility of dominance of the CSMI contribution to powering the SN~2012au light curve. The absence of narrow H Balmer lines in the late-time spectral observation (6.2 years after the explosion) also contradicts the CSMI as a dominant powering mechanism for SN~2012au \citep{Milisavljevic2018}. On the other hand, the MAG model fitted the SN~2012au very well (see upper-right panel of Fig.~\ref{fig:minim}) and gives reasonable values of $M_{\rm ej}$ ($\approx 4.72 \pm 1.03$ M$_\odot$) and $v_{\rm exp}$ ($\sim\,11.66 \pm 0.58 \times 10^3$ km s$^{-1}$), closer to those obtained from the photometric and spectral analysis. Hence, the MAG model has to be considered as the most suitable for SN~2012au. The value of $B$ ($\sim$8 $\times$ $10^{14}$ G) computed in this study for SN~2012au is consistent with that estimated for another Type Ib SN~2005bf by \cite{Maeda2007}. However, $P_i$ ($\sim$18 ms) for SN~2012au is larger in comparison with SN~2005bf ($\sim$10 ms). In addition, \cite{Milisavljevic2018} also suggested a pulsar wind nebula remnant for SN~2012au, which is in favour of a central engine powering source.

\section{Early-phase imaging Polarimetry of SN~2012au}
\label{sec:polarimetry}

We also conducted imaging polarimetric observations of SN~2012au in the $R$ band at six epochs, using the ARIES Imaging Polarimeter mounted at the Cassegrain focus of ST-1.04m. This polarimeter consists of a half-wave plate (HWP) modulator and a Wollaston prism beam-splitter, and the images are captured with a CCD camera (Tektronix 1024 pixels $\times$ 1024 pixels). Detailed information on the instrument can be found in \citet*{Rautela2004}. At each position of HWP position angle, multiple sets of frames were obtained, with individual exposure, times being between 10 and 20 minutes. Images at each position of the HWP were combined to obtain a good signal-to-noise ratio. Table~\ref{Ap:table3} gives the complete log of the observations. 

The ordinary and extraordinary fluxes were extracted by performing standard aperture photometry using {\sc IRAF}. The Stokes parameters were estimated at different apertures (between 2 and 8 pixels), and finally $P$ (degree of polarization) and $\theta$ (polarization angle) were computed for the apertures that fitted best with minimum chi-square. The detailed procedures used to estimate $P$ and $\theta$ can be found in \citet{Ramaprakash1998}, \cite{Rautela2004}, and \cite{Kumar2014,Kumar2016}. The polarization angle was corrected for zero-point polarization by observing various polarized standard stars from \citet*{Schmidt1992}. It should be noted that the instrumental polarization of ARIES imaging polarimeter ($AIMPOL$) has generally been found to be $\sim$0.1 per cent \citep[][and references therein]{Biman2007, Pandey2009, Eswaraiah2012, Kumar2014, Kumar2016, Kumar2019, Srivastav2016}. In this study, the degree of polarization has been corrected by this offset value.

\subsection{Estimation of SN~intrinsic polarization}\label{isp_cor}
The directional extinction arising from dust grains located in the SN~line of sight is a primary contaminating source that might influence the incoming SN~polarization signal. 
To understand the intrinsic polarization properties of a SN, the unwanted interstellar polarization (ISP) must be subtracted from the observed values. There is no direct formulation to estimate ISP precisely; however, it can be constrained by careful analysis of the nearby region.

\begin{figure}
\begin{centering}
\includegraphics[width=\columnwidth]{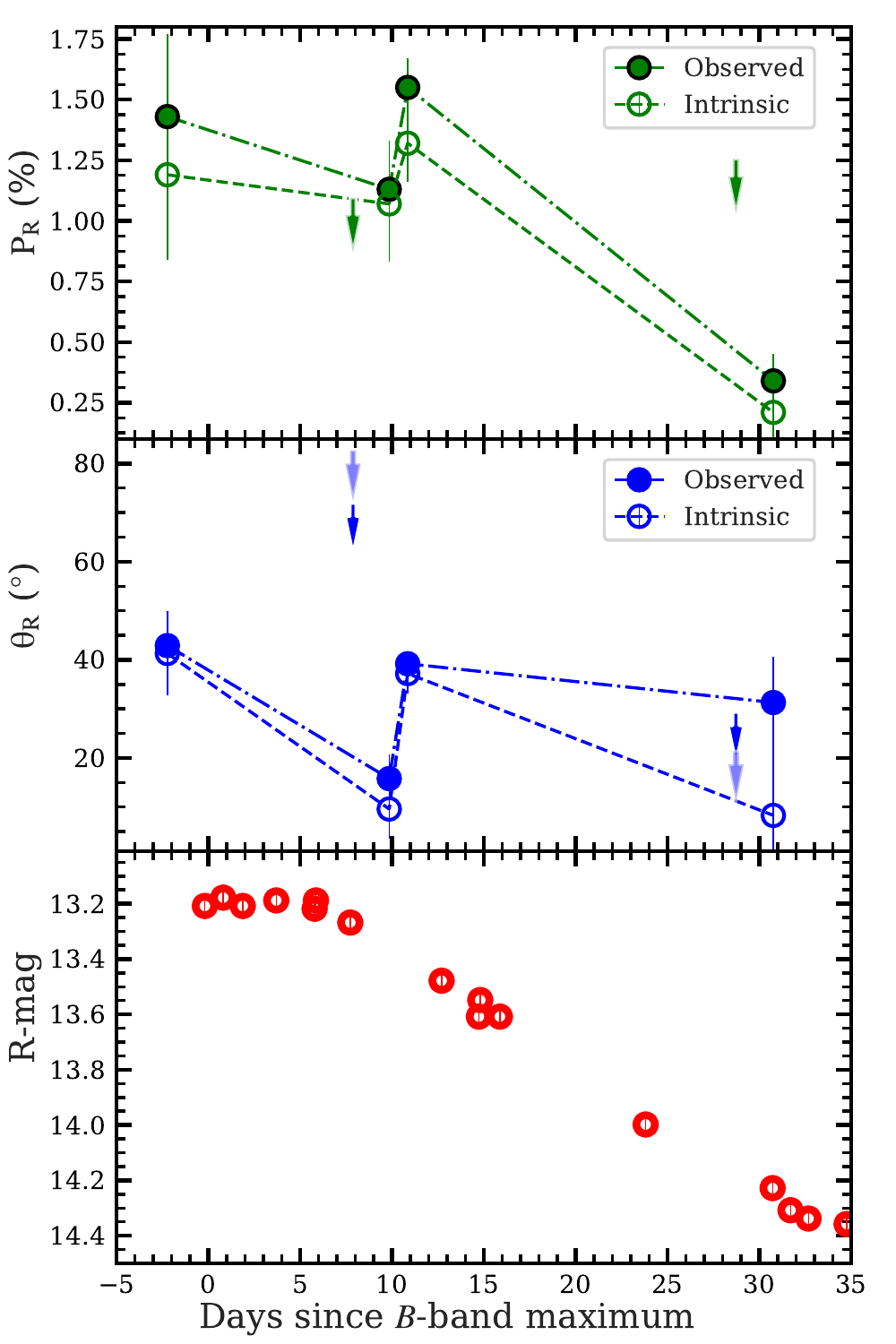}
\caption{Temporal evolution of the polarization parameters of SN~2012au is shown. The degrees of polarization and the polarization angles are plotted in the upper and lower panels, respectively. The limiting values at two more epochs are shown with arrows. The lower panel shows the evolution of the apparent $R$-band magnitude during the same temporal bin.}
\label{p_pa_evol}
\end{centering}
\end{figure}

To determine the Galactic ISP towards SN~2012au, we followed the procedures described in \citet{Kumar2014,Kumar2016,Kumar2019}. We performed $R$-band polarimetric observations of nine isolated field stars within a 10$\degr$ radius of the SN on 2013 January 20. The reference stars were selected from the SIMBAD database from among those that do not show variability and spectral emission lines. The distance information for the nine stars was collected from {\it Hipparcos} parallax \citep{Van2007} and {\it Gaia} data \citep[DR2;][]{GaiaCollaboration2018}. The observational details ($P_{R}$, $\theta_{R}$, and distance) are listed in Table~\ref{Ap:table4}. The $P_{R}$ and $\theta_{R}$ values of these stars were converted to their respective Stokes parameters (i.e., $Q$ and $U$), and weighted linear fits were performed on the distance versus $Q$ and $U$. To estimate the Stokes parameters arising due to ISP, the slopes and intercepts were fitted at the most distant ($\sim$870 pc) field star HD~112325. The corresponding $P_{ISP}$ and $\theta_{ISP}$ values are computed as $<P_{ISP}>$ = 0.23\,$\pm$\,0.01 percent and $<\theta_{ISP}>$ = 127.72$\degr$\,$\pm$\,1.09$\degr$, respectively.

According to \citet{Serkowski1975}, the ISP and Milky Way (MW) reddening might be correlated with the mean and maximum polarization efficiency relation, i.e. $P_{mean}= 5\times E(B-V)$ and $P_{max} = 9\times E(B-V)$. The $E(B-V)$ towards SN~2012au is 0.063 mag (see Section~\ref{sec:Multiband}). If we assume that the dust along SN~2012au follows a mean polarization efficiency, it corresponds to $P_{mean}$\,=\,0.32 per cent. This is consistent with the $P_{ISP}$ value estimated with the nine field stars. We also used the three-dimensional map of dust reddening for the MW as described in \citet{Green2019}. The map is based on Pan-STARRS and 2MASS photometry and the {\it Gaia} DR2 parallaxes. The best fit towards SN~2012au provides $E(B-V)$ = 0.03 mag and remains constant beyond a distance of 400 pc. Using this value, the polarization efficiency relations yield $P_{mean}$ and $P_{max}$ as 0.15 and 0.27 per cent, respectively. From the above exercises, it is inferred that the ISP upper limit is 0.32 per cent and the ISP estimated from the field stars is considered for further analysis in this work.

At each epoch, the intrinsic polarization parameters of SN~2012au were estimated through a vectorial subtraction of the Stokes parameters ($Q_{int}$\,=\,$Q_{obs}$--$Q_{ISP}$, $U_{int}$\,=\,$U_{obs}$--$U_{ISP}$). The resulting intrinsic Stokes parameters were converted to $P_{int}$ and $\theta_{int}$ using the relation given in \citet{Kumar2019}. The intrinsic (ISP-corrected) polarization parameters for SN~2012au are listed in Table~\ref{Ap:table3}.

\subsection{Polarimetric results and comparisons}

Polarimetric investigations of SESNe are important to constrain the level of asymmetry in the ejecta \citep{Tanaka2017}. Polarization in SNe emerges due to incomplete cancellation of the directional components of electrical vectors. However, if the source is spherically symmetric, these vectors cancel each other and result in zero polarization \citep[see][]{Leonard2005, Wang2008}. SNe~Ib/c events exhibit a higher degree of polarization than SNe~II, which indicates a departure from spherical symmetry \citep{Chugai1992, Hoflich2001, Maund2007, Tanaka2008, Stevance2019}. The geometry of the circumstellar material may also contribute to the polarization evolution \citep{Leonard2000, Hoffman2008, Mauerhan2014, Mauerhan2017}. Variation in the polarization parameters has been observed at different evolutionary phases of SESNe. Therefore, multi-epoch polarimetric observations are crucial to understand the SN geometry and the underlying physics better. Imaging polarimetric and spectro-polarimetric techniques are generally used to estimate the linear polarization, which includes both continuum and line polarization. Imaging polarimetry provides a general picture of the explosion geometry or ejecta behaviour; on the other hand, spectro-photometric analysis and interpretations can be more conclusive in constraining polarization of various lines and the 3D geometry of the SN ejecta. In this work, we mainly focus on imaging polarimetric observations of SN~2012au and other similar SNe and their comparisons.

\begin{figure}
\begin{centering}
\includegraphics[width=\columnwidth]{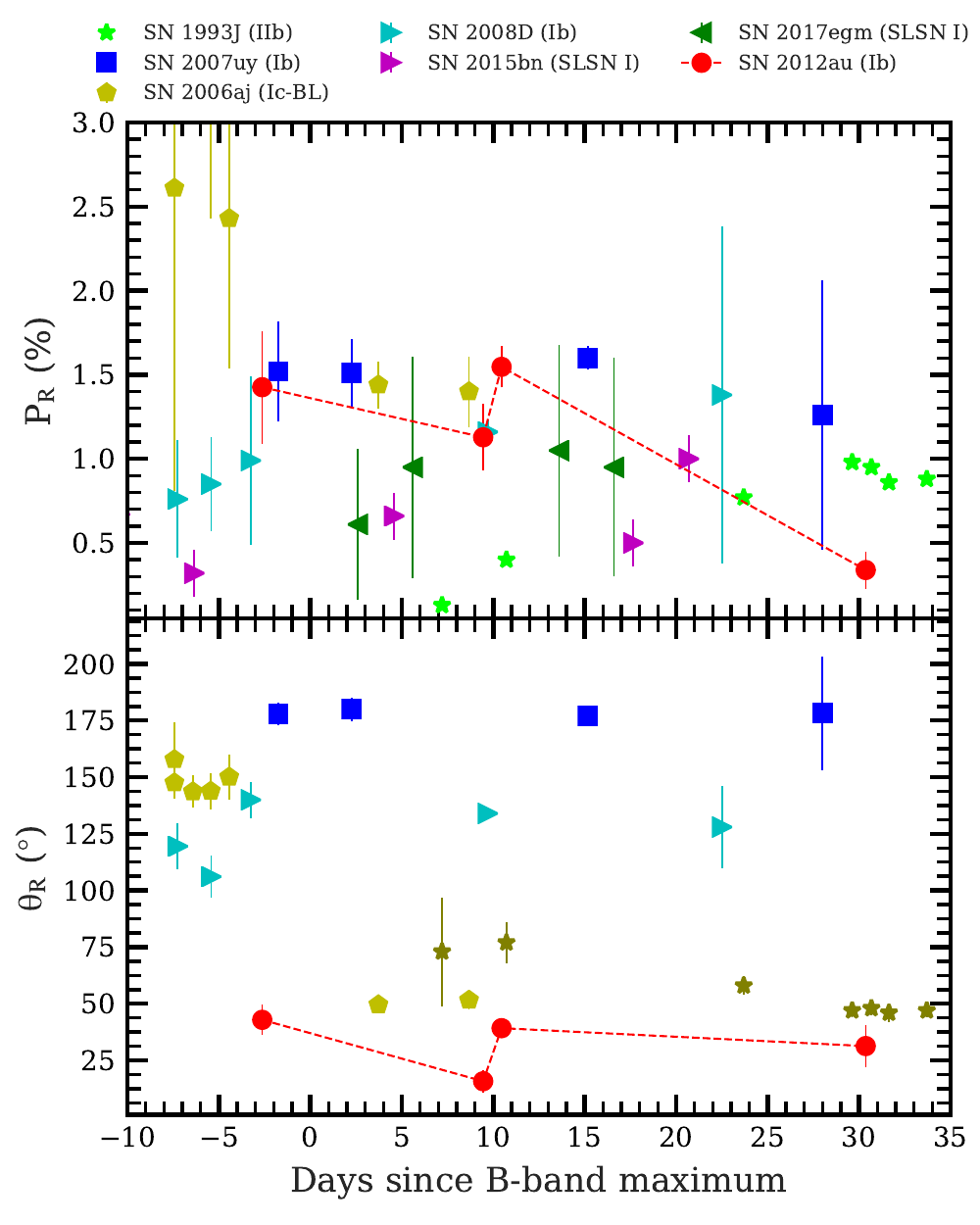}
\caption{Evolution of the degrees of polarization of SN~2012au, compared with other SNe~IIb, Ib, and SLSNe~I from the literature.}
\label{p_pa_evol_comp}
\end{centering}
\end{figure}

\begin{figure}
\begin{centering}
\includegraphics[width=\columnwidth]{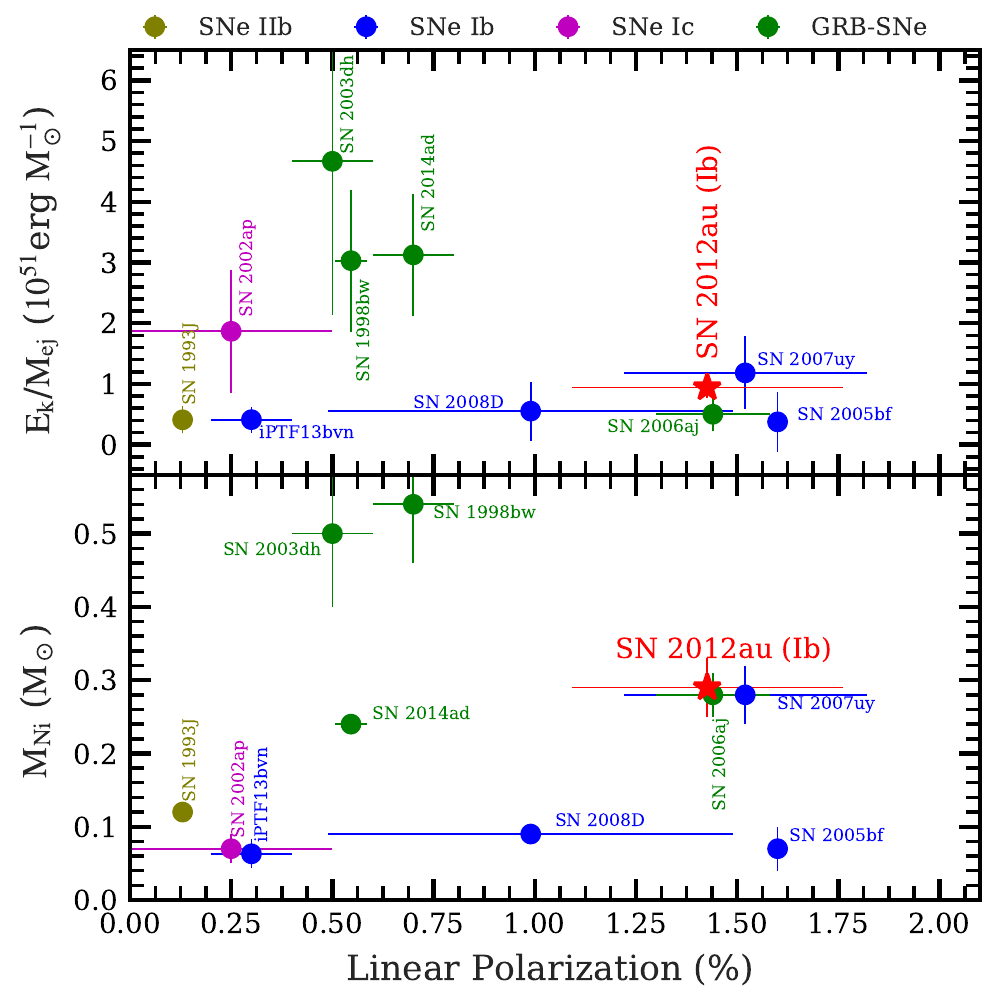}
\caption{Top panel: linear polarization versus $E_k$/$M_{ej}$ of SN~2012au compared with other SNe~Ib, Ic, and Ic-BL in the literature. Bottom panel: linear polarization versus $M_{Ni}$ of SN~2012au compared with the same sample.}
\label{MniMejEk}
\end{centering}
\end{figure}

The imaging polarimetric evolution of SN~2012au in the $R$ band (covering He\,{\sc i} $\lambda\lambda$5876, 6678, and $\lambda$7065) is shown in Fig.~\ref{p_pa_evol}. The follow-up covers almost the whole photospheric phase of the light curve (between $\sim$14 and 47 d from the explosion). During this period, the SN~displays a maximum degree of polarization of 1.32 per cent (at $\sim$27 d), fading to 0.13 per cent (at $\sim$47 d). A variation in the intrinsic polarization angle is also seen during this period, but it remains below 50$\degr$. Such evolution in the polarization parameters of SN~2012au indicates that the ejecta were aspherical and/or clumpy during the early phases. Using spectropolarimetric data at several epochs for this event, \citet{Hoffman2014,Hoffman2017} also noticed similar evolution in the continuum position angle.

The overall observational properties of SN~2012au show that it is a transitional event between SESNe and SLSNe~I \citep[cf.][]{Milisavljevic2013, Takaki2013}. Therefore, we also compared (see Fig.~\ref{p_pa_evol_comp}) the linear imaging polarization properties of SNe~Ib (SN~2007uy and SN~2008D: \citealt{Maund2009, Gorosabel2010}), IIb (SN~1993J: \citealt{Doroshenko1995}), GRB-SN (SN~2006aj: \citealt{Gorosabel2006}), and SLSNe~I (SN~2015bn: \citealt{Leloudas2017} and SN~2017egm: \citealt{Maund2019}) from the literature with those determined for SN~2012au. Here, we consider only events with multi-epoch observations. This analysis reveals that, among SNe~Ib, SN~2008D and SN~2012au show a major variation in polarization parameters, whereas the SN~2007uy evolution is minimal. Similarly, SN~2015bn exhibits an increasing trend in the degree of polarization, while SN~2017egm remains below 1 per cent, without any significant change. Pre-maximum, SN~2006jf exhibits high values of liner polarization in comparison with all SNe presented; however, after the peak it shares polarization values closer to SN~2012au.

\subsubsection{Evolution of $E_k$/$M_{ej}$ and $M_{Ni}$ with degree of polarization}
\label{sec:P_Mej_Ek_Mni}

In this section, $\%P$ versus $E_k$/$M_{ej}$ and $M_{Ni}$ of SN~2012au are compared with those of SNe~IIb, Ib, Ic, and GRB-SNe (see the upper and lower panels of Fig.~\ref{MniMejEk}, respectively) from the literature: SN~1993J \citep{Doroshenko1995, Lyman2016}, SN~1998bw \citep{Patat2001}, SN~2002ap \citep{Wang2003, Mazzali2002, Pandey2003}, SN~2003dh \citep{Kawabata2003, Covino2003, Cano2017}, SN~2005bf\citep{Maund2007}, SN~2006aj \citep{Gorosabel2006, Cano2017}, SN~2007uy \citep{Gorosabel2010}, SN~2008D \citep{Gorosabel2010}, iPTF13bvn \citep{Reilly2016}, and SN~2014ad \citep{Stevance2017, Sahu2018}. The polarimetric values for comparison were taken near the optical maximum ($\sim-$10 to +10 d). With the exception of SN~1998bw (from 4000--7000 \AA), SN~2002ap (from continuum), SN~2005bf ($\sim$3000 \AA), iPTF13bvn (from continuum), and SN~2014ad ($V$-band) for all other events $R$-band polarimetric values were adopted.

From Fig.~\ref{MniMejEk}, SN~2012au appears to have a near-peak $\%P$ value closer to those of SN~2006aj and SN~2007uy. SN~1998bw and SN~2003dh (GRB-SNe) present higher values of $E_k$/$M_{ej}$ and $M_{Ni}$, but exhibit lower $\%P$ values in comparison with SN~2012au. Among SNe~Ib, SN~2012au exhibits a higher $\%P$ value in comparison with SN~2008D and iPTF13bvn (lowest values of $\%P$), closer to SN~2007uy and lower than SN~2005bf. In our limited sample, SN~2005bf (Type Ib/c) exhibits the highest value of linear polarization with a very low value of $M_{Ni}$, whereas SN~1993J shares the lowest value of \%P. SLSNe~I are nearly 2--3 mag brighter than normal SESNe, which corresponds to M$_{Ni}$ $\gtrsim$5 M$_\odot$ \citep{Gal-Yam2012}. Therefore, due to comparatively high $M_{Ni}$, SLSNe~I will show a different class in Fig.~\ref{MniMejEk}, hence they are not included.

\section{Spectroscopic analysis}
\label{sec:spectra}

We present spectroscopic studies of SN~2012au using unpublished spectra: 21 at optical and 2 at NIR wavelengths, spanning the range from $-$5 to +391 d. Spectra used here were obtained using a few telescopes during 2012-2013 (see Table~\ref{Ap:table5} for the spectroscopic observation log) as part of the present work. Out of 21 optical spectra, 15 were obtained using HCT-2m, one using the Galileo-1.22m telescope\footnote{\label{note1}\url{http://www.astro.unipd.it/inglese/observatory/telescopio_en.html}} (Asiago, Italy), one using the CAHA-2.2m, two using the NTT-3.58m, and two using the BTA-6m telescope. On the other hand, both the NIR spectra of SN~2012au (at $-$5 and +21.8 d since maximum) were obtained using SOFI on NTT-3.58m. Spectroscopic data reduction was performed in a standard manner as described in \cite{Kumar2020,Kumar2021}. The spectra of SN~2012au are subdivided into three different phases: early photospheric phase (from $\sim-5$ to +50 d; 14 spectra), late photospheric phase (from $\sim+60$ to +108 d; 7 spectra), and nebular phase (at +391 d; 1 spectrum). The early photospheric phase spectra, along with two modelled spectra (in red), are plotted in Fig.~\ref{fig:photospec}. The modelled spectra (at $-$2 and +36 d; in red) are reproduced using {\tt SYNAPPS} \citep[an automated version of the {\tt SYN++} code:][]{Thomas2011} and are presented with individual ion contributions in Fig.~\ref{fig:singleion}. The late photospheric phase spectra are shown in Fig.~\ref{fig:nebspec}, whereas the nebular phase spectrum at +391 d is shown along with two published nebular spectra \citep[at +275 and +323 d;][]{Milisavljevic2013} in Fig.~\ref{fig:latenebspec}. The remaining unpublished NIR spectra of SN~2012au (at $-$5 and +21.8 d), along with three published ones (at +32, +82, and +319 d) taken from \cite{Milisavljevic2013} are also presented. All spectra were corrected for total extinction (Galactic + host) of $E(B-V)$ = 0.063 mag and also shifted to rest-frame wavelengths.

\subsection{Optical spectroscopic evolution at photospheric phase}
\label{sec:spectroscopy}

\begin{figure}
\includegraphics[width=\columnwidth]{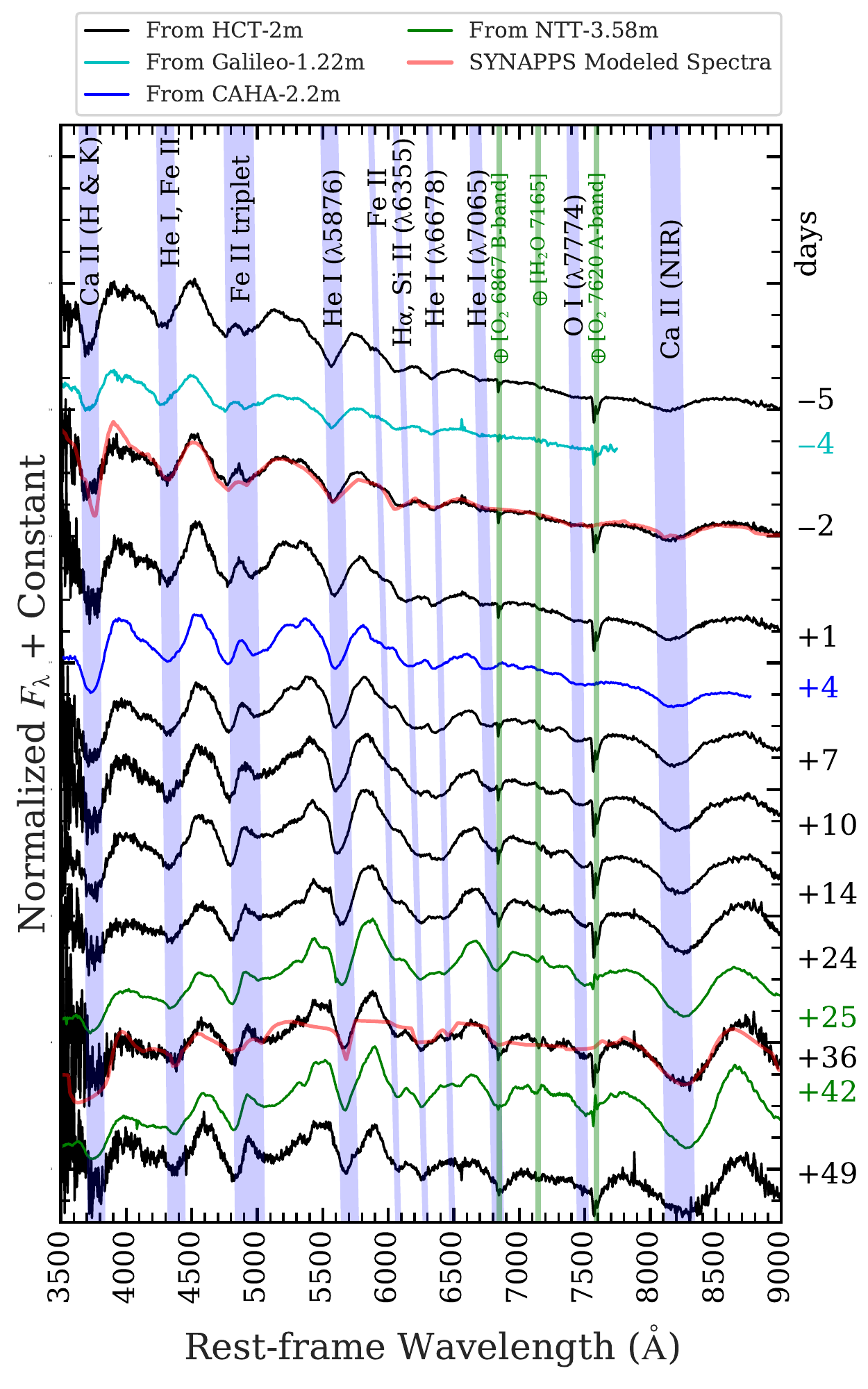}
\caption{Early photospheric phase spectra of SN~2012au, from $\sim-$5 to +49 d. The gray bands trace the evolution of absorption troughs of various elements, and indicate decreasing velocity with time. The elements and their rest-frame wavelengths are written at the top: telluric features are also shown, as bands with green colour.}
\label{fig:photospec}
\end{figure}

\begin{figure}
\includegraphics[width=\columnwidth]{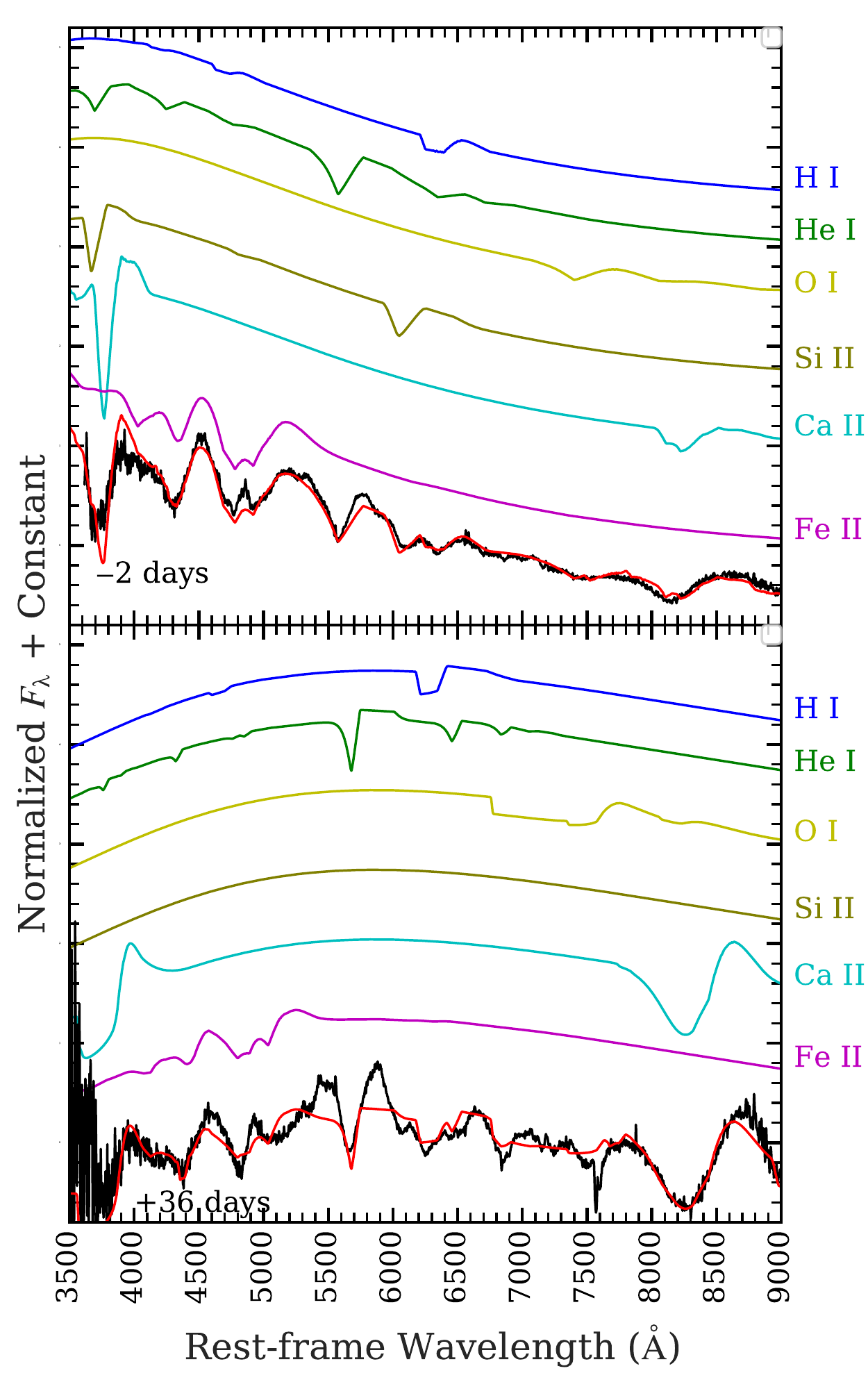}
\caption{Observed and synthetic spectra of SN~2012au at --2 and +36 d are shown. The contribution of individual ions to generate the best-matching synthetic spectrum using the {\tt SYNAPPS} \citep{Thomas2011} code is plotted and accounts for H$_\alpha$, He\,{\sc i}, O\,{\sc i}, Si\,{\sc ii}, Ca\,{\sc ii}, and Fe\,{\sc ii} elements.}
\label{fig:singleion}
\end{figure}

\begin{figure}
\includegraphics[width=\columnwidth]{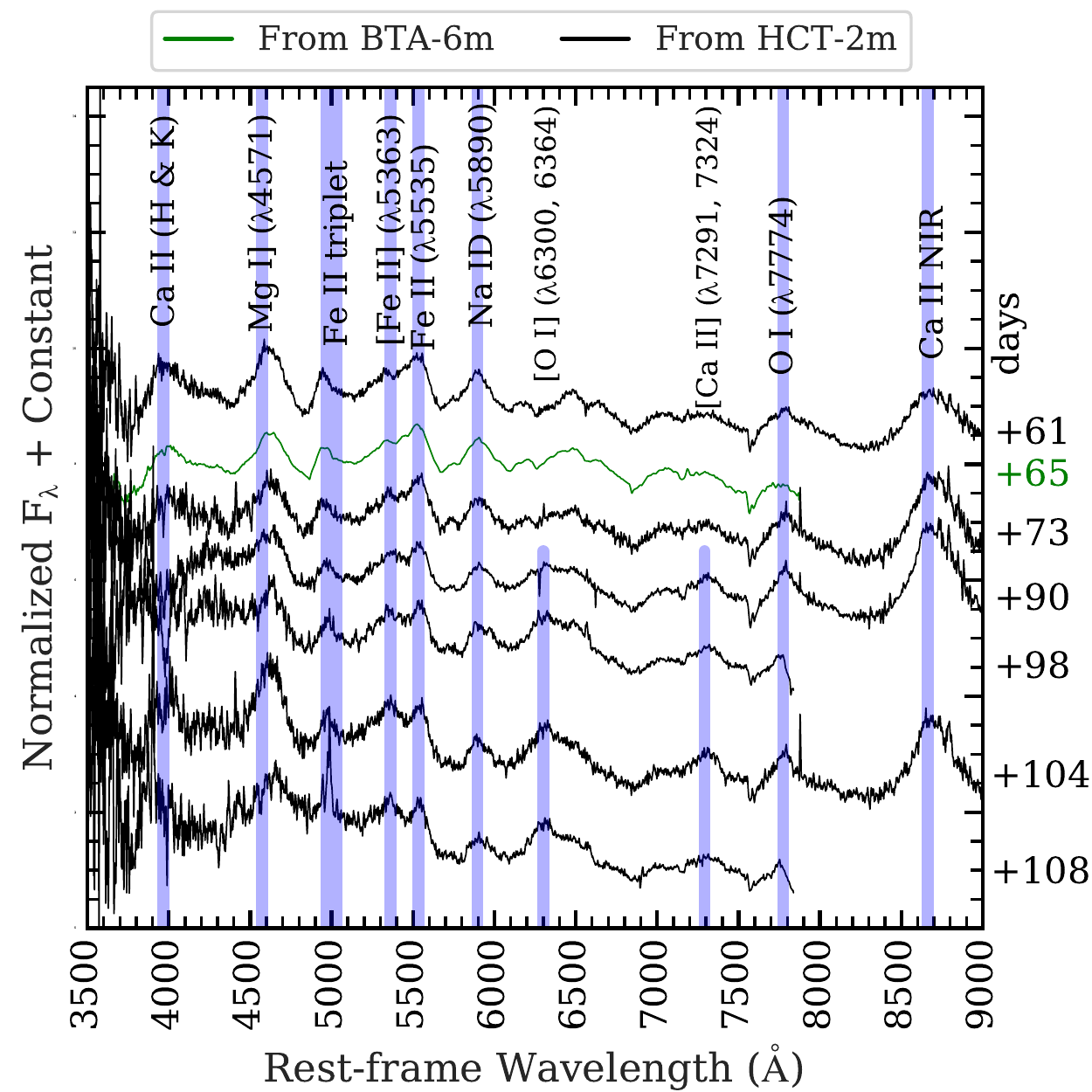}
\caption{The late photospheric phase spectra of SN~2012au from $\sim$+60 to +108 d are plotted. The gray colour bands represent the rest-frame wavelengths (written on the top) of various elements. The [O\,{\sc i}] doublet and [Ca\,{\sc ii}] start little later in comparison with other emission features.}
\label{fig:nebspec}
\end{figure}

Early and late photospheric phase spectra of SN~2012au are plotted in Figs.~\ref{fig:photospec} and~\ref{fig:nebspec}, respectively. The first spectrum of SN~2012au was observed at $-$5 d, showing a blue continuum with a temperature of $\approx 9000K$. The lines are identified following \cite{Milisavljevic2013}, \cite{Takaki2013}, and {\tt SYNAPPS} spectral matching \citep{Thomas2011}. To construct the basic chemical composition, we performed {\tt SYNAPPS} spectral matching for the photospheric phase spectra at $-$2 and +36 d (in red). Thereafter, {\tt SYN++} has been used to get the contribution of individual ions from {\tt SYNAPPS} best-matching synthetic spectra (see Fig.~\ref{fig:singleion}). To generate the best-matched synthetic spectrum at $-$2 d, we used a $v_{ph}$ of $\approx 14 100$ km s$^{-1}$ and photospheric temperature of $\approx 7000$ K. At this epoch, He\,{\sc i}, Ca\,{\sc ii}, and Fe\,{\sc ii} are found to be the most prominent features, with minor contributions from H$_\alpha$, O\,{\sc i}, and Si\,{\sc ii}: see the top panel of Fig.~\ref{fig:singleion}. The conspicuous absorption near 6200 \AA\, in the early photospheric phase spectra of SN~2012au is possibly H$_\alpha$, as observed in a sample of SNe~Ib by \cite{Elmhamdi2006}. This favours considerable mixing of H in the He envelope \citep{Maurer2010}. The absorption minima of H$_\alpha$, He\,{\sc i}, and Si\,{\sc ii} are usually fitted with larger velocities than other features. To generate the spectrum at +36 d using {\tt SYNAPPS}, similar ions to those used for the spectrum at $-2$ d were used (see lower panel of Fig.~\ref{fig:singleion}). We obtained $v_{ph}$ of $\approx 8000$ km $s^{-1}$ and a photospheric temperature of $\approx 4750$ K to regenerate the spectrum at +36 d.

Early photospheric spectra of SN~2012au have absorption features of He\,{\sc i} ($\lambda\lambda$5876, 6678, and $\lambda$7065), O\,{\sc i} ($\lambda$7774), Si\,{\sc ii} ($\lambda\lambda$3838, 6355), Ca\,{\sc ii} (H\&K and NIR), and Fe\,{\sc ii} ($\lambda\lambda\lambda$4549, 4925, 5018, and $\lambda$5169), however He\,{\sc i} is the most prominent one; see Figs.~\ref{fig:photospec} and ~\ref{fig:singleion}. Spectra up to $\sim$+50 d exhibit a clear absorption feature of O\,{\sc i} $\lambda$7774, but its profile is contaminated by the telluric $\oplus$ [O$_2$ $\lambda$7620] band. The feature at around $\sim$3700 \AA\, is attributed as a blend of Ca\,{\sc ii} H\&K and Si\,{\sc ii} $\lambda$3838. The Ca\,{\sc ii nir} triplet is possibly blended with O\,{\sc i} $\lambda$8446 and appears to evolve, becoming more prominent with time (see Fig.~\ref{fig:singleion}). Throughout the early photospheric phase, the $\sim4000$--5500\,\AA\ window is primarily dominated by Fe\,{\sc ii} ions. The Fe\,{\sc ii} ($\lambda\lambda$4925, 5018, and $\lambda$5169) triplet is prominent, while the absorption feature at $\sim$4300 \AA\, seems to be a blend between Fe\,{\sc ii} $\lambda$4549 and He\,{\sc i} $\lambda$4471. The blue wing of the H$_\alpha$ line is contaminated by a weak feature that could be Si\,{\sc ii} $\lambda$ 6355, observed at around 6050 \AA\, (Fig.~\ref{fig:singleion}). In the spectrum at $\sim$ --4 d, absorption troughs of all elements are blueshifted and show $v_{ph}$ of nearly $\sim15 000$ km s$^{-1}$. However, with time the absorption features move towards their rest-frame wavelengths (indicating decreasing velocity), as shown by the gray coloured inclined bands.

As spectra evolve from early to late photospheric phase, the absorption features start to disappear and emission-line components emerge (see Fig.~\ref{fig:nebspec}). The emission lines are indicated with vertical gray bands. O\,{\sc i} $\lambda$7774, Na\,{\sc id}, Mg\,{\sc i}], Ca\,{\sc ii} H\&K and NIR, and the Fe\,{\sc ii} triplet appear earlier than the [O\,{\sc i}] $\lambda\lambda$6300, 6363 doublet and [Ca\,{\sc ii}] $\lambda\lambda$7291, 7324 (hereafter [Ca\,{\sc ii}] $\lambda$7300). The emission feature at $\sim$5500 \AA~ is possibly a blend of [O\,{\sc i}] $\lambda$5577 and [Fe\,{\sc ii}] $\lambda$5363. Initially, the [O\,{\sc i}] doublet appears to be flat-topped, whereas in later epochs it becomes asymmetric, with a suppressed redder component. 

\begin{figure*}
\includegraphics[angle=0,scale=0.75]{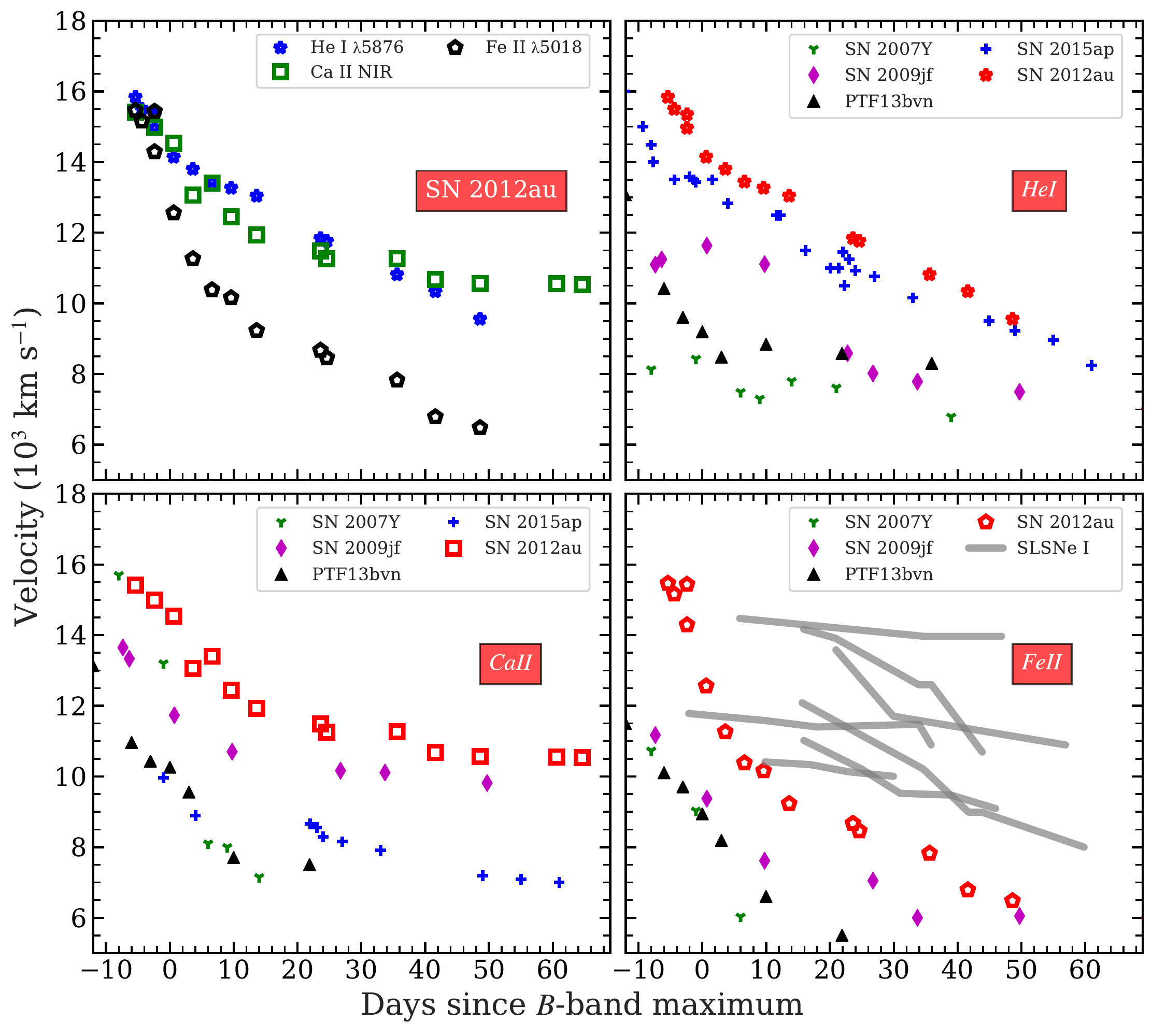}
\caption{Upper left: He\,{\sc i}, Ca\,{\sc ii nir}, and the Fe\,{\sc ii} line velocities of SN~2012au. The line velocities of SN~2012au are also compared with those of SN~2007Y \citep{Stritzinger2009}, SN~2009jf \citep{Sahu2011}, iPTF13bvn \citep{Srivastav2014}, and SN~2015ap \citep{Prentice2019, Aryan2021} in the other three panels.}
\label{fig:ionvelocity}
\end{figure*}

\begin{figure}
\includegraphics[width=\columnwidth]{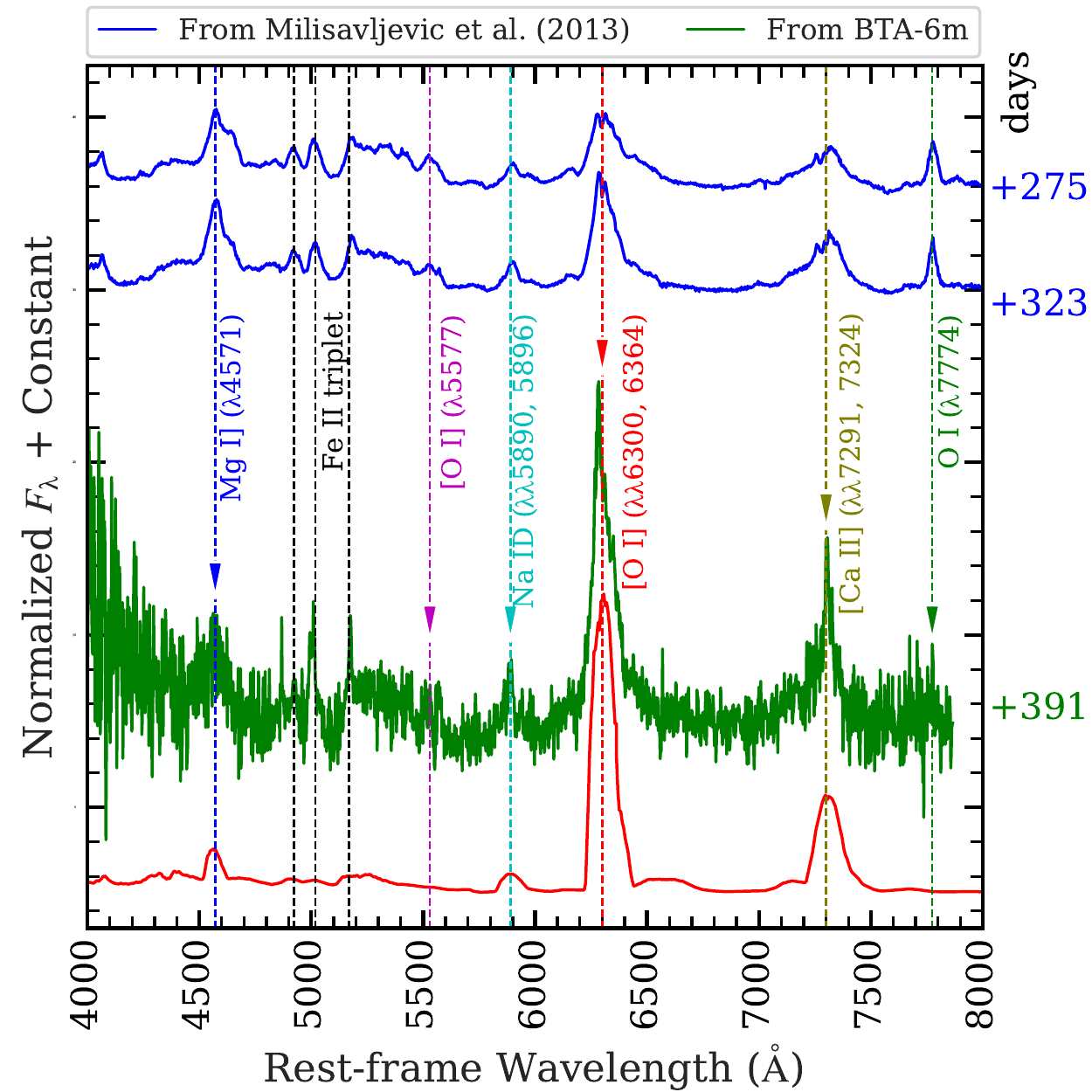}
\caption{The nebular phase spectrum of SN~2012au at +391 d is plotted along with two published spectra at +275 and +323 d (taken from \citealt{Milisavljevic2013}). The dotted vertical lines represent the rest-frame wavelengths of various elements. The modelled spectrum generated using $M_{ZAMS}$ = 17 M$_\odot$ is also shown in red for comparison \citep{Jerkstrand2015}.}
\label{fig:latenebspec}
\end{figure}

\begin{figure*}
\includegraphics[angle=0,scale=0.42]{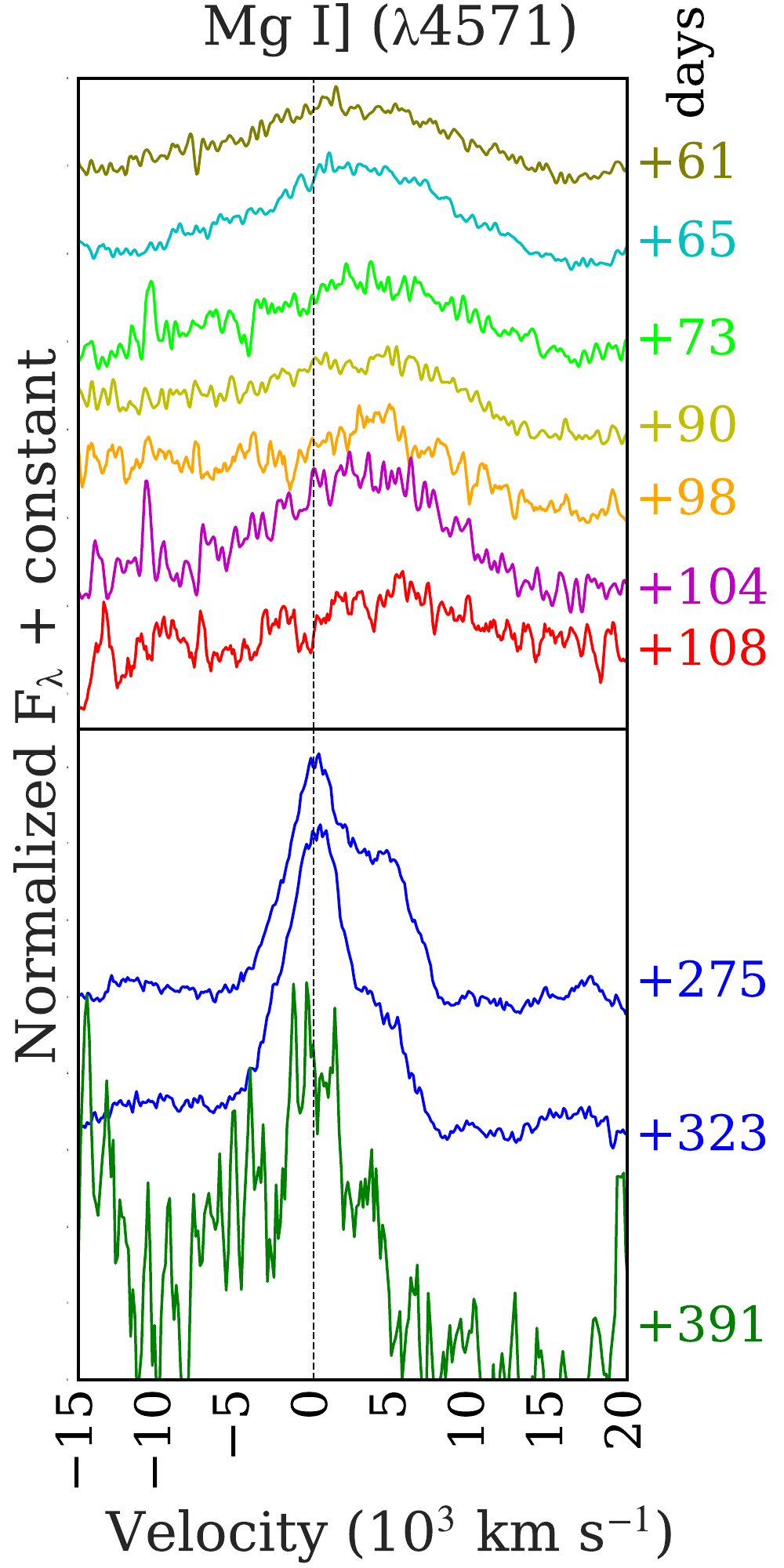}
\includegraphics[angle=0,scale=0.42]{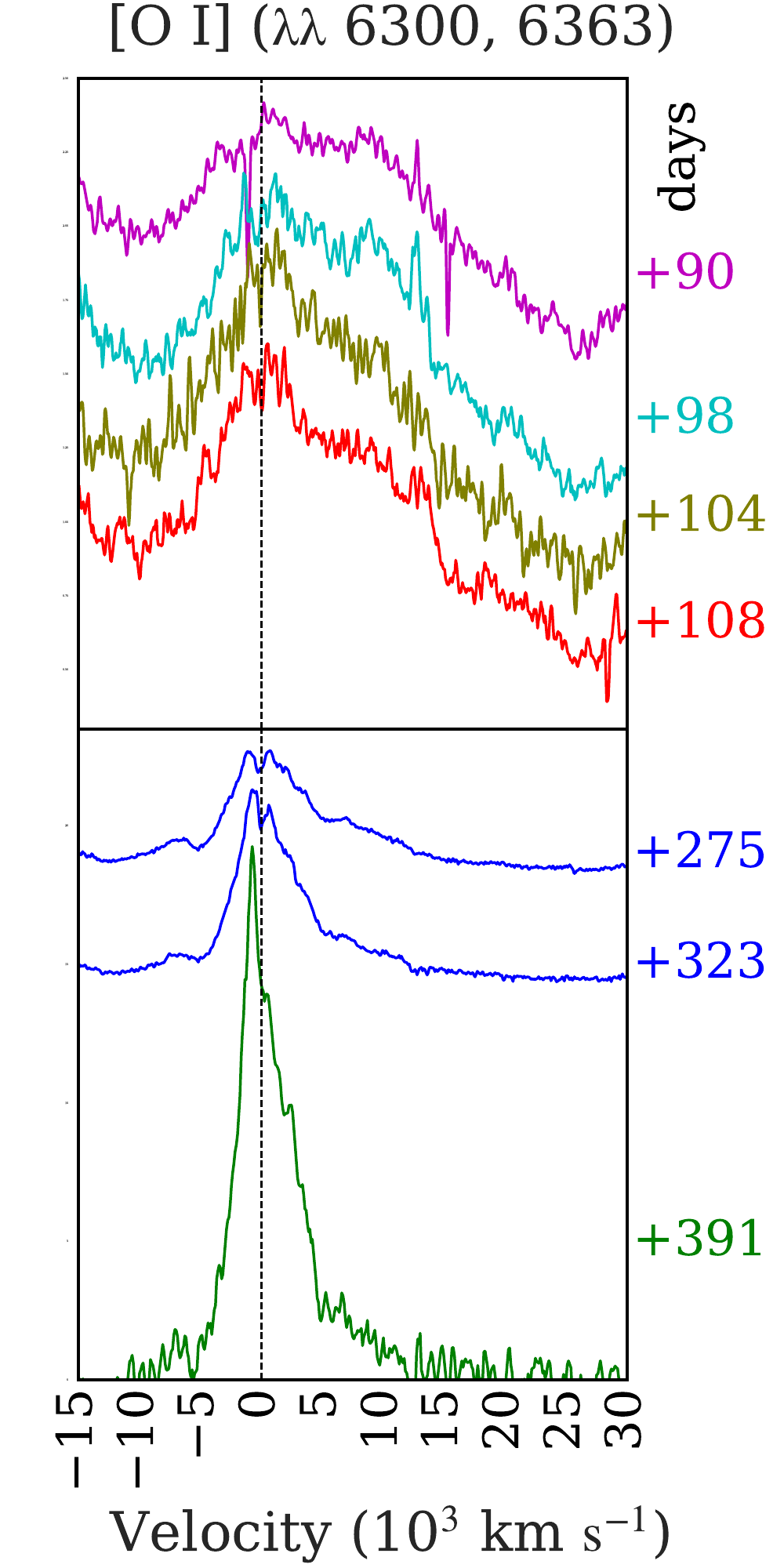}
\includegraphics[angle=0,scale=0.42]{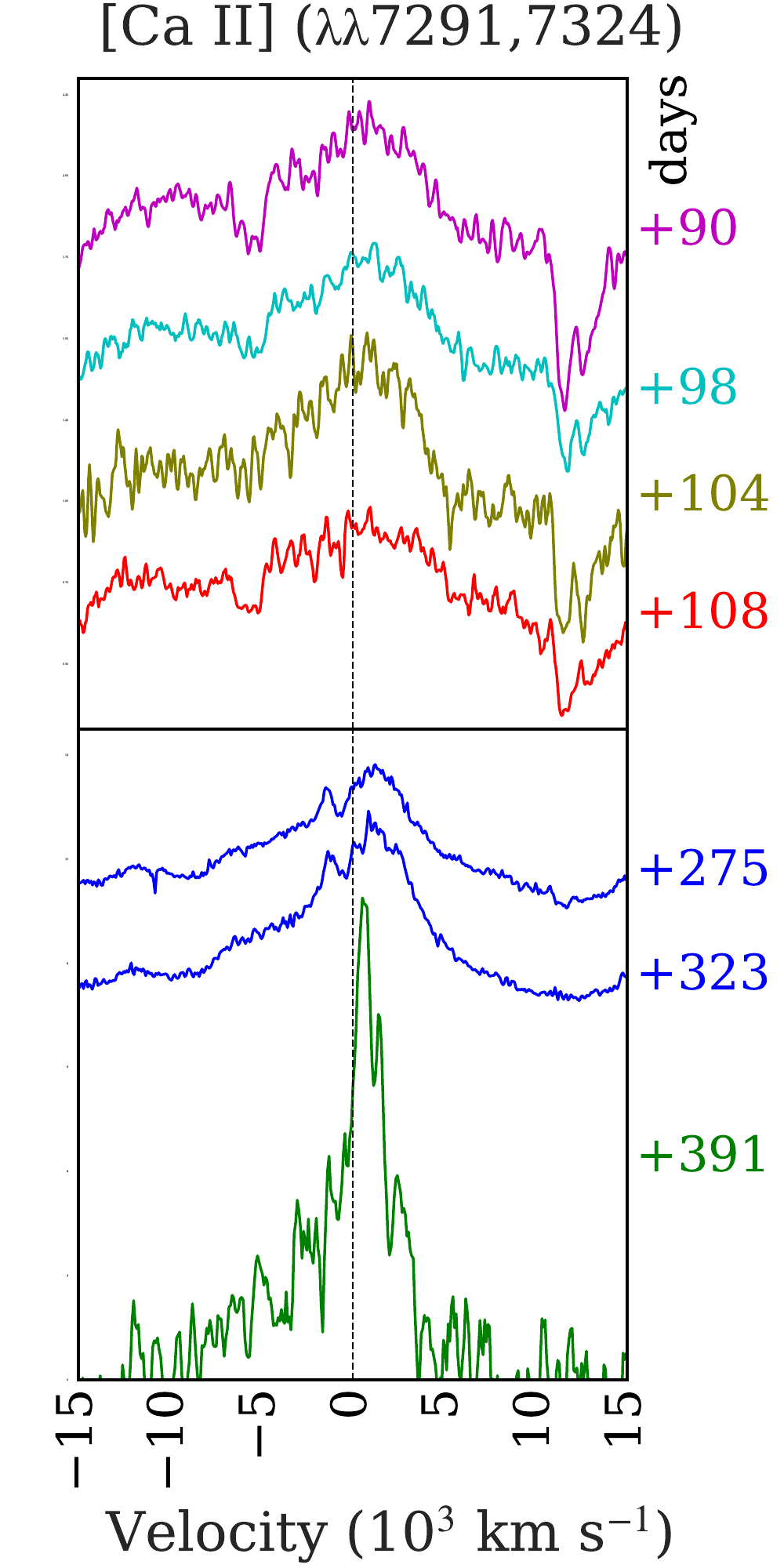}
\includegraphics[angle=0,scale=0.42]{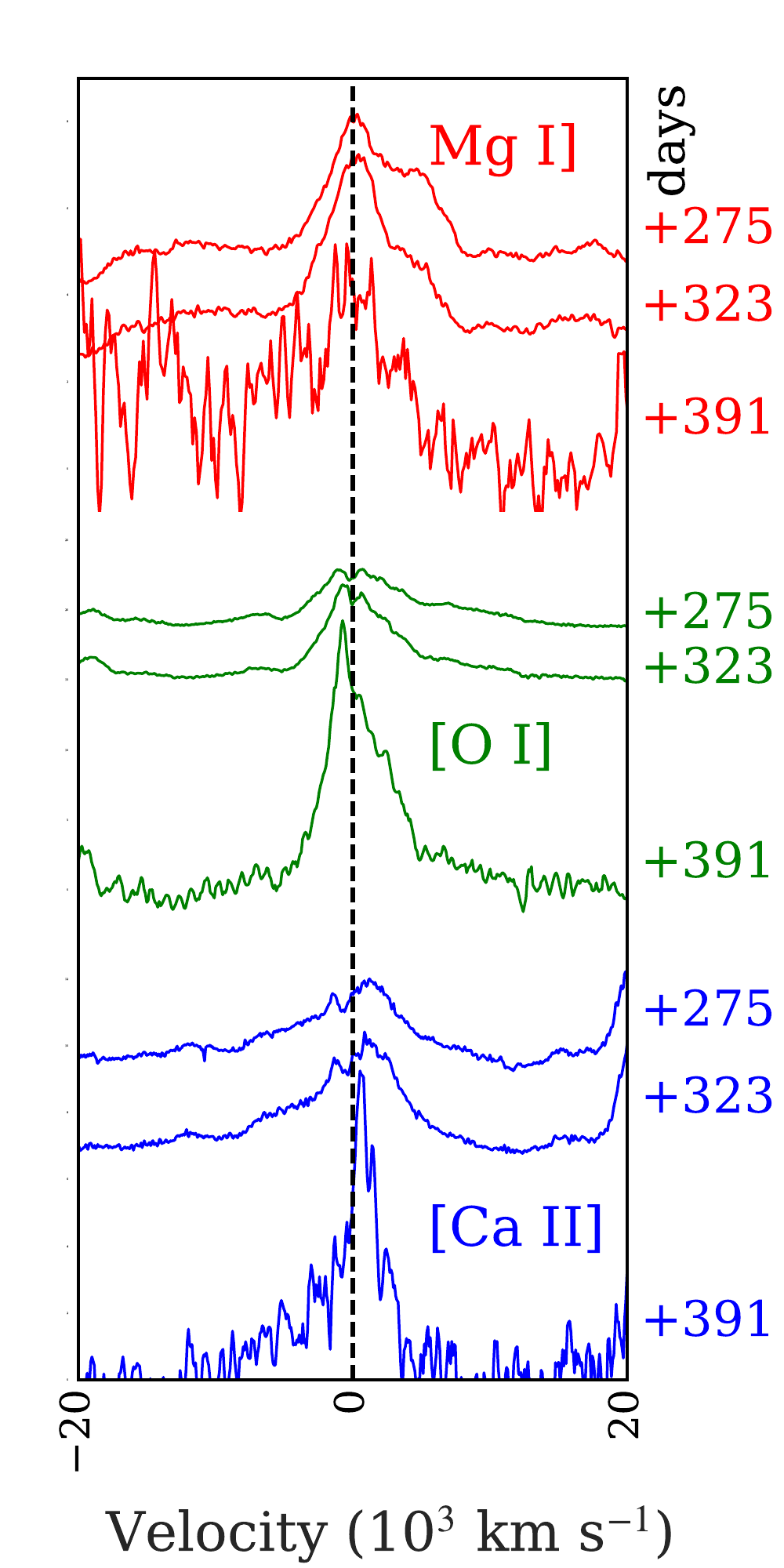}
\caption{Evolution of Mg\,{\sc i}], [O\,{\sc i}] doublet, and [Ca\,{\sc ii}] lines plotted in the velocity domain, with their zero velocities being taken at $\lambda$4571, $\lambda$6300, and $\lambda$7291, respectively.}
\label{fig:indivievolu}
\end{figure*}

\begin{figure*}
\includegraphics[angle=0,scale=0.7]{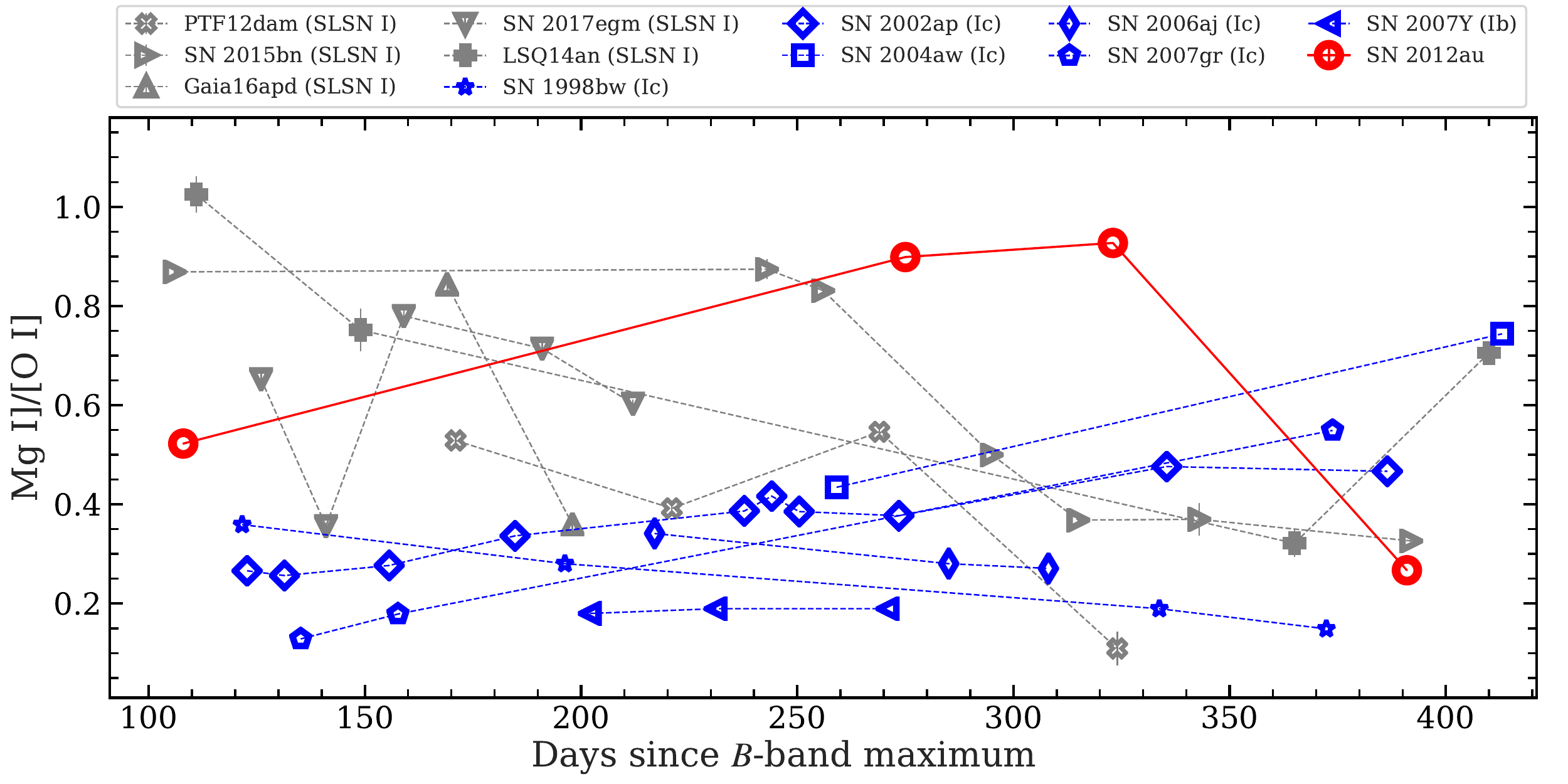}
\caption{Evolution of the Mg\,{\sc i}]/[O\,{\sc i}] ratio of SN~2012au compared with those found for SLSNe~I (taken from \citealt{Nicholl2019} and references therein) and other Type Ib/c SNe (taken from \citealt{Hunter2009} and references therein).}
\label{fig:Mg_OI_ratio}
\end{figure*}

\subsubsection{Ejecta mass, photospheric radius, and optical depth}
\label{sec:spectrophpara}

The photospheric radius around the peak ($r_{ph}$), total optical depth below the photosphere around maximum ($\tau_{total}$), and $M_{ej}$ of the SN can be constrained using $v_{ph}$ and the time since the explosion. Assuming homogeneous expansion, $r_{ph}$ can be estimated using $v_{ph}$ and the time of explosion ($t_{exp}$) as

\begin{equation}\label{eq:rphot}
r_{ph} = v_{ph} \times \frac{(t - t_{exp})}{(1 + z)}
\end{equation}

where $t$ is the time of the observation and $t_{exp}$ is the time of explosion. We also adopt

\begin{equation}
\label{eq:tau}
\tau_{total} \approx 3c/v_{ph}
\end{equation}

where c is the speed of light; the details are discussed in \cite{Konyves-Toth2020}, see also \cite{Arnett1996} and \cite{Branch2017}. Finally, to estimate $M_{\rm ej}$ using $v_{ph}$ and $\tau_{total}$, equation 8 of \cite{Konyves-Toth2020} has been used:
\begin{equation}\label{eq:Mej}
M_{ej} = \frac{4 \pi}{3} \frac{{v^2_{ph}} \times (t - t_{exp})^2}{(1 + z)^2} \frac{\tau_{total}}{\kappa}.
\end{equation}

To calculate $M_{ej}$ using the above equation, a value of $\kappa$ = 0.05 cm$^2$ g$^{-1}$ was chosen. For SN~2012au, the near-peak spectrum at $-2$ d (14.5 d since explosion) was modelled with a $v_{ph}$ of $\sim$14 100 km s$^{-1}$. Using these equations, the values of $r_{phot}$, $\tau_{total}$, and $M_{ej}$ are $\sim$1.8 $\times$ $10^{15}$ cm, $\sim$63.8, and $\sim$8.3 $M_{\odot}$. Values of $v_{ph}$ and the time since explosion discussed here correspond to $E_k$ $\sim$5.4 $\times$ 10$^{51}$ erg. For SN~2012au, the values of $r_{ph}$ and $M_{ej}$ are lower and $\tau_{total}$ is higher than for SLSNe~I: SN 2010kd \citep{Kumar2020} and SN~2019neq \citep{Konyves-Toth2020}.

\subsubsection{Line velocities}\label{sec:velocomp}

The evolution of line velocities in the rest-frame spectra of SN~2012au is estimated by fitting a Gaussian profile to the absorption components. The evolution of the velocity for He\,{\sc i} $\lambda$5876, Ca\,{\sc ii nir} triplet (rest-frame wavelength is taken at $\lambda$8571), and Fe\,{\sc ii} $\lambda$5169 lines is shown in Fig.~\ref{fig:ionvelocity} (upper left panel). The velocities of the He\,{\sc i} and Fe\,{\sc ii} lines are measured only until $\sim$+50 d, because at later phases their profiles have blended with other lines. Before the $B$-band maximum (at --4 d), the three features have nearly the same velocity ($\sim 15 000$ km s$^{-1}$). The $v_{ph}$ estimated using {\tt SYNAPPS} in the --2 d spectrum ($\sim 14 100$ km s$^{-1}$) is in good agreement with the value obtained from measuring the position of the Fe\,{\sc ii} absorption trough. For this reason, we consider the Fe\,{\sc ii} velocities as representative for the evolution of $v_{ph}$. The Fe\,{\sc ii} line velocity declines to $\sim$\,12 500 km s$^{-1}$ at maximum, and exhibits lower velocities than He\,{\sc i} and Ca\,{\sc ii} at later phases. We note that the velocities estimated in this study for the different ions in the spectra of SN~2012au are closer to those reported by \cite{Takaki2013}.

From peak to $\sim$+60 d, the velocity of He\,{\sc i} decreases from $\sim 15 000$ to 8000 km s$^{-1}$. In the case of SN~2012au, it remains higher than the average He\,{\sc i} velocities estimated for a sample of SNe Ib by \cite{Fremling2018}. Until $\sim$\,+45 d, the Ca\,{\sc ii nir} triplet seems to have a similar velocity to the He\,{\sc i} line, but it decays more slowly at later phases, at a nearly constant velocity of $\sim$\,11 000 km s$^{-1}$. He\,{\sc i} and the Ca\,{\sc ii nir} triplet show higher velocities than Fe\,{\sc ii}, suggesting that the former lines are generated in the outer parts of the ejecta, while Fe\,{\sc ii} lines form in the inner layers. We also note that, up to $\sim$\,+25 d, the $v_{ph}$ (Fe\,{\sc ii} line velocity) of SN~2012au is found to be higher than the average $v_{ph}$ (8000 $\pm$ 2000 km s$^{-1}$) estimated at maximum light for a sample of SNe~Ib/c \citep{Cano2013}.

In the other three panels of  Fig.~\ref{fig:ionvelocity}, we compare the He\,{\sc i}, Ca\,{\sc ii nir}, and Fe\,{\sc ii} line velocities of SN~2012au (red line) with those of SN~2007Y \citep[green:][]{Stritzinger2009}, SN~2009jf \citep[magenta:][]{Sahu2011}, iPTF13bvn \citep[black:][]{Srivastav2014}, and SN~2015ap \citep[blue:][]{Prentice2019, Aryan2021}. The Fe\,{\sc ii} line velocity for SN~2015ap is not available because of contamination by host galaxy lines \citep{Prentice2019}. Across the entire evolution, the He\,{\sc i} and Ca\,{\sc ii} line velocities of SN~2012au remain higher than those of the other SNe~Ib discussed, whereas the Fe\,{\sc ii} line velocity of SN~2012au is higher only until $\sim$+30 d, and thereafter it is nearly equal to that of SN~2009jf. In the lower right panel of Fig.~\ref{fig:ionvelocity}, the Fe\,{\sc ii} ion velocities of SNe~Ib along with SN~2012au have been compared with those estimated for SLSNe~I (in gray) by \cite{Nicholl2015}. Similarly to Type Ic in \cite{Nicholl2015}, SN~2012au and other plotted SNe~Ib exhibit faster decaying  Fe\,{\sc ii} velocities in comparison with SLSNe~I.

\subsection{Optical spectroscopic evolution in the nebular phase}
\label{sec:spectroscopy}

In the nebular phase, ejecta become optically thin and the deeper layers are probed, and the available spectral features are used to investigate the geometry of the ejecta and other physical parameters \citep{Taubenberger2009}. Therefore, to trace the evolution of emission lines at late phases, the spectrum of SN~2012au at +391 d (in green colour) along with two publicly available spectra at +275 and +323 d (in blue colour; taken from \citealt{Milisavljevic2013}) is investigated and plotted in Fig.~\ref{fig:latenebspec}. The nebular spectra of SN~2012au are dominated by the semi-forbidden Mg\,{\sc i}] $\lambda$4571, forbidden [O\,{\sc i}] $\lambda\lambda$6300,6364 doublet and [Ca\,{\sc ii}] $\lambda\lambda$7300 features, along with weaker Fe\,{\sc ii} triplet, [O\,{\sc i}] $\lambda$5577, Na\,{\sc id}, and O\,{\sc i} $\lambda$7774 lines. All the features discussed above are marked with vertical dotted lines in Fig.~\ref{fig:latenebspec}. The Mg\,{\sc i}], Na\,{\sc i} D, and [Ca\,{\sc ii}] lines do not evolve significantly between +323 and +391 d, in contrast to the [O\,{\sc i}] doublet which shows major evolution. While at +275 and +323 d the [O\,{\sc i}] $\lambda$5577 and O\,{\sc i} $\lambda$7774 features are evident, those become weaker at +391 d.

The nebular phase spectrum of SN~2012au at +391 d is also compared with the modelled nebular spectra published by \cite{Jerkstrand2015} for SNe~IIb. After $\sim$+150 d, the spectra published by \cite{Jerkstrand2015} can also be compared with SNe~Ib because at these late phases the influence of the H envelope is negligible. We tried to match the spectrum of SN~2012au at +391 d with all modelled spectra published by \cite{Jerkstrand2015}. However, we found that the modelled spectrum generated for $M_{ZAMS}$ of 17 M$_\odot$ considering strong mixing and dust (in red) matched well with the +391 d spectrum of SN~2012au (see Fig.~\ref{fig:latenebspec}). The prominent Mg\,{\sc i}], [O\,{\sc i}] doublet, and [Ca\,{\sc ii}] emission lines in the synthetic spectrum matched the observations well.

\subsection{Emission-Line study}
\label{sec:evolutionindi}

The geometry of the SN ejecta can be constrained from the emission-line profiles and observed fluxes in nebular phase spectra, mainly using the isolated [O\,{\sc i}] doublet \citep{Taubenberger2009, Fang2019}. In Fig.~\ref{fig:indivievolu}, we present the evolution of Mg\,{\sc i}], [O\,{\sc i}] doublet, and [Ca\,{\sc ii}] lines in the velocity domain. The zero velocities for Mg\,{\sc i}], [O\,{\sc i}] doublet, and [Ca\,{\sc ii}] lines are taken at $\lambda$4571, $\lambda$6300, and $\lambda$7291, respectively. Up to +108 d, Mg\,{\sc i}] has a redshifted maximum, but from $\gtrsim$+275 d it appears to be at zero velocity. The [Ca\,{\sc ii}] emission lines are redshifted throughout the nebular phase ($\sim$500-1200 km s$^{-1}$, see Fig.~\ref{fig:indivievolu}), though a possibility of [Ca\,{\sc ii}] $\lambda$ blending with [O\,{\sc ii}] $\lambda\lambda$7320, 7331 also exists. As discussed earlier, the [O\,{\sc i}] doublet profile is flat-topped at around $\sim$+90 d, attributed to the blending of $\lambda$6300 and $\lambda$6363 features. However, at later phases (from +275 d), the core becomes narrower, superposed on a broader base. At +275 d, the [O\,{\sc i}] doublet profile shows two peaks of nearly equal intensity with $\delta \lambda\sim$40 \AA, although at +323 d the redder component of the double peak appears suppressed and $\delta \lambda$ decreases to $\sim$30 \AA. The suppression of the redder component increases with time, as can be determined from the spectra at +391 and +2270 d \citep{Milisavljevic2018}. This suppression of the redder component may point to large-scale clumps in the oxygen ejecta at late phases. On the other hand, the blue wing of the [O\,{\sc i}] doublet seems blueshifted by about $\sim$1100 km s$^{-1}$ at +275 d, whereas, with time it approaches zero velocity and presents blueshiftings of $\sim$750 and 700 km s$^{-1}$ in the spectra at +391 and +2270 d, respectively \citep{Milisavljevic2018}. The blueshifted [O\,{\sc i}] doublet profile indicates that photons were emitted near the side of the ejecta \citep{Maeda2008}. Plausible reasons behind the blueshifted narrow peak of the [O\,{\sc i}] doublet on a broader base are large-scale clumping, a unipolar jet, or a single massive blob moving towards the observer \citep{Mazzali2001, Maeda2002, Maeda2006, Taubenberger2009}. The rightmost panel of Fig.~\ref{fig:indivievolu} shows that in the spectral profiles, Mg\,{\sc i}], [O\,{\sc i}] doublet, and [Ca\,{\sc ii}] features at +275, +323, and +391 d are asymmetric around zero velocity, which indicates clearly that synthesized elements are being distributed asymmetrically or clumping occurs in the ejecta of SN~2012au \citep{Taubenberger2009}.

\subsubsection{[O\,{\sc i}] $\lambda$5577 and [O\,{\sc i}] $\lambda\lambda$6300, 6364 doublet}

The [O\,{\sc i}] emission lines in the nebular spectra are used to constrain $M_{O}$, $M_{ZAMS}$, and $M_{He}$. We derived $M_{O}$ in units of M$_\odot$ using the following formula from \cite{Uomoto1986}:

\begin{equation}\label{eq:oxygenmass}
M_{O} = 10^8 \times D^2 \times F([O~I]) \times exp(2.28/T_4).
\end{equation}
 
Here, D represents the host galaxy distance in Mpc (23.5 for NGC~4790; \citealt{Milisavljevic2013}), F([O\,{\sc i}]) is the total flux of the [O\,{\sc i}] doublet in units of erg s$^{-1}$ cm$^{-2}$, and T$_4$ is the temperature of the oxygen-emitting region in 10$^4$ K. The formula discussed above is applicable in the high-density and low-temperature regime, and the ejecta of SNe~Ib favour these conditions \citep{Leibundgut1991, Elmhamdi2004}. The flux ratio of [O\,{\sc i}] $\lambda$5577 and the [O\,{\sc i}] doublet is dependent on the temperature and optical depth \citep{Osterbrock1989}, hence  the assumption of an optically thin regime is used to calculate the O\,{\sc i} temperature. In the case of SN~2012au, [O\,{\sc i}] $\lambda$5577 is not detected clearly in the spectrum at +391 d; however, the spectrum at +323 d \citep[taken from][]{Milisavljevic2013} exhibits significant [O\,{\sc i}] $\lambda$5577 emission and can be used to estimate the O\,{\sc i} temperature. Before calculating the flux values, the spectrum has been scaled to the photometric flux. We infer F([O\,{\sc i}] $\lambda$5577) $\approx$(1.18 $\pm$ 0.56) $\times$ 10$^{-14}$ and F([O\,{\sc i}] doublet) $\approx$(1.12 $\pm$ 0.07) $\times$ 10$^{-13}$ erg s$^{-1}$ cm$^{-2}$. Using the flux values estimated above and equation 2 of \cite{Jerkstrand2014}, we confer an O\,{\sc i} temperature of $\approx$4098.39 $\pm$ 309.15 K assuming $\beta_{ratio}$ = 1.5 \citep{Jerkstrand2014}. This corresponds to $M_{O} \approx$ 1.62 $\pm$ 0.15 M$_\odot$, which is higher than the values for Type Ib SNe tabulated in Table~\ref{tab:tablecomp}, and also higher than the M$_{O}$ range (0.1 to 1.4 M$_\odot$) estimated by \cite{Elmhamdi2004} for a sample of SESNe.

From $M_{O}$, we can also derive $M_{ZAMS}$ and $M_{He}$. As proposed by \cite{Nomoto2006}, for $M_{O} \approx$0.16, 0.77, 1.05, 2.35, 3.22, and 7.33 M$_\odot$, $M_{ZAMS}$ will be 15, 18, 20, 25, 30, and 40 M$_\odot$, respectively, whereas, $M_{ZAMS}$ of 13, 15, and 25 M$_\odot$ produce $M_{He}$ of 3.3, 4, and 8 M$_\odot$, respectively \citep{Thielemann1996}. As a consequence, a $M_{ZAMS}$ of $\sim$20-25 M$_\odot$ and $M_{He}$ of $\sim$4-8 M$_\odot$ are inferred for SN~2012au.

\subsubsection{[Ca\,{\sc ii}] $\lambda$7300 and [O\,{\sc i}] $\lambda\lambda$6300, 6364 doublet}

$M_{ZAMS}$ can also be constrained using the flux ratio between [Ca\,{\sc ii}] $\lambda$7300 and the [O\,{\sc i}] doublet. The [Ca\,{\sc ii}]/[O\,{\sc i}] flux ratio is weakly dependent on density and temperature, whereas its lower value indicates a higher core mass and hence a higher $M_{ZAMS}$ \citep{Fransson1989, Elmhamdi2004, Fang2018, Fang2019}. In the case of SN~2012au, [Ca\,{\sc ii}]/[O\,{\sc i}] increases from $\approx$0.5 to 0.7 spanning between +108 and +323 d, thereafter decreasing to $\sim$0.3 at +391 d. These [Ca\,{\sc ii}]/[O\,{\sc i}] values for SN~2012au are close to those estimated for SN~1998bw \citep{Patat2001, Kuncarayakti2015}, SN~2002ap \citep{Modjaz2014, Kuncarayakti2015}, and SN~2009jf \citep{Sahu2011}. In the case of SN~2012au, the [Ca\,{\sc ii}]/[O\,{\sc i}] ratio of $\sim$0.3 at +391 d indicates a relative abundance of Ca\,{\sc ii}/O\,{\sc i} $\approx$(0.3 to 1.0) $\times$ 10$^{-3}$, as suggested by \cite{Fransson1989}. The theoretical values of [Ca\,{\sc ii}]/[O\,{\sc i}] were estimated by \cite{Fransson1989} for $M_{He}$ and $M_{ZAMS}$ values from 2.68-5.83 M$_\odot$ and 15-25 M$_\odot$, respectively. For SN~2012au, the range of [Ca\,{\sc ii}]/[O\,{\sc i}] supports $M_{He}$ $\approx$5.83 M$_\odot$ and $M_{ZAMS}$ $\approx$25 M$_\odot$ under Model 1b of \cite{Fransson1989}. Throughout the evolution (from +108 to +391 d), the [Ca\,{\sc ii}]/[O\,{\sc i}] ratio values for SN~2012au are lower in comparison with those estimated for most of the SLSNe~I and also suggest a $M_{He}$ of $\approx$5.9 M$_\odot$ via comparison with the ratio from the SESN models of \citealt{Jerkstrand2015} (see fig. 20 of \citealt{Nicholl2019}). Based on the [Ca\,{\sc ii}]/[O\,{\sc i}] values, \cite{Kuncarayakti2015} conjectured a demarcation of SNe progenitors as binary (if [Ca\,{\sc ii}]/[O\,{\sc i}] $>$0.7) or a single WR star (if [Ca\,{\sc ii}]/[O\,{\sc i}] $<$0.7). The range of [Ca\,{\sc ii}]/[O\,{\sc i}] flux ratio for SN~2012au is in agreement with the range of [Ca\,{\sc ii}]/[O\,{\sc i}] ($\sim$0.3-0.7) for SNe~Ib/c that are possibly produced by the explosion of a single massive WR star \citep{Kuncarayakti2015}.

\subsubsection{Mg\,{\sc i}] $\lambda$4571 and [O\,{\sc i}] $\lambda\lambda$6300, 6364 doublet}

Fig.~\ref{fig:Mg_OI_ratio} shows the evolution of the Mg\,{\sc i}] and [O\,{\sc i}] doublet flux ratios estimated using the nebular spectra of SN~2012au from $\sim$+100 to +390 d. The Mg\,{\sc i}]/[O\,{\sc i}] evolution of SN~2012au is also compared with that of Type Ib/c SNe (taken from \citealt{Hunter2009} and references therein) and some of the well-studied SLSNe~I (taken from \citealt{Nicholl2019} and references therein); see Fig.~\ref{fig:Mg_OI_ratio}. A higher Mg\,{\sc i}]/[O\,{\sc i}] ratio is indicative of a higher degree of outer envelope stripping, as more of the O--Ne--Mg layer is exposed \citep{Foley2003}. From $\sim$+100 to +275 d, the Mg\,{\sc i}]/[O\,{\sc i}] ratio of SN~2012au increases from $\sim$0.52 to 0.9, which is high in comparison with all Type Ib/c SNe and comparable to the sample of SLSNe~I presented. However, at +323 d, SN~2012au exhibits the highest value of Mg\,{\sc i}]/[O\,{\sc i}] ratio ($\sim$0.93). The increasing trend of Mg\,{\sc i}]/[O\,{\sc i}] ratio of SN~2012au up to +323 d may be attributed to an Mg\,{\sc i}/O\,{\sc i} abundance effect, symmetric ejecta, or non-symmetric ejecta viewed equatorially \citep{Foley2003}. On the other hand, from +323 to +391 d, the Mg\,{\sc i}]/[O\,{\sc i}] ratio decreases to $\sim$0.27, which is comparable to that of the other SESNe presented (see Fig.~\ref{fig:Mg_OI_ratio}). The plausible reasons behind the decreasing Mg\,{\sc i}]/[O\,{\sc i}] trend observed at later epochs could be attributed to lack of high-density enhancements due to clumping, mixing, or asymmetry in the ejecta \citep{Jerkstrand2015}. However, dust formation and blending of Mg\,{\sc i}] with Fe\,{\sc ii} ions may also be other possible reasons.

\begin{figure*}
\includegraphics[angle=0,scale=0.7]{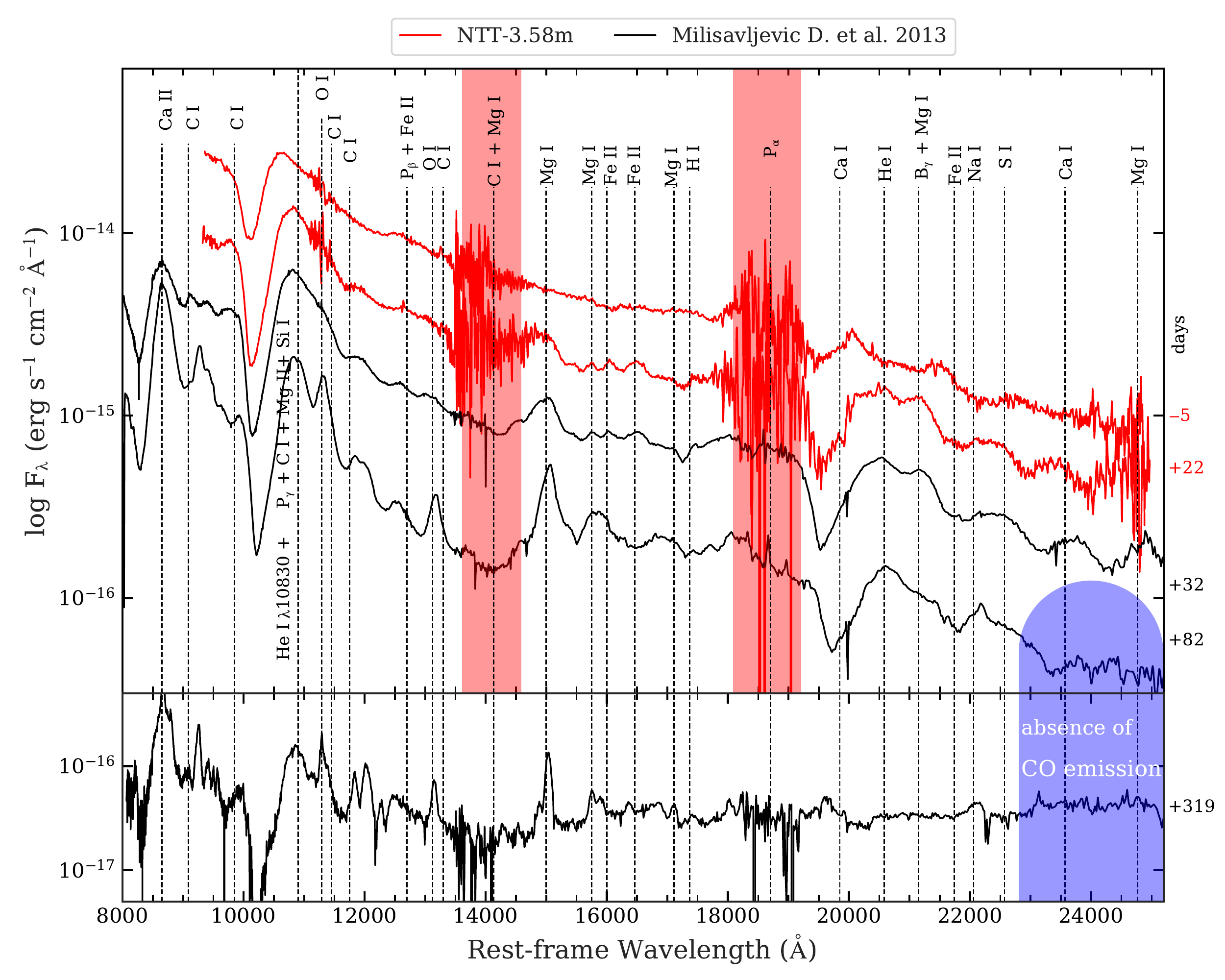}
\caption{$NIR$ spectra of SN~2012au obtained using SOFI on NTT-3.58m (at $-$5.1 and +21.8 d) and taken from \citealt{Milisavljevic2013} (from +32 to +319 d) are presented. Prominent features are shown with vertical dotted lines, whereas vertical red bands denote regions of poor atmospheric transparency. Throughout the evolution, a prominent feature of He\,{\sc i} $\lambda$10830 and absence of the first overtone of CO can be seen.}
\label{fig:spec_nir}
\end{figure*}

\subsection{$NIR$ spectroscopic evolution}\label{sec:spec_nir}

NIR spectroscopy and related studies have been performed for a good number of Type II SNe (e.g., SN~1998dl and SN~1999em: \citealt{Spyromilio2001}), but only for a handful of SESNe so far (SN~2000ew (Ic): \citealt{Gerardy2002}; SN~2007gr (Ic): \citealt{Hunter2009}; LSQ13abf (Ib): \citep{Stritzinger2020}; iPTF13bvn (Ib): \citep{Fremling2016}; SN~2013ge (Ib/c): \citealt{Drout2016}; SN~2016adj (IIb/Ib): \citealt{Banerjee2018}; SN~2020oi (Ic) and SN~2020bvc (Ic): \citealt{Rho2021}). In this section, we present NIR spectroscopy of SN~2012au based on spectra at two epochs (at $-$5 and +21.8 d since maximum) observed using SOFI on NTT-3.58m (see Table~\ref{Ap:table5} for the spectroscopic observations log) along with three published NIR spectra (at +32, +82, +319 d) taken from \cite{Milisavljevic2013}; see Fig.~\ref{fig:spec_nir}. The spectra presented are in rest-frame wavelengths and also corrected for extinction using $E(B - V)$ = 0.063 mag \citep{Milisavljevic2013}, exhibiting a decrease in flux with wavelength, typically observed in SESNe. Line identifications of NIR spectra were done following \cite{Gerardy2002}, \cite{Hunter2009}, \cite{Banerjee2018}, and \cite{Rho2021}. The vertical red bands denote regions of poor atmospheric transparency. The first spectrum (at $-$5 d) appears featureless with few lines due to intermediate-mass elements such as He\,{\sc i}, C\,{\sc i}, Mg\,{\sc i}, O\,{\sc i}, Na\,{\sc i}, whereas, in the post-peak spectra, lines due to iron-group elements begin to appear. As the continuum flux decreases in late-time spectra, prominent features of the hydrogen Paschen series, H\,{\sc i}, He\,{\sc i}, Na\,{\sc i}, O\,{\sc i}, Mg\,{\sc i}, Mg\,{\sc ii}, Si\,{\sc i}, S\,{\sc i}, and Ca\,{\sc i} are traced, highlighted with vertical dotted lines in Fig.~\ref{fig:spec_nir}. Early spectra (up to +32 d) are mainly dominated by P cygni profiles superimposed on a blue continuum, whereas in late-time spectra, broad absorption features start appearing as the continuum fades; however, the spectrum at +319 d is mainly dominated by emission features. 

In the $J$ band ($\sim$10 000--14 000 \AA) region of all the spectra presented, the most prominent absorption feature near $\sim$10 100 \AA~ is identified as very high-velocity He\,{\sc i} $\lambda$10830, see Fig.~\ref{fig:spec_nir}. From $-$5 to +82 d, He\,{\sc i} $\lambda$10830 absorption features correspond to $v_{ph}$ values from $\sim$20 000--17 000 km s$^{-1}$, respectively. The He\,{\sc i} $\lambda$10830 velocity is higher for SN~2012au in comparison with that observed for SN~2016adj ($\sim$14000 km s$^{-1}$) by \cite{Banerjee2018}. However, we caution here that a small contribution from P$_\gamma$, C\,{\sc i} $\lambda$10686, Mg\,{\sc ii} $\lambda$10926, and Si\,{\sc i} $\lambda$10991 to the He\,{\sc i} $\lambda$10830 feature could also be present, as also discussed by \cite{Milisavljevic2013}; see also \cite{Meikle1996}, and \cite{Wheeler1998}. In the $J$-band region, all the spectra presented share nearly similar features except O\,{\sc i} $\lambda$11291. O\,{\sc i} $\lambda$11291 is clearly observed in the late-time spectra at +82 and +319 d, whereas it is absent in the spectra up to +32 d. In the $H$-band ($\sim$15 000--18 000 \AA) region, the pre-maximum spectrum of SN~2012au is nearly featureless, whereas prominent features of Mg\,{\sc i} $\lambda\lambda\lambda$15025, 15040, 15048, and $\lambda$15750, Fe\,{\sc ii} $\lambda\lambda$16000 and 16440, a blended feature of C\,{\sc i} $\lambda$16895 and Mg\,{\sc i} $\lambda$17110, and H\,{\sc i} $\lambda$17370 are present in the post-maximum spectra. In the $K$-band ($\sim$19 000--24 000 \AA) region also, the pre-maximum spectrum is almost featureless except for the prominent Ca\,{\sc i} $\lambda$19850 line; on the other hand, the post-maximum spectra are dominated by He\,{\sc i} $\lambda$20581, a blended feature of B$_\gamma$ and Mg\,{\sc i} $\lambda$21150, Na\,{\sc i} $\lambda$22070, S\,{\sc i} $\lambda$22570, Ca\,{\sc i} $\lambda$23572, and Mg\,{\sc i} $\lambda$24820 elements. 

\subsubsection{Absence of CO emission}

Among SESNe, the first overtone of CO between $\sim$22 900 and 24 000 \AA\, was first detected by \citet{Gerardy2002} in the case of SN~2000ew (Type Ic) and later on in many others, including SN~2016adj (a Type IIb/Ib SN), showing CO features as early as  $\sim$58 d since discovery \citep{Banerjee2018}. In the NIR spectra of SN~2012au at +82 and +319 d, a significant rising of the continuum has not been observed in the $K$ band (see Fig.~\ref{fig:spec_nir}), which indicates an absence of CO emission. The time-scale of CO formation depends on the synthesized C/O mass, metal depletion, and level of mixing in the ejecta \citep{Cherchneff2008}. Moreover, based on previous studies and limited examples (SN 2000ew, SN 2007gr, SN~2016adj, and SN~2020oi), the time-scales of CO formation in the case of SESNe appear to be shorter than those of Type IIP SNe \citep{Banerjee2018, Sarangi2018}.

The absence of CO molecule formation in the case of SN~2012au could be because of the high temperature of the ejecta (above molecule formation threshold) even at late phases. Another possible reason could be the higher mixing of ionized helium between ejecta layers, which can also hinder CO formation, as CO can be quickly destroyed by the presence of ionized helium \citep{Gearhart1999, Gerardy2000, Cherchneff2010}. The presence of ionized helium in Type Ib SNe might therefore also have hampered CO formation, whereas this is not the case with Type Ic SNe \citep{Sarangi2013}, as CO formation is a density-dependent process and a lower C/O velocity is a requirement for CO formation. Hence it is possible that, due to a higher velocity, atmosphere density decreases, resulting in the prevention of CO formation \citep{Gerardy2002}. As higher $v_{ph}$ is observed in SN2012au than in other presented SESNe (see Fig.~\ref{fig:ionvelocity}), this may also be a possible reason behind the absence of CO emission in the NIR spectra of SN 2012au. CO emission also appears to be absent in the +79 d spectrum of iPTF13bvn \citep{Fremling2016}.

\section{Spectral comparison of SN~2012au with other SESNe}
\label{sec:comwithother}

\begin{figure}
\includegraphics[width=\columnwidth]{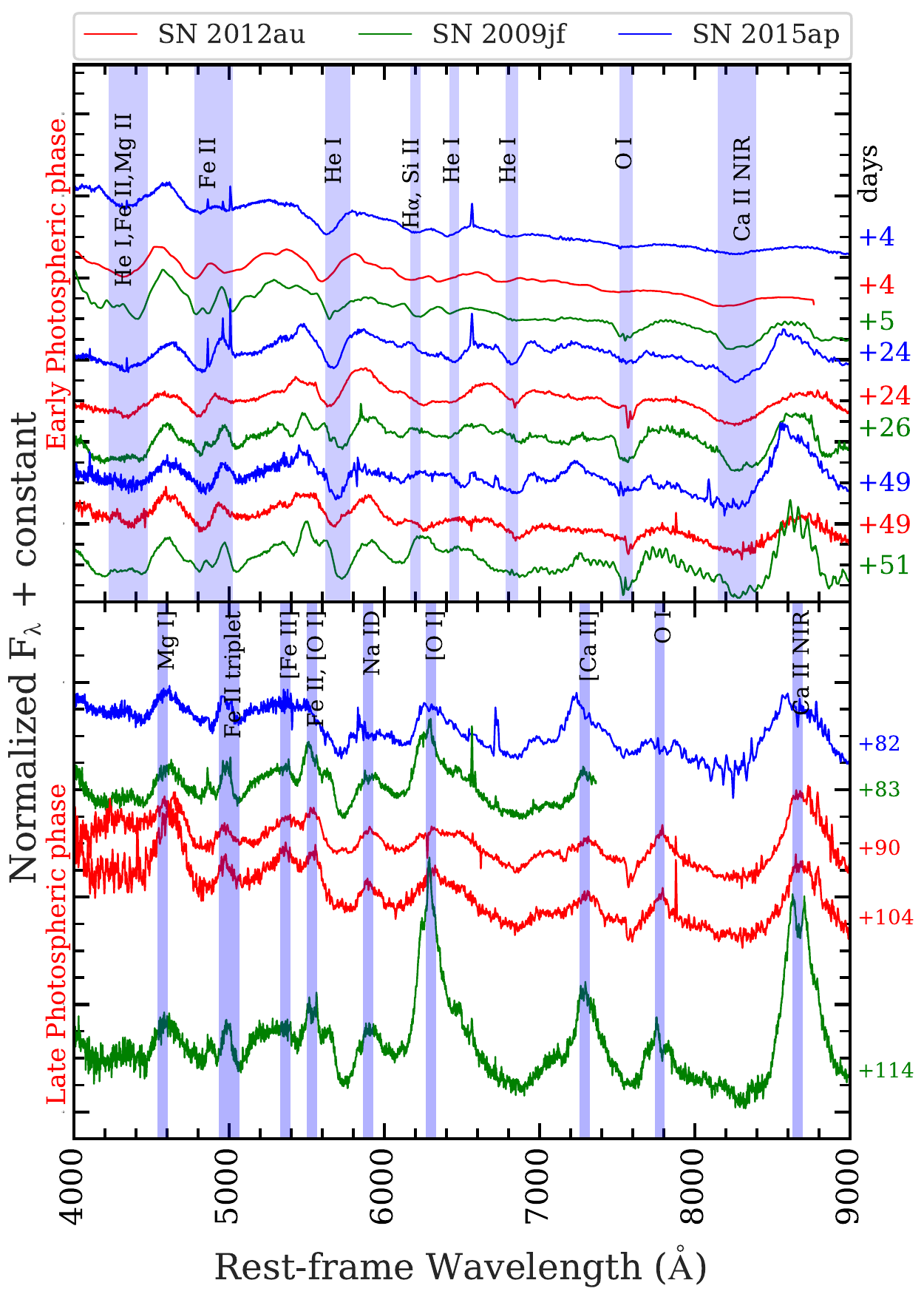}
\caption{Early (from peak to $\sim$+50 d) and late (after $\sim$+80 d) photospheric spectra of SN~2012au compared with SN~2009jf and SN~2015ap at similar epochs in the upper and lower panels , respectively.}
\label{fig:spectracomp}
\end{figure}

Early and late photospheric spectra of SN~2012au (from +4 to +104 d, red colour) are compared with two well-studied SNe~Ib: SN~2009jf \citep[in green:][]{Sahu2011,Valenti2011} and SN~2015ap \citep[in blue:][]{Aryan2021}, see Fig.~\ref{fig:spectracomp}. The spectral evolution of the three SNe~Ib is similar, in particular the P Cygni profile and strength of He\,{\sc i} at $\sim$\,5600 \AA. Throughout the early photospheric phase, the absorption troughs of all elements in the SN~2012au spectra are highly blueshifted (higher velocities) compared with those of SN~2009jf and SN~2015ap. On the other hand, SN~2009jf seems to have narrower though faster-evolving features \citep{Sahu2011}. He\,{\sc i} $\lambda\lambda$6678 and 7065 features are present in the spectra of SN~2012au and SN~2015ap, whereas in the case of SN~2009jf He\,{\sc i} $\lambda$6678 is absent and He\,{\sc i} $\lambda$7065 is weaker. Overall, the early photospheric spectra of SN~2012au closely match those of SN~2015ap. The late photospheric spectra of the three SNe~Ib exhibit the same forbidden lines; see the lower panel of Fig.~\ref{fig:spectracomp}. The main difference is in the evolution of [O\,{\sc i}] $\lambda\lambda$6300, 6363 and [Ca\,{\sc ii}] $\lambda\lambda$7291, 7324, which are weaker in SN~2012au. The [O\,{\sc i}] doublet and the [Ca\,{\sc ii}] lines in the spectra of SN~2009jf and SN~2015ap at $\sim$\,+80 d seem to be stronger in comparison with SN~2012au at $\sim$\,+90 d. This indicates that the photometrically slow-evolving SN~2012au also evolves spectroscopically with longer time-scales.

\begin{figure}
\includegraphics[width=\columnwidth]{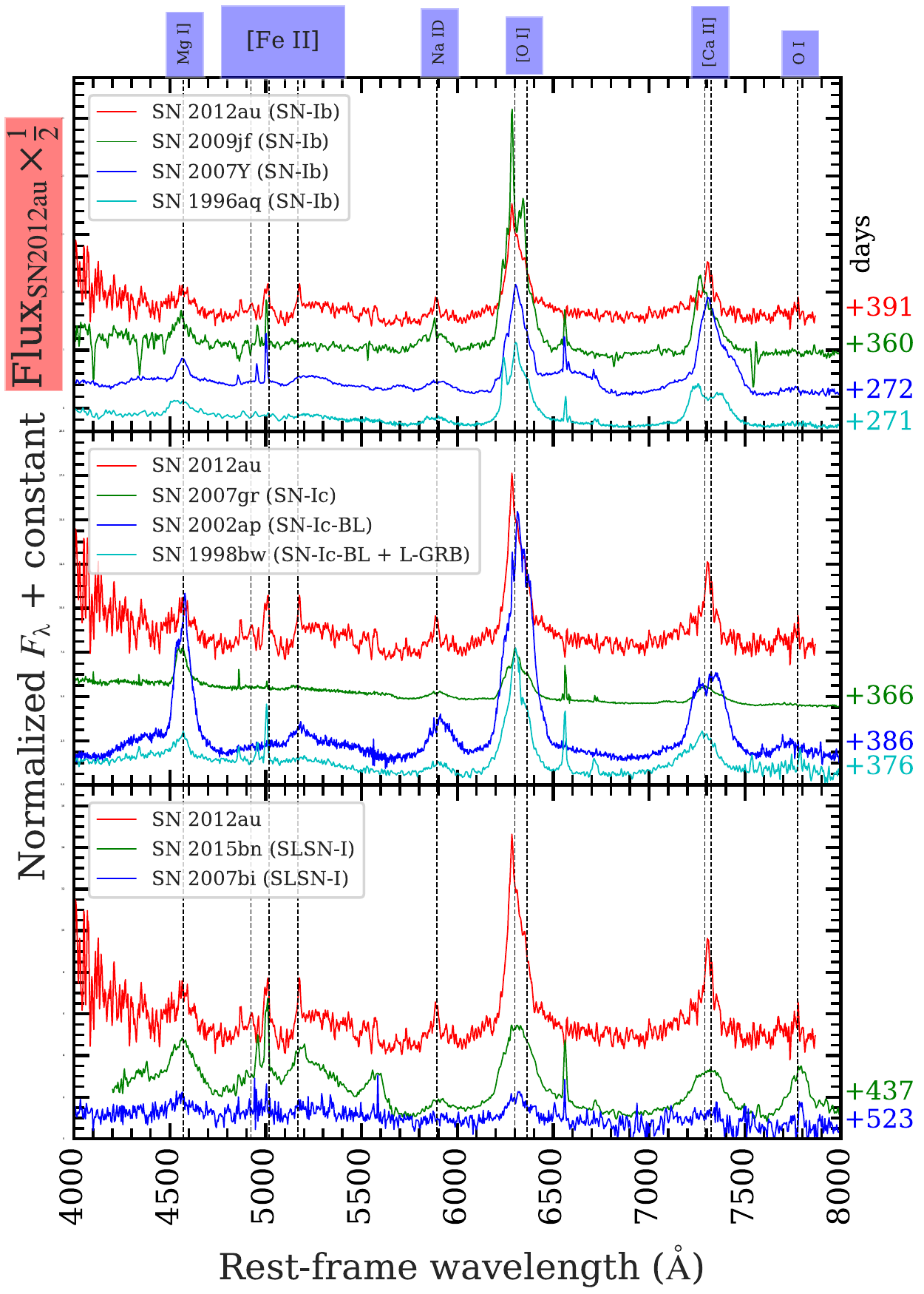}
\caption{The late nebular spectrum (at +391 d) of SN~2012au (in red) is compared with some SNe~Ib, Ic, and slow-decaying SLSNe~I. The comparison spectra of SNe~Ib and Ic are chosen at similar phases to SN~2012au, whereas the spectra of SLSNe~I are taken at later epochs ($>$+400 d).}
\label{fig:nebcomp}
\end{figure}

In Fig.~\ref{fig:nebcomp}, we compare the late nebular phase spectrum of SN~2012au (at +391 d) with some SNe~Ib, Ic, and slow-decaying SLSNe~I. Among SNe~Ib, we chose SN~1996aq (Asiago archive), SN~2007Y \citep{Stritzinger2009}, and SN~2009jf \citep{Sahu2011,Valenti2011}; in SNe~Ic, we chose SN~1998bw \citep[Ic-BL + GRB:][]{Patat2001}, SN~2002ap \citep[Ic-BL:][]{Foley2003}, and SN~2007gr \citep[Ic:][]{Shivvers2019} and from the slow-decaying SLSNe~I, SN~2007bi \citep{Gal-Yam2009} and SN~2015bn \citep{Nicholl2016} have been chosen, based on the availability of very late nebular spectra. 

In the upper panel of Fig.~\ref{fig:nebcomp}, the nebular spectrum of SN~2012au is compared with those of the three SNe~Ib mentioned above. Notably, the [O\,{\sc i}] and [Ca\,{\sc ii}] emission lines of SN~2012au match closely those of SN~2007Y. The spectra of SN~2012au and SN~2007Y show a single-peaked [O\,{\sc i}], unlike SN~1996aq and SN~2009jf, the [O\,{\sc i}] features of which have a double peak with different intensities. The [Ca\,{\sc ii}] emission peak of SN~2009jf appears to be blueshifted on the other hand SN~2012au presents a redshifted single emission peak. The [O\,{\sc i}] $\lambda$7774 feature is missing at these late phases in all SNe~Ib considered here. We also compare the nebular spectrum of SN~2012au with three SNe~Ic, see the middle panel of Fig.~\ref{fig:nebcomp}. The three SNe~Ic show similar lines to SN~2012au, although these lines are strongest in SN~2002ap (Ic-BL). SNe~Ic exhibit weak emission lines of O\,{\sc i} $\lambda$7774, which is nearly absent in the case of SN~2012au. In the lower panel of Fig.~\ref{fig:nebcomp}, we compare the nebular spectrum of SN~2012au with two slow-decaying SLSNe~I. At such late phases, SLSNe~I appear to share similar spectral features to SNe~Ib/c, as also suggested by \cite{Pastorello2010} and \cite{Quimby2018}. SN~2012au and SLSNe~I appear to have similar emission lines, although they are broader in SLSNe~I. The emission lines evolve faster in the case of SN~2012au than in SN~2007bi and SN~2015bn, which have stronger O\,{\sc i} $\lambda$7774.

\begin{figure*}
\centering
\includegraphics[width=\textwidth]{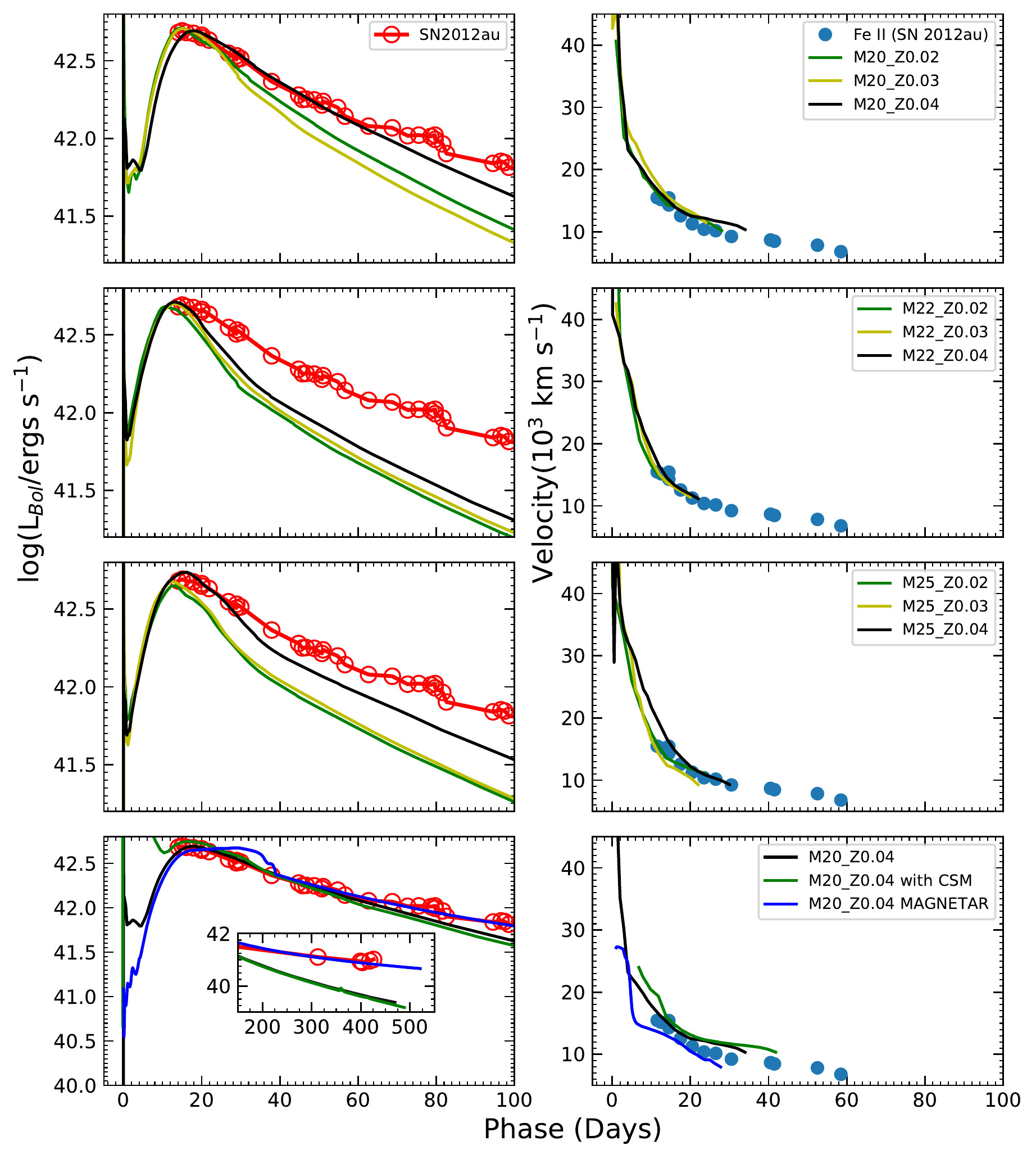}
\caption{The panels in the left column represent various light curves obtained with progenitor masses 20 M$_{\odot}$ (top panel), 22 M$_{\odot}$ (second panel from top), and 25 M$_{\odot}$ (third panel from top), respectively, and with different metalicities. The leftmost bottom panel shows the 20 M$_{\odot}$ model with Z = 0.04 that best explains the observed light curve.\ All these models are generated using {\tt STELLA}, except the case where {\tt SNEC} is used to employ the magnetar powering mechanism. The inset plot in the bottom light-curve plot shows the improvement achieved due to inclusion of the magnetar model over other models. The panels in the right column compare Fe\,{\sc ii} line velocities of our models with those obtained through the absorption troughs of the Fe\,{\sc ii} lines in the observed spectra.}
\label{fig:STELLA}
\end{figure*}

\section{Progenitor modelling using {\tt MESA}}\label{sec:MESA}

{\tt MESA} is a hydrodynamical simulation code that evolves a certain ZAMS progenitor up to the stage where the core of the modelled progenitor is about to collapse. Further, the output of {\tt MESA} is fed as input to {\tt STELLA} and {\tt SNEC}, which solve the radiative transfer equations and simulate the synthetic explosion. As a result of the synthetic explosion, supernova light curves and $v_{ph}$ evolution, along with many other parameters, are generated. The light curves and $v_{ph}$ evolution are then matched with the observed ones. Starting from the ZAMS,  we are able to match the light curves and the photospheric velocities, so we have replicated the actual supernova explosion. Here, the explosions using {\tt STELLA} and {\tt SNEC} are complemented by the parameters obtained using {\tt MINIM}. With a certain ZAMS mass progenitor, the explosion parameters in {\tt STELLA} and {\tt SNEC} are kept close to those obtained using {\tt MINIM}. Thus, {\tt MESA} is very useful to put important constraints on progenitor mass and other progenitor properties like metallicity, rotation, etc.

\begin{table*}
\caption{{\tt MESA} model explosion parameters.}
\label{tab:MESA_MODELS}
\begin{center}
{\scriptsize
\begin{tabular}{ccccccccccccc} 
\hline\hline
Model name	& M$_{ZAMS}$	& Z	&  ($\nu$/$\nu_{c})_{ZAMS}$ $^{a}$ & f$_{ov}$ $^{b}$ &	$M_{\mathrm{f}}$ $^{c}$	&	$M_\mathrm{c}$ $^{d}$	&	$M_{\mathrm{ej}}$	&	$M_{\mathrm{Ni}}$ &	$E_{\mathrm{exp}}$ $^{e}$ &	$t_{\mathrm{CSM}}$ $^{f}$ & $\dot{M}y^{-1}$	 &	$v_{\mathrm{CSM}}$ $^{g}$	\\
	&	(M$_{\odot}$)	&	 &  &  & (M$_{\odot}$)	&	(M$_{\odot}$) 	&	(M$_{\odot}$) &	(M$_{\odot}$) &	(10$^{51}$ erg) &	(y) &	($M_{\mathrm{\odot}}y^{-1}$) &	(km s$^{-1}$)	\\
\hline
\hline
M20$\_$Z0.02    &	20  	&	0.02   & 0.0    & 0.01   &	6.74   & 1.78  & 4.96	& 0.23  & 15 & 0.0  & 0.0	& 0.0 	\\
M20$\_$Z0.03    &	20  	&	0.03   & 0.0    & 0.01   &	6.85   & 1.77  & 5.08	& 0.23  & 15 & 0.0  & 0.0	& 0.0	\\
M20$\_$Z0.04    &	20  	&	0.04   & 0.0    & 0.01   &	7.07   & 2.11  & 4.96	& 0.23  & 15 & 0.0  & 0.0	& 0.0	\\
M20$\_$Z0.04$\_$CSM    &	20  	&	0.04   & 0.0    & 0.01   &	6.74   & 1.78  & 4.96	& 0.23  & 15 & 1.3  & 0.0001	& 12.0 	\\
M20$\_$Z0.04 MAGNETAR    &	20  	&	0.04   & 0.0    & 0.01   &	6.74   & 1.44  & 5.30	& 0.00  & 5.0 & 0.0  & 0.0	& 0.0 	\\
M22$\_$Z0.02    &	22  	&	0.02   & 0.0    & 0.01   &  7.35   & 2.35  & 5.00	& 0.21  & 15 & 0.0  & 0.0	& 0.0	\\
M22$\_$Z0.03    &	22  	&	0.03   & 0.0    & 0.01   &	7.50   & 1.86  & 5.14	& 0.22  & 15 & 0.0  & 0.0	& 0.0	\\
M22$\_$Z0.04    &	22  	&	0.04   & 0.0    & 0.01   &  7.71   & 1.64  & 6.07	& 0.20  & 15 & 0.0  & 0.0	& 0.0	\\
M25$\_$Z0.02    &	25  	&	0.02   & 0.0    & 0.01   &	8.50   & 1.97  & 6.53	& 0.21  & 15 & 0.0  & 0.0	& 0.0	\\
M25$\_$Z0.03    &	25  	&	0.03   & 0.0    & 0.01   &	8.63   & 1.99  & 6.64	& 0.25  & 15 & 0.0  & 0.0	& 0.0	\\
M25$\_$Z0.04    &	25  	&	0.04   & 0.0    & 0.01   &	9.35   & 1.85  & 7.5	& 0.20  & 15 & 0.0  & 0.0	& 0.0	\\
\hline\hline
\end{tabular}}
\end{center}
{$a$ initial rotation, $b$ overshooting parameter, $c$ final mass, $d$ mass of the central remnant}\\
{$e$ explosion energy. Parameters controlling the extent of CSMI; $f$ wind duration, $g$ CSM velocity. }\\
\end{table*}

For SN~2012au, our analysis supports a progenitor mass of $\sim$20-25 $M_{\odot}$. Following \citet{Milisavljevic2013} and \citet{Kamble2013} where the metallicity (Z) was determined using the methods of \citet{Sanders2012}, and also using the N2 diagnostic of \citet{Pettini2004}, we assume the metallicity at the site of SN~2012au to be around 1--2 Z$_{\odot}$. We modelled three non-rotating progenitor scenarios with $M_{ZAMS}$ of 20 M$_{\odot}$, 22 M$_{\odot}$, and 25 M$_{\odot}$, each of them at three different Z values of 0.02 Z$_{\odot}$, 0.03 Z$_{\odot}$, and 0.04 Z$_{\odot}$. In all, we obtain a total of nine models. The model designation contains the mass and the Z value. For instance, the M20$\_$Z0.02 model represents a progenitor of $M_{ZAMS}$ = 20 M$_{\odot}$ and with Z = 0.02 Z$_{\odot}$, while the M25$\_$Z0.04 model indicates a progenitor of $M_{ZAMS}$ 25 M$_{\odot}$, with Z = 0.04 Z$_{\odot}$. One additional model, designated with the CSM mark, also includes the contribution of CSMI in the light curve. Using {\tt MESA} \citep[version 11701:][]{Paxton2011, Paxton2013, Paxton2015, Paxton2018}, the models are evolved up to the stage of onset of rapid infall of the iron core. After excising the central core mass and when the shock is near breakout, we provide the corresponding output as input to the public version of {\tt STELLA} \citep{Blinnikov1998, Blinnikov2000, Blinnikov2006} included in {\tt MESA}. {\tt STELLA} evolves the model through shock breakout and beyond, generating the light curves and velocity evolution. Here, we briefly summarize the methods and assumptions for the different models.

Adopting the Ledoux criterion, we model the convection adopting the mixing theory of \citet{Henyey1965}. The mixing length parameter is set to be $\alpha = 3.0$ in the region where the mass fraction of H is greater than 0.5, and 1.5 in other regions. Following \citet{Langer1985}, semi-convection is modelled with an efficiency parameter of $\alpha_{\mathrm{sc}} = 0.01$. We follow \citet{Kippenhahn1980} for thermohaline mixing, with an efficiency parameter of $\alpha_{\mathrm{th}} = 2.0$. Convective overshooting is modelled with the diffusive approach of \citet{Herwig2000}, with $f = 0.01$ and $f_0 = 0.004$ for both the convective core and shells. The {\tt DUTCH} \citep{Vink2001, Nugis2000} scheme is used for the stellar wind, with a scaling factor of 1.0. SNe~Ib are assumed to originate from massive stripped-envelope stars, which have lost their outer H envelope via binary interactions \citep{Yoon2010, Dessart2012, Eldridge2016, Ouchi2017} or because of strong stellar winds \citep[e.g.,][]{Gaskell1986, Eldridge2011, Groh2013}. To strip the H envelope artificially, the model evolves until the exhaustion of helium; we then impose an artificial $\dot{M} \gtrsim 10^{-4} M_{\odot} \mathrm {yr}^{-1}$ until the total H mass of the star decreases to 0.01 M$_{\odot}$. On reaching the specified H-mass limit, we switch off artificial mass loss, and the model evolves until the onset of core collapse. The final parameters for all models are listed in Table \ref{tab:MESA_MODELS}. Fig.~\ref{fig:STELLA} shows the {\tt STELLA/SNEC} results obtained for the models mentioned above. Our models with explosion energy of 15$\times10^{51}$ erg and $M_{\rm Ni}$ in the range 0.2 -- 0.3 M$_{\odot}$ along with CSMI can explain the observed peak luminosity fairly well, but fails to explain the overall light-curve shape beyond the peak light of SN~2012au. These models could satisfactorily explain the photospheric velocities (inferred from Fe\,{\sc ii} line velocities) as shown in the right panels of Fig.~\ref{fig:STELLA}. The semi-analytical light curve modelling using {\tt MINIM} shows the possibility of SN~2012au being powered by a magnetar. Following \citet[][]{Metzger2015}, the magnetar spin-down luminosity is given by
\[
L_{\mathrm{sd}} = L_{\mathrm{sd_i}} (1 + t/t_{\mathrm{sd}})^{-2}.
\]
\noindent
Here, $L_{\mathrm{sd_i}}$ is the spin-down luminosity at $t=0$, and $t_{\mathrm{sd}}$ is the initial spin-down time. This is the luminosity that we inject into the whole ejecta above the mass cut uniformly in mass. The explosion using the magnetar model is simulated using another publicly available code {\tt SNEC} \citep[][]{Morozova2015} by following the methods of \citet[][]{Aryan2021}. For the initial spin-down luminosity, we assume $L_{\mathrm{sd{_{i}}}} = 2.2 \times 10^{43}$\,erg\,s$^{-1}$, while for the initial spin-down time we assume $t_{\mathrm{sd}} = 25$\,d (close to the value obtained from {\tt MINIM}). The effects of Ni heating are ignored in this model. Following \citet[][]{Metzger2015} (their equations 2 and 3), corresponding to $L_{\mathrm{sd{_{i}}}} = 2.2 \times 10^{43}$\,erg\,s$^{-1}$ and $t_{\mathrm{sd}} = 25$\,d, we obtain $B$ $\sim$ $2.0 \times 10 ^{14}$\,G and $P_i$ $\sim$ 23\,ms for the modelled magnetar. These values of $B$ and $P_i$ are close to those obtained using {\tt MINIM}. We see that a 20 M$_{\odot}$ ZAMS progenitor with Z = 0.04 could explain the observed bolometric luminosity and photospheric velocities nicely, assuming the magnetar powering mechanism. Thus, the inclusion of the magnetar powering mechanism greatly improves light-curve matching even in late phases, as shown in the inset plot of the leftmost bottom panel of Fig.~\ref{fig:STELLA}.

\section{Discussion and Results}
\label{sec:Descussion}

In this work, we present well-calibrated optical photometric ($-$0.2 to +413 d), polarimetric ($-2$ to +31 d) and optical ($-$5 to +391 d), NIR ($-$5 to +22 d) spectroscopic studies of SN~2012au, based on data obtained using many observational facilities around the globe. Analysis based on our photometric observations suggests that SN~2012au appears to be one of the most luminous SNe~Ib (M$_{B,peak}$ = $-$18.06 $\pm$ 0.12 mag), though fainter than the threshold limit of SLSNe~I \citep[M$_g$ $<-$19.8 mag;][]{Quimby2018}. The M$_{R,peak}$ ($\sim$ --18.67\,$\pm$\,0.11 mag) of SN~2012au is brighter than the average values of SNe~Ib and Ic, but closer to those reported for SNe Ic-BL \citep{Drout2011}. Similarly, the peak bolometric luminosity of SN~2012au ($\sim$\,[6.56 $\pm$ 0.70] $\times$ 10$^{42}$ erg s$^{-1}$) is higher than the mean peak luminosities of SNe~Ib and Ic, but still lower than those of SNe~Ic-BL \citep{Lyman2016}. Using the early bolometric light curve of SN~2012au, the estimated values of $M_{ej}$, $E_{k}$, $M_{Ni}$, and $T_0$ are $\sim 5.1 \pm$ 0.7 M$_\odot$, $\sim$\,(4.8 $\pm$ 0.6) $\times$ 10$^{51}$ erg, $\sim$0.27-0.30 M$_\odot$, and $\sim 66.0 \pm$ 9.4 d, respectively. These physical parameters of SN~2012au are close to those inferred for SN~2009jf \citep[a bright SN~Ib:][]{Sahu2011} and-on average-larger than for classical SNe~Ib/c but smaller than for some SNe~Ic-BL. SN~2012au manifests larger M$_{ej}$ and M$_{Ni}$ in comparison with most of the SNe~IIb, Ib, and Ic, which may be the prime reason behind the luminous peak of SN 2012au, as seen in the case of SLSNe~I \citep{Nicholl2015}. On the other hand, light-curve decline rates of SN~2012au (at phases $\geq$+40 d) in all the optical bands are shallower than typically observed in the case of SNe~Ib and slow-decaying SLSNe~I, and theoretically predicated for \isotope[56]{Co} $\rightarrow$ \isotope[56]{Fe} decay. As SN~2012au exhibits comparatively larger M$_{ej}$, a larger optical depth resulting in a larger diffusion time-scale (for the trapped energy to cross the outer envelope) could broaden the light curve. Therefore, high trapping of gamma-rays at late phases or higher opacity of massive ejecta are among the plausible interpretations for the modest luminosity decline rate of SN~2012au in comparison with other SNe~Ib \citep{Clocchiatti1997}. However, smoothly distributed circumstellar media up to a larger radius could be another possibility behind the late time shallower decay rate for SN~2012au, but an absence of the CSMI in the late-time spectra ruled out this scenario \citep{Milisavljevic2018}. The late-time bolometric light curve of SN~2012au is better constrained by $L~\varpropto$ $t^{-2}$, a conventional magnetic dipole equation. Hence for SN~2012au the shallower decay of the late-time light curve might be a potential indicator of a central engine powering source that is accelerating the inner ejecta.

The analytical light-curve modelling of SN~2012au using the {\tt MINIM} code infers that a spin down millisecond magnetar plausibly powers the observed luminosity of SN~2012au. This outcome is also consistent with the late-time light-curve reproduced by the equation of a magnetic dipole. Some of the physical parameters obtained using the {\tt MINIM}/MAG model are $M_{\rm ej} \approx 4.72 \pm$ 1.03 M$_\odot$, $v_{\rm exp} \sim(11.66 \pm 0.58) \times 10^3$ km s$^{-1}$, $B \sim(8.05 \pm 0.15) \times 10^{14}$ G, and $P_{\rm i} \sim 18.26 \pm 0.01$ ms. The MAG model gives reasonable values of $M_{\rm ej}$ and $v_{\rm exp}$, closer to those obtained from the present photometric and spectral analysis. For the magnetar powering source of SN~2012au, $B$ is closer and $P_{\rm i}$ is higher than observed in the case of Type Ib SN~2005bf \citep{Maeda2007}. On the other hand, for SN~2012au $P_i$ is higher and $B$ lies at the top of the range of $P_i$ and $B$ values ($\sim$1-8 ms and $\sim$1-8 $\times$ 10$^{14}$ G, respectively) compared with those observed for SLSNe~I (see fig.10 of \citealt{Kumar2021}).

Following the results obtained using semi-analytical light curve modelling code {\tt MINIM}, a magnetar powering mechanism is built through {\tt MESA} and {\tt SNEC}. After implementing the magnetar powering mechanism, a 20\,M$_{\odot}$ ZAMS progenitor with Z = 0.04 could reproduce the observed bolometric luminosity and $v_{ph}$ reasonably. The $^{56}$Ni $\rightarrow$ $^{56}$Co decay and CSM interaction powering mechanisms using {\tt MESA} and {\tt STELLA}, adopting an explosion energy of 15$\times10^{51}$ erg and $M_{\rm Ni}$ of $\sim$ 0.2 M$_{\odot}$, produced a rather poorer match to the observed bolometric light curve. We could see that the inclusion of the  magnetar powering mechanism could significantly improve the fit even in the later phases (+300 to +400 d). These results enhance the probability of the SN~2012au explosion being powered by a magnetar.

Our imaging polarimetric data of SN~2012au show signatures of asphericity in the ejecta. The observed polarization values of SN~2012au are significantly higher than those of the field stars lying within a $10\degr$ radius of the SN~location. Similar to SN~2008D, we observe a variation in the polarization parameters of SN~2012au. Near the peak, SN~2012au resembles a \%P-value closer to those of SN~2006aj and SN~2007uy. Among Type Ib SNe, SN~2012au presents higher \%P, M$_{Ni}$, and E$_{k}$/M$_{ej}$ values in comparison with SN~2008D and iPTF13bvn, but closer to those observed in the case of SN 2007uy.

Spectroscopic observations of SN~2012au reveal that it shares spectral features with a typical SNe~Ib. The absence of the prominent W-shaped O\,{\sc ii} features in the near-peak spectra of SN~2012au confirm its spectral divergence from SLSNe~I \citep{Quimby2011, Quimby2018}. The emergence of O\,{\sc ii} features generally requires a photospheric temperature of $\gtrsim$12 000 K, which could be obtained very easily in the hot ejecta of SLSNe~I. On the other hand, near the peak, SN~2012au presents a photospheric temperature of $\approx$9000 K for the near-peak spectra. In the hot photospheric phase, SN~2012au, like other classical SNe~Ib, exhibits prominent He\,{\sc i} ($\lambda\lambda$5876, 6678, and $\lambda$7065) features from the start of our spectral observations (from $-$5 d). Near the maximum light of SN~2012au, expansion velocities of He\,{\sc i} and Ca\,{\sc ii nir} are $\sim 14 500$ km s$^{-1}$, whereas Fe\,{\sc ii} displays slightly lower velocity ($\sim$12 500 km s$^{-1}$), while these line velocities are higher than those of typical SNe~Ib. The photospheric velocity (Fe\,{\sc ii} ion velocity) of SN~2012au decays faster than that of SLSNe~I \citep{Nicholl2015}. Using the $v_{ph}$ obtained from the {\tt SYNAPPS} spectral fitting to the near-peak spectrum of SN~2012au, we constrained values of $r_{ph}$  $\sim$1.8 $\times$ $10^{15}$ cm, $\tau_{total}$ $\sim$63.8, $M_{ej}$ $\sim$8.3 M$_{\odot}$, and $E_k$ $\sim$5.4 $\times$ 10$^{51}$ erg.

The [O\,{\sc i}] $\lambda\lambda$6300, 6363 doublet and [Ca\,{\sc ii}] $\lambda\lambda$7291, 7324 emerge later (from +90 d) in the nebular phase compared with O\,{\sc i} $\lambda$7774, Na\,{\sc i} D, Mg\,{\sc i}], Ca\,{\sc ii} H\&K, and NIR, and the Fe\,{\sc ii} triplet features can be seen from +60 d. In the spectra at +275, +323, and +391 d, profiles of Mg\,{\sc i}], the [O\,{\sc i}] doublet, and [Ca\,{\sc ii}] features are asymmetric which is indicative of synthesized elements being distributed asymmetrically or mixing and clumps in the ejecta of SN~2012au. The plausible reasons behind the asymmetric blueshifted [O\,{\sc i}] line profile as a narrow peak on a broader base are large-scale clumping, a unipolar jet, or a single massive blob moving towards the observer. These findings about asymmetry are in conformity with those obtained using the imaging polarimetric observations discussed above. The nebular spectrum of SN~2012au at +391 d is closely matched with the modelled spectrum produced for a progenitor with $M_{ZAMS}$ of 17 M$_\odot$ by \cite{Jerkstrand2015}, considering strong mixing and dust in the ejecta. From the spectrum at +323 d of SN~2012au and equation 2 of \cite{Jerkstrand2014}, we infer F([O\,{\sc i}] $\lambda$5577) $\approx$(1.18 $\pm$ 0.56) $\times$ 10$^{-14}$ and F([O\,{\sc i}] doublet) $\approx$(1.12 $\pm$ 0.07) $\times$ 10$^{-13}$ erg s$^{-1}$ cm$^{-2}$, which confers O\,{\sc i} temperature $\approx$4098.39 $\pm$ 309.15 K and $M_{O} \approx$ 1.62 $\pm$ 0.15 M$_\odot$. The M$_{O}$ of SN~2012au is higher than the one observed for SN~2009jf \citep[1.34 M$_\odot$:][]{Sahu2011} and also higher than the M$_{O}$ range (0.1-1.4 M$_\odot$) estimated for a sample of SESNe by \cite{Elmhamdi2004}. Using this $M_{O}$, we also constrained values of $M_{ZAMS}$ ($\sim$20-25 M$_\odot$) and $M_{He}$ ($\sim$4-8 M$_\odot$). The $M_{ZAMS}$ of SN~2012au is higher than those for well-studied SNe~Ib (except SN~2009jf) but lower than for the SLSNe~I tabulated in Table~\ref{tab:tablecomp}.

In nebular spectra of SN~2012au, there are no signatures of CSMI up to +391 d. Additionally, \cite{Milisavljevic2018} also did not notice any signatures of CSMI in a very late-time (at +2270 d) spectrum of SN~2012au. On the other hand, a pulsar wind nebula as a heating source that could be generated by the spin-down power of a central pulsar was proposed by \cite{Milisavljevic2018}, which is in agreement with the results from analytical light-curve modelling as described above. Based on the [Ca\,{\sc ii}]/[O\,{\sc i}] ratio of SN~2012au and the findings of \cite{Kuncarayakti2015}, we suggest that the progenitor of SN~2012au is a single WR star with $M_{ZAMS}$ $\approx$25 M$_\odot$. SN~2012au displays a higher Mg\,{\sc i}]/[O\,{\sc i}] flux ratio in comparison with typical Type Ib/c events, though comparable to some of the well-studied SLSNe~I. A higher value of Mg\,{\sc i}]/[O\,{\sc i}] ratio of SN~2012au up to +323 d indicates a higher degree of outer envelope stripping \citep{Foley2003}. From +320 to +390 d, the Mg\,{\sc i}] and [O\,{\sc i}] flux ratio of SN~2012au decreases sharply to $\sim$0.27, which indicates a lack of high-density enhancements due to clumping, mixing, and asymmetry in the ejecta \citep{Jerkstrand2015}. However, dust formation and blending of the Mg\,{\sc i}] from Fe\,{\sc ii} ions may be other plausible reasons.

In addition to optical spectral observations, our first NIR spectrum at $-$5 d appears featureless with a few lines of He\,{\sc i} $\lambda$10800 (the most prominent one), C\,{\sc i}, Mg\,{\sc i}, O\,{\sc i}, and Na\,{\sc i}. The strength of absorption features reduces over time and is superseded by other prominent emission lines. The late-time spectra up to +319 d exhibit prominent features of the hydrogen Paschen series, H\,{\sc i}, He\,{\sc i}, Na\,{\sc i}, O\,{\sc i}, Mg\,{\sc i}, Mg\,{\sc ii}, Si\,{\sc i}, S\,{\sc i}, Ca\,{\sc i}, and Fe\,{\sc ii}. The first overtone of CO between $\sim$22 900 and 24 000 \AA\ is absent in the NIR spectra of SN~2012au (up to +319 d). The CO emission was also absent in the +79 d spectrum of iPTF13bvn \citep{Fremling2016}. The absence of CO molecule formation in the case of SN~2012au could be because of the temperature being higher than the molecule formation threshold, higher mixing of ionized helium between ejecta layers, or a higher $v_{ph}$ of SN~2012au. The absence of CO emission in the spectra of SN~2012au up to +319 d confirms the lack of emission signatures by heated dust.

In both hot and cool photospheric phases, SN~2012au shares an overall spectral similarity with SN~2015ap, though, absorption features in the spectra SN~2012au have a higher blueshift than in the case of SN~2009jf and SN~2015ap, which is in agreement with the comparatively higher $v_{ph}$ of SN~2012au. In the nebular phase, some of the spectral features, in particular, the [O\,{\sc i}] and the [Ca\,{\sc ii}] doublets, evolve at later times than those observed in other SNe~Ib. In the late phases ($>$+250 d), the [O\,{\sc i}] and [Ca\,{\sc ii}] emission lines of SN~2012au match closely those observed in the case of SN~2007Y. Overall, the spectral comparison performed in the present study reveals that the photometrically slow-decaying SN~2012au also evolves spectroscopically on longer time-scales.

\section{Conclusion}
\label{sec:CONclusion}

In this study, we present well-calibrated photometric, polarimetric, and spectroscopic studies of SN~2012au spanning from 5 d before the $B$-band maximum to nearly one year post-maximum. SN~2012au exhibits higher peak luminosity in comparison with typical Type IIb, Ib, and Ic but lower than those observed for Type Ic-BL and SLSNe~I. The peak bolometric luminosity of SN~2012au implies a synthesis of $\sim$0.27-0.30 M$_\odot$ of \isotope[56]{Ni} during the explosion. For SN~2012au, M$_{ej}$ values constrained using photometric light-curve analysis ($\sim$ 5.1 M$_\odot$), semi-analytical modelling ($\sim$ 4.7 M$_\odot$), and spectral analysis ($\sim$ 8.3 M$_\odot$) hinted a range of $M_{\rm ej}$ $\sim$4.7 -- 8.3 M$_\odot$ for SN~2012au. The M$_{ej}$ ($\sim$4.7 -- 8.3 M$_\odot$) and $E_{k}$ ($\sim$\,[4.8 -- 5.4] $\times$ 10$^{51}$ erg) of SN~2012au are also higher in comparison with most Type IIb, Ib, and Ic SNe, but lower than those of Ic-BL/GRB-SNe and SLSNe~I. Therfore, based on the peak brightness and inferred physical parameters, SN~2012au appears more like a bridge between normal Ib/c and Ic-BL, rather than Ib/c and SLSNe~I. Comparatively higher values of $M_{\rm Ni}$ and $M_{\rm ej}$ could explain the rather luminous peak of SN~2012au. SN~2012au presents the most shallow post-peak decay rate in comparison with Type Ib and slow-decaying SLSNe~I, which indicates high trapping of gamma-rays or higher opacity of the massive ejecta.

Analytical light-curve modelling insinuates a spin-down millisecond magnetar with $B \approx(8.05 \pm 0.15) \times 10^{14}$ G and $P_{\rm i} \approx18.26 \pm 0.01$ ms as a likely powering source for SN~2012au. The late-time light curve of SN~2012au is flatter than the theoretical \isotope[56]{Co} $\rightarrow$ \isotope[56]{Fe} decay curve and consistent with the equation of the standard magnetic dipole, also favouring a central engine powering source. Additionally, results from {\tt MESA} and {\tt SNEC} support a magnetar powering mechanism with a progenitor having M$_{ZAMS}$ $\sim$20\,M$_{\odot}$ and Z = 0.04. However, $^{56}$Ni $\rightarrow$ $^{56}$Co decay and CSM interaction powering mechanisms produced a comparatively poorer match in {\tt MESA}. Based on the absence of narrow H Balmer lines in the late-time spectrum (at +2270 d), \cite{Milisavljevic2018} exclude the possibility of CSMI as the primary powering source of SN~2012au. In addition, \cite{Milisavljevic2018} propose a pulsar wind nebula as a heating source of SN~2012au that could be generated by the spin-down power of a central pulsar. Also, a central engine powering source is very likely to explain observed asphericity, mixing, and the comparatively higher $v_{ph}$ of SN~2012au ejecta \citep{Chen2020}. These results together strengthen the probability of the SN~2012au explosion being powered by a magnetar.

Imaging polarization values of SN~2012au are significantly higher than those of field stars, favouring asphericity in the ejecta. Also, in the nebular spectra of SN~2012au ($\geq$270 d), asymmetric profiles of Mg\,{\sc i}], the [O\,{\sc i}] doublet, and [Ca\,{\sc ii}] features indicate synthesized elements being distributed asymmetrically and/or clumpy ejecta. However, the shape of the [O\,{\sc i}] line profile is a narrow peak on a broader base that is blueshifted and suggests large-scale clumping, a unipolar jet, or a single massive blob moving towards the observer. The sharply decreasing behaviour of the Mg\,{\sc i}]/[O\,{\sc i}] flux ratio of SN~2012au from +323 to +391 d indicates clumping, mixing, and asymmetry in the ejecta. The first overtone of CO is absent in the NIR spectra of SN~2012au, indicative of high temperature and strong mixing of ionized helium between ejecta layers, and shows a lack of emission signatures by heated dust. Overall, these observational signatures favour mixing and asymmetry in the ejecta of SN~2012au.

The spectral evolution of SN~2012au is similar to that of other typical SNe~Ib. However, spectral comparison reveals that the photometrically slow-decaying SN~2012au also evolves more slowly spectroscopically. The near-peak spectra of SN~2012au show a clear absence of W-shaped O\,{\sc ii} features, commonly observed in SLSNe~I. SN~2012au shows higher $v_{ph}$ than other SNe~Ib but evolves faster than SLSNe~I. Using the fluxes of the [O\,{\sc i}] doublet and [Ca\,{\sc ii}], we propose that the progenitor of SN~2012au is a single WR star, with $M_{O}$ and $M_{He}$ of $\sim$1.62 $\pm$ 0.15 M$_\odot$ and $\sim$4-8 M$_\odot$, respectively. M$_{ZAMS}$ values constrained using the +391 d spectrum matching those of the modelled spectrum of \cite{Jerkstrand2015} ($\sim$ 17 M$_\odot$), from $M_{O}$ and the [Ca\,{\sc ii}]/[O\,{\sc i}] doublet flux ratio ($\sim$25 M$_\odot$), and also those constrained using {\tt MESA} and {\tt SNEC} modelling ($\sim$20 M$_\odot$), suggest a range of M$_{ZAMS}$ $\sim$17 -- 25 M$_\odot$ for SN~2012au. The above physical parameters are close to those inferred for some SNe~Ic-BL (e.g., SN~1998bw and SN~2002ap), but lower than those of  SLSNe~I (e.g., PTF12dam and SN~2015bn), favouring SN~2012au as a bridge between normal Ib/c and Ic-BL.

\section*{Data Availability}
The data used in this work can be made available on request to the corresponding authors.

\section*{Acknowledgements}
This study uses data from DOT-3.6m, HCT-2.0m, DFOT-1.3m, and ST-1.04m, and the authors of this work are highly grateful to the observers at the Aryabhatta Research Institute of Observational Sciences (ARIES) and Indian Astronomical Observatory (IAO), Hanle for their valuable time and support for the observations of this event. This study also uses data from the BTA-6.0m at SAO Russia. Observations with the SAO RAS telescopes are supported by the Ministry of Science and Higher Education of the Russian Federation (including agreement No05.619.21.0016, project ID RFMEFI61919X0016). This article is partially based on observations collected at the Galileo 1.22m telescope operated by DFA University of Padova (Asiago, Italy). For the present study, the spectroscopic observations have also been taken in the framework of the European supernova collaboration involved in ESO--NTT large program 184.D-1140 led by Stefano Benetti. SBP, RG, AA and KM acknowledge BRICS grant DST/IMRCD/BRICS/Pilot call/ProFCheap/2017(G) and DST/JSPS grant DST/INT/JSPS/P/281/2018 for this work. BK, GCA and DKS acknowledge BRICS grant DST/IMRCD/BRICS/PilotCall1/MuMeSTU/2017(G) for the present work. AA also acknowledges funds and assistance provided by the Council of Scientific \& Industrial Research (CSIR), India. JV is supported by the project ``Transient Astrophysical Objects'' GINOP 2.3.2-15-2016-00033 of the National Research, Development, and Innovation Office (NKFIH), Hungary, funded by the European Union. Research by SV is supported by NSF grants AST-1813176 and AST-2008108. The authors are thankful to the anonymous referee for constructive comments and suggestions to improve the overall analysis presented in this work. SBP and AK are highly grateful to Professor J. Craig Wheeler for his consistent support and guidance in learning many aspects of the frontiers of CCSNe physics. AK and BK also acknowledge C. Eswaraiah, Kaushal Sharma, and Raya Dastidar for valuable discussions on various aspects and S. Bose for observing the event with the ST 1.04m. BK thanks M. Yamanaka and K. S. Kawabata for sharing the data. This research has utilized the NED, which is operated by the Jet Propulsion Laboratory, California Institute of Technology, under contract with NASA. We acknowledge the use of NASA's Astrophysics Data System Bibliographic Services. This research also made use of the Open Supernova Catalog (OSC), currently maintained by James Guillochon and Jerod Parrent. The work was partially performed as part of the government contract of the SAO RAS approved by the Ministry of Science and Higher Education of the Russian Federation. Observations at the SAO RAS telescopes are supported by the Ministry of Science and Higher Education of the Russian Federation.


\appendix
\setcounter{table}{0}
\renewcommand{\thetable}{A\arabic{table}}
\setcounter{figure}{0}
\renewcommand{\thefigure}{A\arabic{figure}}

\section*{Appendix}

\begin{table*}
\centering
\scriptsize
\caption{Calibrated magnitudes (DOT-3.6m and HCT-2m) of the secondary standard stars in the SN~2012au field, as shown in Fig.~\ref{fig:find-chart}.}
\label{Ap:table1}
\begin{tabular}{c c c c c c}
\hline
Star &  $U$             &   $B$            &   $V$            &   $R$            &   $I$            \\
ID   & (mag)            &  (mag)           &  (mag)           &  (mag)           &  (mag)           \\ \hline \hline
1    & 16.85 $\pm$ 0.03 & 15.95 $\pm$ 0.01 & 14.98 $\pm$ 0.01 & 14.40 $\pm$ 0.01 & 13.89 $\pm$ 0.01 \\
2    & 16.40 $\pm$ 0.02 & 15.50 $\pm$ 0.01 & 14.50 $\pm$ 0.01 & 13.89 $\pm$ 0.01 & 13.37 $\pm$ 0.01 \\
3    & 15.38 $\pm$ 0.03 & 15.35 $\pm$ 0.01 & 14.68 $\pm$ 0.01 & 14.27 $\pm$ 0.01 & 13.88 $\pm$ 0.01 \\
4    & --               & 18.61 $\pm$ 0.04 & 17.71 $\pm$ 0.03 & 17.14 $\pm$ 0.02 & 16.60 $\pm$ 0.03 \\
5    & 17.20 $\pm$ 0.04 & 16.12 $\pm$ 0.01 & 14.74 $\pm$ 0.01 & 13.87 $\pm$ 0.01 & 12.97 $\pm$ 0.01 \\
\hline
\end{tabular}
\end{table*}

\begin{table*}
\scriptsize
 \begin{center}
  \begin{threeparttable}
    \caption{Photometric data  of SN~2012au in \textit{U}, \textit{B}, \textit{V}, \textit{R}, and \textit{I} bands along with one-epoch observations in $J$, $H$, and $K$ bands.}
    \label{Ap:table2}
    \addtolength{\tabcolsep}{5pt}
    \begin{tabular}{c c c c c c c c c c}
\hline

    MJD & Phase $^a$ & \textit{U}  & \textit{B}  &\textit{V}  &\textit{R}  &\textit{I}  & Telescope Used \\
     $ $  & d  & mag  & mag  & mag  & mag & mag \\         
    \hline
56005.81 &$-$0.19 &$ 13.71 \pm  0.02$& $ 14.06 \pm    0.02$& $ 13.67  \pm   0.02$& $ 13.37 \pm   0.02$& $ 13.30 \pm   0.02$&  ST-1.04m    \\   
56006.81 &0.81  &  $ 13.69 \pm  0.03$&          ---     & $ 13.58  \pm   0.02$& $ 13.34 \pm   0.02$& $ 13.29 \pm   0.02$&  ST-1.04m    \\   
56007.88 &1.88  &           ---      & $ 14.11 \pm    0.02$& $ 13.58  \pm   0.02$& $ 13.37 \pm   0.02$& $ 13.24 \pm   0.02$&  HCT-2m   \\   
56009.70 &3.70  &  $ 13.97 \pm  0.03$& $ 14.18 \pm    0.02$& $ 13.60  \pm   0.02$& $ 13.35 \pm   0.02$& $ 13.22 \pm   0.02$&  ST-1.04m    \\   
56010.98 &4.98  &  $ 14.34 \pm  0.05$& $ 14.27 \pm    0.10$& $ 13.60  \pm   0.04$& $ 13.36 \pm   0.08$& $ 13.21 \pm   0.03$&  CAHA-2.2m    \\  
56011.79 &5.79  &  $ 14.50 \pm  0.02$& $ 14.53 \pm    0.03$& $ 13.72  \pm   0.03$& $ 13.38 \pm   0.02$& $ 13.25 \pm   0.04$&  HCT-2m   \\   
56011.86 &5.86  &           ---     &          ---     & $ 13.62  \pm   0.02$& $ 13.35 \pm   0.02$& $ 13.23 \pm   0.03$&  ST-1.04m    \\   
56013.73 &7.73  &           ---     & $ 14.58 \pm    0.02$& $ 13.79  \pm   0.02$& $ 13.43 \pm   0.02$& $ 13.23 \pm   0.03$&  HCT-2m   \\   
56018.70 &12.7  &  $ 14.95 \pm  0.05$& $ 15.15 \pm    0.02$& $ 14.06  \pm   0.02$& $ 13.64 \pm   0.02$& $ 13.39 \pm   0.03$&  ST-1.04m    \\   
56020.80 &14.8  &  $ 15.34 \pm  0.04$& $ 15.34 \pm    0.03$& $ 14.25  \pm   0.03$& $ 13.71 \pm   0.02$& $ 13.41 \pm   0.02$&  HCT-2m   \\   
56020.73 &14.73 &           ---     & $ 15.45 \pm    0.03$& $ 14.31  \pm   0.03$& $ 13.77 \pm   0.02$& $ 13.46 \pm   0.03$&  ST-1.04m    \\   
56021.87 &15.87 &           ---     &          ---     & $ 14.21  \pm   0.02$& $ 13.77 \pm   0.02$& $ 13.42 \pm   0.04$&  ST-1.04m    \\   
56025.68 &19.67 &  $ 15.33 \pm  0.07$&          ---     &   ---            &   ---             &          ---     &  ST-1.04m    \\   
56029.81 &23.81 &  $ 15.80 \pm  0.03$& $ 15.94 \pm    0.02$& $ 14.72  \pm   0.02$& $ 14.16 \pm   0.02$& $ 13.83 \pm   0.03$&  ST-1.04m    \\   
56036.72 &30.72 &  $ 16.14 \pm  0.07$& $ 16.11 \pm    0.04$& $ 14.91  \pm   0.02$& $ 14.39 \pm   0.03$& $ 14.08 \pm   0.02$&  ST-1.04m    \\   
56037.70 &31.7  &           ---     & $ 16.17 \pm    0.03$& $ 15.05  \pm   0.02$& $ 14.47 \pm   0.03$& $ 14.12 \pm   0.03$&  HCT-2m   \\   
56038.68 &32.68 &           ---     & $ 16.12 \pm    0.03$& $ 14.96  \pm   0.02$& $ 14.50 \pm   0.03$& $ 14.12 \pm   0.03$&  ST-1.04m    \\   
56040.75 &34.75 &  $ 16.51 \pm  0.04$& $ 16.20 \pm    0.04$& $ 15.06  \pm   0.02$& $ 14.52 \pm   0.03$& $ 14.18 \pm   0.04$&  HCT-2m   \\   
56042.71 &36.71 &           ---     & $ 16.25 \pm    0.04$& $ 15.18  \pm   0.02$& $ 14.61 \pm   0.03$& $ 14.22 \pm   0.03$&  HCT-2m   \\   
56043.01 &37.01 &           ---     & $ 16.12 \pm    0.04$& $ 15.17  \pm   0.02$& $ 14.55 \pm   0.03$& $ 14.19 \pm   0.02$&  BTA-6m   \\   
56046.76 &40.76 &  $ 16.76 \pm  0.09$& $ 16.28 \pm    0.02$& $ 15.19  \pm   0.02$& $ 14.68 \pm   0.02$& $ 14.31 \pm   0.02$&  ST-1.04m    \\   
56048.64 &42.64 &           ---     & $ 16.40 \pm    0.02$& $ 15.32  \pm   0.02$& $ 14.85 \pm   0.02$& $ 14.46 \pm   0.02$&  HCT-2m   \\   
56054.69 &48.69 &           ---     & $ 16.47 \pm    0.03$& $ 15.45  \pm   0.02$& $ 15.04 \pm   0.02$& $ 14.41 \pm   0.03$&  ST-1.04m    \\   
56060.80 &54.8  &           ---     & $ 16.58 \pm    0.02$& $ 15.50  \pm   0.02$& $ 15.00 \pm   0.02$& $ 14.56 \pm   0.03$&  ST-1.04m    \\   
56064.73 &58.73 &           ---     & $ 16.60 \pm    0.02$& $ 15.52  \pm   0.03$& $ 15.11 \pm   0.02$& $ 14.72 \pm   0.04$&  ST-1.04m    \\   
56067.61 &61.61 &           ---     & $ 16.67 \pm    0.02$& $ 15.57  \pm   0.02$& $ 15.10 \pm   0.02$& $ 14.66 \pm   0.03$&  HCT-2m   \\   
56070.72 &64.72 &           ---     &          ---     & $ 15.62  \pm   0.02$& $ 15.12 \pm   0.02$&   ---             &  DFOT-1.3m  \\   
56071.78 &65.78 &           ---     & $ 16.59 \pm    0.02$& $ 15.52  \pm   0.04$& $ 15.11 \pm   0.03$& $ 14.55 \pm   0.03$&  BTA-6m   \\   
56071.78 &65.79 &           ---     & $ 16.67 \pm    0.03$& $ 15.57  \pm   0.02$& $ 15.15 \pm   0.02$& $ 14.74 \pm   0.03$&  HCT-2m   \\   
56073.75 &67.75 &           ---     & $ 16.78 \pm    0.02$& $ 15.65  \pm   0.03$& $ 15.19 \pm   0.02$& $ 14.91 \pm   0.02$&  ST-1.04m    \\   
56074.70 &68.7  &           ---     & $ 16.69 \pm    0.03$& $ 15.64  \pm   0.02$& $ 15.37 \pm   0.03$& $ 14.99 \pm   0.03$&  ST-1.04m    \\   
56086.67 &80.67 &           ---     & $ 16.81 \pm    0.03$& $ 15.74  \pm   0.02$& $ 15.50 \pm   0.03$& $ 15.14 \pm   0.03$&  ST-1.04m    \\   
56088.71 &82.71 &           ---     & $ 16.83 \pm    0.03$& $ 15.81  \pm   0.04$& $ 15.43 \pm   0.03$&          ---     &  DFOT-1.3m  \\   
56089.69 &83.69 &           ---     & $ 16.77 \pm    0.03$& $ 15.77  \pm   0.03$& $ 15.46 \pm   0.03$&          ---     &  DFOT-1.3m  \\   
56090.76 &84.76 &           ---     & $ 16.76 \pm    0.03$& $ 15.88  \pm   0.02$& $ 15.55 \pm   0.03$&          ---     &  HCT-2m   \\   
56103.70 &97.7  &           ---     & $ 16.83 \pm    0.04$& $ 16.05  \pm   0.02$& $ 15.69 \pm   0.03$& $ 15.22 \pm   0.04$&  ST-1.04m    \\   
56104.61 &98.61 &           ---     & $ 16.90 \pm    0.05$& $ 16.12  \pm   0.03$& $ 15.65 \pm   0.02$& $ 15.25 \pm   0.03$&  HCT-2m   \\   
56120.63 &114.63&           ---     & $ 17.03 \pm    0.05$& $ 16.37  \pm   0.03$& $ 16.06 \pm   0.02$& $ 15.53 \pm   0.03$&  HCT-2m   \\   
56128.64 &122.64&           ---     &          ---     & $  ---              $& $ 16.37 \pm   0.09$& $ 15.80 \pm   0.12$&  HCT-2m   \\   
56305.90 &299.9 &           ---     & $ 17.83 \pm    0.05$& $ 17.31  \pm   0.03$& $ 17.39 \pm   0.06$& $ 16.76 \pm   0.05$&  ST-1.04m    \\   
56392.72 &386.72&           ---     &          ---     & $ 18.08  \pm   0.04$& $ 17.99 \pm   0.04$& $ 17.02 \pm   0.05$&  ST-1.04m    \\   
56396.73 &390.73&           ---     &          ---     & $ 18.00  \pm   0.04$& $ 18.09 \pm   0.04$& $ 17.09 \pm   0.05$&  ST-1.04m    \\   
56397.93 &391.93&           ---     &          ---     & $ 18.18  \pm   0.05$&   ---               &          ---     &  BTA-6m   \\   
56411.78 &405.78&           ---     &          ---     &   ---                & $ 18.12 \pm   0.05$& $ 17.19 \pm   0.08$&  ST-1.04m    \\   
56419.67 &413.67&           ---     &          ---     &   ---                & $ 18.15 \pm   0.08$& $ 17.18 \pm   0.08$&  ST-1.04m    \\
\hline    
    MJD &Phase & \textit{J}  & \textit{H}  &\textit{K}  &   &   & Telescope Used \\
     $ $  & d  & mag  & mag  & mag  &  &  \\         
\hline
56001.40 & $-$5 & $13.40  \pm 0.06$ & $13.33  \pm 0.10$ &  $12.69  \pm 0.20$ & & & NTT-3.58m \\
      \hline 
    \end{tabular}
    \begin{tablenotes}[para,flushleft]
    $^{a}$ with reference to d since \textit{B}-band maximum. \\
    \end{tablenotes}
  \end{threeparttable}
  \end{center}
\end{table*}

\begin{table*}
\centering
\caption{Log of polarimetric observations using the AIMPOL instrument and estimated parameters of SN~2012au during early phases.
\label{Ap:table3}}
\addtolength{\tabcolsep}{10pt}
\begin{tabular}{lcc ll ll ll}
\hline \hline
\textsc{ut} Date & MJD      & Phase$^{a}$ & \multicolumn{2}{c}{Observed} & \multicolumn{2}{c}{Intrinsic (ISP subtracted)} \\
(2012)           &          & (d)      & $P_{R} \pm \sigma_{P_{R}}$   & $ \theta{_R} \pm \sigma_{\theta{_R}}$ & $P_{R} \pm \sigma_{P_{R}}$ & $\theta{_R} \pm \sigma_{\theta{_R}}$ \\
                 &          &             & ($\%$)                       & ($^\circ)$                            & ($\%$)                     & ($^\circ$)    \\ \hline
March 16         & 56003.77 &$-$2.24         & 1.43$\pm$0.34                &42.90$\pm$6.67                         &1.20$\pm$0.34               & 43.90$\pm$7.90 \\
March 26$^{*}$   & 56013.87 & +7.86           & 1.02 --                   & 68.60  --                              &1.02 --                     & 79.40 --        \\              
March 28         & 56015.85 & +9.84         & 1.13$\pm$0.20                &15.79$\pm$5.04                         &0.98$\pm$0.20               &11.10$\pm$ 5.80  \\
March 29         & 56016.85 & +10.84         & 1.55$\pm$0.12                &39.22$\pm$2.29                         &1.32$\pm$0.12               &39.50$\pm$ 2.70  \\
April 16$^{*}$   & 56034.73 & +28.72          & 1.18 --                       &26.00 --                                 &1.18 --                      &18.10 --         \\ 
April 18           & 56036.76 & +30.70         & 0.34$\pm$0.11                &31.32$\pm$9.32                         &0.13$\pm$0.11               &19.50$\pm$25.20  \\
\hline
\end{tabular}  \\
$^{a}$ with reference to d since \textit{B}-band maximum. \\
$^{*}$ The estimated parameters on these dates are limiting values. \\ 
\end{table*}

\begin{table*}
\centering
\caption{Observational detail of nine isolated field stars selected to subtract the interstellar polarization. Observations of all field stars were performed on 2013 January 20 in $R$-band with the ST-1.04m. All these stars were selected with known distances and within 10$\degr$ radius around SN~2012au. The distance mentioned 
in the column 6 and 7 have been taken from \citet{Van2007} and \textit{Gaia} DR2 catalogue \citep{GaiaCollaboration2018}, respectively.
\label{Ap:table4}}
\addtolength{\tabcolsep}{10pt}
\begin{tabular}{lcccllll}
\hline \hline
Star     &  RA (J2000) & Dec (J2000)  & $P_{R} \pm \sigma_{P_{R}}$ & $ \theta{_R} \pm \sigma_{\theta{_R}}$ &  Distance$^{a}$ &  Distance$^{b}$\\
id       & ($^\circ$)  & ($^\circ$)   & $\%$                       & ($^\circ$)                            &  (in Plx)       &  (in Plx)      \\
\hline
HD~111384 &  192.252621 & --09.220328   & 0.25 $\pm$0.07& 40.18 $\pm$ 7.26 & 4.37$\pm$ 0.67                 &  5.53 $\pm$ 0.06                 \\
HD~111595 &  192.643571 & --10.571688   & 0.16 $\pm$0.12& 82.29 $\pm$21.64 & 6.22$\pm$ 1.00                 &  6.42 $\pm$ 0.05                 \\
HD~111678 &  192.787005 & --11.267007   & 0.10 $\pm$0.12&109.26 $\pm$33.23 & 9.32$\pm$ 1.50                 &  8.92 $\pm$ 0.06                 \\
HD~111999 &  193.331293 & --10.921184   & 0.18 $\pm$0.06& 58.98 $\pm$ 9.07 & 4.78$\pm$ 1.09                 &  3.70 $\pm$ 0.04                 \\
HD~112050 &  193.423944 & --09.618172   & 0.24 $\pm$0.04& 54.36 $\pm$ 5.16 & 2.69$\pm$ 1.03                 &  1.81 $\pm$ 0.09                 \\
HD~112464 &  194.241065 & --10.882960   & 0.34 $\pm$0.03& 72.62 $\pm$ 2.54 & 8.12$\pm$ 0.95                 &  6.87 $\pm$ 0.05                 \\
HD~112708 &  194.712469 & --11.504578   & 0.47 $\pm$0.07& 77.47 $\pm$ 4.01 & 2.58$\pm$ 1.61                 &  3.80 $\pm$ 0.04                 \\
HD~112806 &  194.878196 & --11.548831   & 0.26 $\pm$0.01& 30.77 $\pm$ 1.08 & 5.26$\pm$ 1.79                 &  1.89 $\pm$ 0.09                 \\
HD~112325 &  194.011930 & --11.309629   & 0.22 $\pm$0.05& 53.03 $\pm$ 6.01 & 2.36$\pm$ 1.27                 &  1.15 $\pm$ 0.06                 \\
\hline
\end{tabular}  \\
$^{a}$ \citet{Van2007} distance. \\ 
$^{b}$ Gaia distance. \\
\end{table*}

\begin{table*}
\scriptsize
 \begin{center}
  \begin{threeparttable}
    \caption{Optical and NIR spectroscopic data of SN~2012au obtained using the Galileo-1.22m, HCT-2m, CAHA-2.2m, NTT-3.58m, and BTA-6m.}
    \label{Ap:table5}
    \addtolength{\tabcolsep}{30pt}
    \begin{tabular}{c c c c c c c c c c}
\hline
Date & MJD & Phase & Spectral Range  & Telescope used \\
\hline
20120315&   56001.0  &   $-$5 &     3484--9284   &        HCT-2m\\
20120315&   56001.4  &   $-$5 &     9400--24990  &        NTT-3.58m\\
20120316&   56002.0  &   $-$4 &     3270--7780   &        Galileo-1.22m\\
20120318&   56004.0  &   $-$2 &     3484--9284   &        HCT-2m\\
20120321&   56007.0  &   +1   &     3484--9286   &        HCT-2m\\
20120324&   56010.0  &   +4   &     3335--8766   &        CAHA-2.2m\\
20120327&   56013.0  &   +7   &     3484--9268   &        HCT-2m\\
20120330&   56016.0  &   +10  &     3484--9284   &        HCT-2m\\
20120403&   56020.0  &   +14  &     3484--9282   &        HCT-2m\\
20120411&   56028.2  &   +22  &     9365--25080  &        NTT-3.58m\\
20120413&   56030.0  &   +24  &     3484--9284   &        HCT-2m\\
20120414&   56031.0  &   +25  &     3356--9996   &        NTT-3.58m\\
20120425&   56042.0  &   +36  &     3484--9283   &        HCT-2m\\
20120501&   56048.0  &   +42  &     3355--9992   &        NTT-3.58m\\
20120508&   56055.0  &   +49  &     3484--9283   &        HCT-2m\\
20120520&   56067.0  &   +61  &     3484--9283   &        HCT-2m\\
20120524&   56071.0  &   +65  &     3660--7871   &        BTA-6m\\
20120601&   56079.0  &   +73  &     3484--9283   &        HCT-2m\\
20120618&   56096.0  &   +90  &     3484--9281   &        HCT-2m\\
20120626&   56104.0  &   +98  &     3485--7837   &        HCT-2m\\
20120702&   56110.0  &   +104 &     3484--9281   &        HCT-2m\\
20120706&   56114.0  &   +108 &     3484--7836   &        HCT-2m\\
20130415&   56397.0  &   +391 &     3654--7867   &        BTA-6m\\
   \hline 
    \end{tabular}
    \begin{tablenotes}[para,flushleft]
    \end{tablenotes}
  \end{threeparttable}
  \end{center}
\end{table*}

\bsp
\label{lastpage}

\begin{thebibliography}{99}

\bibitem[Arnett(1982)]{Arnett1982} Arnett, W. D. 1982, The Astrophysical Journal, 253, 785

\bibitem[Arnett(1996)]{Arnett1996} Arnett, W. D. 1996, Supernovae and Nucleosynthesis: An Investigation of the History of Matter (Princeton, NJ: Princeton Univ. Press)

\bibitem[Aryan et al.(2021)]{Aryan2021} Aryan, A., Pandey, S.~B., Zheng, W., et al. 2021, MNRAS, 505 2530 

\bibitem[Banerjee et al.(2018)]{Banerjee2018} Banerjee, D.~P.~K., Joshi, V., Evans, A., et al. 2018, MNRAS, 481, 806

\bibitem[Blinnikov et al.(1998)]{Blinnikov1998} Blinnikov, S. I., Eastman, R. Bartunov, O. S., Popolitov V. A., Woosley S. E., 1998, ApJ, 496, 454 

\bibitem[Blinnikov et al.(2000)]{Blinnikov2000} Blinnikov S. I., Lundqvist P., Bartunov O., at al. 2000, ApJ, 532, 1132

\bibitem[Blinnikov et al.(2006)]{Blinnikov2006} Blinnikov S. I., Röpke F. K., Sorokina E. I. et al., 2006, A\&A, 453, 229

\bibitem[Branch et al.(2006)]{Branch2006} Branch, D., Jeffery, D. J., Young, T. R., \& Baron, E. 2006, PASP, 118, 791

\bibitem[Branch \& Wheeler(2017)]{Branch2017} Branch, D., \& Wheeler, J. C. 2017, Supernova Explosions: Astronomy and Astrophysics Library Supernova (Berlin: Springer)

\bibitem[Cano(2013)]{Cano2013} Cano Z., 2013, MNRAS, 434, 1098

\bibitem[Cano(2017)]{Cano2017} Cano, Z., Wang, S. Q., Dai, Z. G., \& Wu, X. F. et al. 2017, AdAst, 2017E, 5C 

\bibitem[Cao et al.(2013)]{Cao2013} Cao, Y., Kasliwal, M.~M., Arcavi, I., et al. 2013, ApJL, 775, L7

\bibitem[Chatzopoulos \& Tuminello(2019)]{Chatzopoulos2019} Chatzopoulos E. \& Tuminello R. 2019, ApJ, 874, 68

\bibitem[Chatzopoulos et al.(2009)]{Chatzopoulos2009} Chatzopoulos, E., Wheeler, J. C., \& Vinko, J. 2009, ApJ, 704, 1251

\bibitem[Chatzopoulos et al.(2012)]{Chatzopoulos2012} Chatzopoulos, E., Wheeler, J. C., Vinko, J., et al. 2012, ApJ, 746, 121

\bibitem[Chatzopoulos et al.(2013)]{Chatzopoulos2013} Chatzopoulos, E., Wheeler, J. C., Vinko, J., et al. 2013, ApJ, 773, 76

\bibitem[Chen et al.(2020)]{Chen2020} Chen, K.-J., Woosley, S.~E., \& Whalen, D.~J. 2020, ApJ, 893, 99

\bibitem[Cherchneff \& Dwek(2010)]{Cherchneff2010} Cherchneff I., Dwek E., 2010, ApJ, 713, 1

\bibitem[Cherchneff \& Lilly(2008)]{Cherchneff2008} Cherchneff I., Lilly S., 2008, ApJ, 683, L123

\bibitem[Chevalier \& Fransson(1994)]{Chevalier1994} Chevalier, R. A., \& Fransson, C. 1994, ApJ, 420, 268

\bibitem[Chugai et al.(1992)]{Chugai1992} Chugai, N. N., 1992, Soviet Astronomy Letters, 18, 168

\bibitem[Clocchiatti \&  Wheeler(1997)]{Clocchiatti1997} Clocchiatti A., Wheeler J. C., 1997, ApJ, 491, 375

\bibitem[Covino et al.(2003)]{Covino2003} Covino, S., et al. 2003, GCN Circ., 2167, 1C

\bibitem[De Cia et al.(2018)]{DeCia2018} De Cia A. et al., 2018, ApJ, 860, 100

\bibitem[Dessart et al.(2012)]{Dessart2012} Dessart, L., Hillier, D.~J., Li, C., et al. 2012, MNRAS, 424, 2139 

\bibitem[Dessart et al.(2015)]{Dessart2015} Dessart, L., Hillier, D.~J., Woosley, S., et al. 2015, MNRAS, 453, 2189 

\bibitem[Doroshenko et al.(1995)]{Doroshenko1995} Doroshenko, V.~T., Efimov, Y.~S., \& Shakhovskoi, N.~M. 1995, Astronomy Letters, 21, 513

\bibitem[Drout et al.(2011)]{Drout2011} Drout, M. R., et al. 2011, ApJ, 741, 97

\bibitem[Drout et al.(2016)]{Drout2016} Drout M. R. et al., 2016, ApJ, 821, 57

\bibitem[Eldridge \& Maund(2016)]{Eldridge2016} Eldridge, J.~J., \& Maund, J.~R. 2016, MNRAS, 461, L117

\bibitem[Elmhamdi et al.(2006)]{Elmhamdi2006} Elmhamdi, A., Danziger, I. J., Branch, D., et al. 2006, A\&A, 450, 305

\bibitem[Elmhamdi et al.(2004)]{Elmhamdi2004} Elmhamdi, A., Danziger, I. J., Cappellaro E., et al. 2004, A\&A, 426, 963

\bibitem[Eldridge et al.(2011)]{Eldridge2011} Eldridge, J.~J., \& Langer, N., Tout, C. A. 2011, MNRAS, 414, 3501

\bibitem[Filippenko(1993)]{Filippenko1993} Filippenko, A. V., Matheson T., Ho L. C. 1993, ApJL, 481, L89

\bibitem[Filippenko(1997)]{Filippenko1997} Filippenko, A. V. 1997, ARA\&A, 35, 309

\bibitem[Fang et al.(2019)]{Fang2019} Fang, Q., Maeda, K., Kuncarayakti, H., et al. 2019, Nature Astronomy, 3, 434

\bibitem[Fang \& Maeda(2019)]{Fang2018} Fang, Q., \& Maeda, K. 2018, ApJ, 864, 47

\bibitem[Folatelli et al.(2016)]{Folatelli2016} Folatelli, G., Van Dyk, Schuyler D., Kuncarayakti, H., 2016, ApJ, 825, 22

\bibitem[Foley et al.(2003)]{Foley2003} Foley R. J. et al., 2003, PASP, 115, 1220

\bibitem[Fransson \& Chevalier(1989)]{Fransson1989} Fransson C., Chevalier R. A., 1989, ApJ, 343, 323

\bibitem[Fremling et al.(2018)]{Fremling2018} Fremling, C., Sollerman, J., Kasliwal, M. M., 2018, A\&A, 618, 37 

\bibitem[Fremling et al.(2014)]{Fremling2014} Fremling, C., Sollerman, J., Taddia, F., et al. 2014, A\&A, 565, 114

\bibitem[Fremling et al.(2016)]{Fremling2016} Fremling, C., Sollerman, J., Taddia, F., 2016, A\&A, 593, 68 

\bibitem[Eswaraiah et al.(2012)]{Eswaraiah2012} Eswaraiah, C., Pandey, A.~K., Maheswar, G., et al. 2012, MNRAS, 419, 2587

\bibitem[Gaia Collaboration(2018)]{GaiaCollaboration2018} Gaia Collaboration, Brown, A. G. A., Vallenari, A., et al. 2018, A\&A, 616, 1 

\bibitem[Gal-Yam(2012)]{Gal-Yam2012} Gal-Yam, A. 2012, Sci, 337, 927

\bibitem[Gal-Yam et al.(2007)]{Gal-Yam2007} Gal-Yam, A., Leonard, D. C., Fox, D. B., et al. 2007, ApJ, 656, 372

\bibitem[Gal-Yam et al.(2009)]{Gal-Yam2009} Gal-Yam, A., Mazzali, P., Ofek, E. O., et al. 2009, Nature, 462, 624

\bibitem[Gal-Yam(2017)]{Gal-Yam2017} Gal-Yam A., 2017, Observational and Physical Classification of Supernovae. p. Springer, Cham, p. 195

\bibitem[Gal-Yam(2019)]{Gal-Yam2019} Gal-Yam A., 2019, ARA\&A, 57, 305

\bibitem[Gangopadhyay et al.(2020)]{Gangopadhyay2020} Gangopadhyay, A., Misra, K., Sahu, D. K., et al. 2020, MNRAS, 497, 3770G 

\bibitem[Gaskell et al.(1986)]{Gaskell1986} Gaskell C. M., Cappellaro E., Dinerstein H. L., et al. 1986, ApJ, 306 L77

\bibitem[Georgy et al.(2009)]{Georgy2009} Georgy C., Meynet G., Walder R., Folini D., Maeder A., 2009, A\&A, 502, 611

\bibitem[Gerardy et al.(2000)]{Gerardy2000} Gerardy, C. L., Fesen, R. A., Höflich, P., et al. 2000, AJ, 119, 2968

\bibitem[Gerardy et al.(2002)]{Gerardy2002} Gerardy, C. L., Fesen, R. A., Nomoto, K., et al. 2002, Publ. Astron. Soc. Jpn, 54, 905

\bibitem[Gearhart et al.(1999)]{Gearhart1999} Gearhart, R.~A., Wheeler, J.~C., \& Swartz, D.~A. 1999, ApJ, 510, 944

\bibitem[Gilkis et al.(2019)]{Gilkis2019} Gilkis, A., Vink, J. S., Eldridge, J. J., \& Tout, C. A. 2019, MNRAS, 486, 4451

\bibitem[Groh et al.(2013)]{Groh2013} Groh, J. H., Maynet, G., Georgy, C., \&, Ekstrom, S. 2013, A\&A, 558, A13

\bibitem[Gorosabel et al.(2010)]{Gorosabel2010} Gorosabel J., de Ugarte Postigo, Castro-Tirado A. J., et al., 2010, A\&A, 522, 14G 

\bibitem[Gorosabel et al.(2006)]{Gorosabel2006} Gorosabel J., Larionov V., Castro-Tirado A. J., et al., 2006, A\&A, 459, L33

\bibitem[Green et al.(2019)]{Green2019} Green G. M., Schlafly E., Zucker C., Speagle J. S., Finkbeiner D., 2019, ApJ, 887, 93

\bibitem[Guti{\'e}rrez et al.(2021)]{Gutierrez2021} Guti{\'e}rrez, C.~P., Bersten, M.~C., Orellana, M., et al.\ 2021, 504, 4907

\bibitem[Hachinger et al.(2012)]{Hachinger2012} Hachinger, S., Mazzali, P. A., Taubenberger, S., et al. 2012,MNRAS, 422, 70

\bibitem[Henyey et al.(1965)]{Henyey1965} Henyey, L., Vardya, M.~S., \& Bodenheimer, P. 1965, ApJ, 142, 841

\bibitem[Herwig(2000)]{Herwig2000} Herwig, F. 2000, A\&A, 360, 952

\bibitem[Hogg et al.(2002)]{Hogg2002} Hogg, D. W., Baldry, I. K., Blanton, M. R., \& Eisenstein, D. J. 2002, arXiv:astro-ph/0210394

\bibitem[Hoffman et al.(2014)]{Hoffman2014} Hoffman J. L., et al., 2014, American Astronomical Society Meeting Abstracts. p. AAS, 223 354.21

\bibitem[Hoffman et al.(2008)]{Hoffman2008} Hoffman, J.~L., Leonard, D.~C., Chornock, R., et al. 2008, ApJ, 688, 1186

\bibitem[Hoffman et al.(2017)]{Hoffman2017} Hoffman, J.~L., Williams, G.~G., Leonard, D.~C., et al.\ 2017, The Lives and Death-Throes of Massive Stars, 329, 54

\bibitem[Hoflich(1991)]{Hoflich1991} Hoflich P., 1991, A\&A, 246, 481

\bibitem[Hoflich et al.(2001)]{Hoflich2001} Höflich P., Khokhlov A., Wang L., 2001, AIPC, 586, 459H

\bibitem[Hunter et al.(2009)]{Hunter2009} Hunter, D.~J., Valenti, S., Kotak, R., et al. 2009, A\&A, 508, 371

\bibitem[Inserra et al.(2013)]{Inserra2013} Inserra C., et al., 2013, ApJ, 770, 128

\bibitem[Inserra(2019)]{Inserra2019} Inserra C., 2019, Nat. Astron., 3, 697

\bibitem[Jerkstrand et al.(2014)]{Jerkstrand2014} Jerkstrand, A., Smartt, S. J., Fraser, M., et al. 2014, MNRAS, 439, 3694

\bibitem[Jerkstrand et al.(2015)]{Jerkstrand2015} Jerkstrand, A., Ergon, M., Smartt, S. J., et al. 2015, A\&A, 573,12 

\bibitem[Jordi et al.(2006)]{Jordi2006} Jordi, K., Grebel, E. K., \& Ammon, K., et al. 2006, A\&A, 460, 339

\bibitem[Heger et al.(2003)]{Heger2003} Heger A., Fryer C. L., Woosley S. E., et al. 2003, ApJ, 591, 288

\bibitem[Kamble et al.(2013)]{Kamble2013} Kamble A., Soderberg M. A., Chomiuk L., et al. 2013, arXiv:1309.3573 

\bibitem[Kamble et al.(2014)]{Kamble2014} Kamble, A., Soderberg, M. A., Chomiuk, L., et al. 2014, ApJ, 797, 2

\bibitem[Kasen \& Bildsten(2010)]{Kasen2010} Kasen, D., \& Bildsten, L. 2010, ApJ, 717, 245

\bibitem[Kawabata et al.(2003)]{Kawabata2003} Kawabata K. S., Deng J., Wang L., et al., 2003, ApJ, 593, L19

\bibitem[Kilpatrick et al.(2018)]{Kilpatrick2018} Kilpatrick C. D., et al., 2018, MNRAS, 480, 2072

\bibitem[Kilpatrick et al.(2021)]{Kilpatrick2021} Kilpatrick, C. D. et al., 2021, MNRAS, 480, 2072

\bibitem[Kippenhahn et al.(1980)]{Kippenhahn1980} Kippenhahn, R., Ruschenplatt, G., \& Thomas, H.-C. 1980, A\&A, 91, 175

\bibitem[Kumar et al.(2020)]{Kumar2020} Kumar, A., Pandey S. B.,  Konyves-Toth, R., et al. 2020, ApJ, 892, 28K

\bibitem[Kumar et al.(2021)]{Kumar2021} Kumar, A., Kumar, B., Pandey, S.~B., et al. 2021, MNRAS, 502, 1678

\bibitem[Kumar et al.(2014)]{Kumar2014} Kumar B., Pandey S. B., Eswaraiah C., Gorosabel J., 2014, MNRAS, 442, 2

\bibitem[Kumar et al.(2016)]{Kumar2016} Kumar B., Pandey S. B., Eswaraiah C., Kawabata K. S., 2016, MNRAS, 456, 3157

\bibitem[Kumar et al.(2019)]{Kumar2019} Kumar B., et al., 2019, MNRAS, 488, 3089

\bibitem[Kumar et al.(2018)]{Kumar2018} Kumar, B., Omar, A., Maheswar, G., et al., 2018, BSRSL, 87, 29

\bibitem[Kuncarayakti et al.(2015)]{Kuncarayakti2015} Kuncarayakti, H., Maeda, K., Bersten, M. C.,  et al. 2015, A\&A, 579, A95

\bibitem[Konyves-Toth et al.(2020)]{Konyves-Toth2020} K{\"o}nyves-T{\'o}th, R., Thomas, B.~P., Vink{\'o}, J., et al. 2020, ApJ, 900, 73

\bibitem[Langer et al.(1985)]{Langer1985} Langer, N., El Eid, M.~F., \& Fricke, K.~J.\ 1985, A\&A, 145, 179

\bibitem[Leibundgut et al.(1991)]{Leibundgut1991} Leibundgut B., Kirshner R. P., Pinto P. A., et al. 1991, ApJ, 372, 531

\bibitem[Leloudas et al.(2017)]{Leloudas2017} Leloudas, G., Maund, J.~R., Gal-Yam, A., et al. 2017, ApJL, 837, L14

\bibitem[Leonard et al.(2006)]{Leonard2006} Leonard, D.C., Filippenko, A.V., Ganeshalingam, M., et al. 2006, Nature, 440, 505

\bibitem[Leonard et al.(2000)]{Leonard2000} Leonard, D.~C., Filippenko, A.~V., Barth, A.~J., et al. 2000, ApJ, 536, 239

\bibitem[Leonard \& Filippenko(2005)]{Leonard2005} Leonard, D.~C. \& Filippenko, A.~V. 2005, ASPC, 342, 330

\bibitem[Liljegren et al.(2020)]{Liljegren2020} Liljegren, S., Jerkstrand, A., \& Grumer, J. 2020, A\&A, 642, A135

\bibitem[Lyman et al.(2016)]{Lyman2016} Lyman, J. D., Bersier, D., James, P. A., et al. 2016, MNRAS, 457, 328

\bibitem[Maeda et al.(2002)]{Maeda2002} Maeda K., Nakamura T., Nomoto K., Mazzali P. A., Patat F., Hachisu I., 2002, ApJ, 565, 405

\bibitem[Maeda et al.(2006)]{Maeda2006} Maeda K., Nomoto K., Mazzali P. A., Deng J., 2006, ApJ, 640, 854

\bibitem[Maeda et al.(2007)]{Maeda2007} Maeda, K., Tanaka, M., Nomoto, K., et al. 2007, ApJ, 666, 1069

\bibitem[Maeda et al.(2008)]{Maeda2008} Maeda, K., Kawabata, K., Mazzali, P.~A., et al. 2008, Science, 319, 1220

\bibitem[Mauerhan et al.(2014)]{Mauerhan2014} Mauerhan, J., Williams, G.~G., Smith, N., et al. 2014, MNRAS, 442, 1166

\bibitem[Mauerhan et al.(2017)]{Mauerhan2017} Mauerhan, J.~C., Van Dyk, S.~D., Johansson, J., et al. 2017, ApJ, 834, 118

\bibitem[Maund et al.(2007)]{Maund2007} Maund J. R., Wheeler J. C., Patat F., Baade D., Wang L., Höflich P., 2007, MNRAS, 381, 201

\bibitem[Maund et al.(2009)]{Maund2009} Maund, J.~R., Wheeler, J.~C., Baade, D., et al. 2009, ApJ, 705, 1139

\bibitem[Maund et al.(2019)]{Maund2019} Maund, J.~R., Steele, I., Jermak, H., et al. 2019, MNRAS, 482, 4057

\bibitem[Maurer et al.(2010)]{Maurer2010} Maurer, I., Mazzali, P.A., Taubenberger, S., et al. 2010, MNRAS, 409, 1441M 

\bibitem[Mazzali et al.(2002)]{Mazzali2002} Mazzali P. A. et al., 2002, ApJL, 572, L61

\bibitem[Mazzali et al.(2005)]{Mazzali2005} Mazzali, P. A., Kawabata, K.S., Maeda K., Nomoto, et al. 2005, Science, 308, 1284

\bibitem[Mazzali et al.(2001)]{Mazzali2001} Mazzali, P. A., Nomoto, K., Patat, F., Maeda K., 2001, ApJ, 559, 1047

\bibitem[McCall(1984)]{McCall1984} McCall M. L., 1984, MNRAS, 210, 829

\bibitem[Medhi et al.(2007)]{Biman2007} Medhi, B.~J., Maheswar, G., Brijesh, K., et al. 2007, MNRAS, 378, 881

\bibitem[Medhi et al.(1996)]{Meikle1996} Meikle, W. P. S., Cumming, R. J., Geballe, T. R., et al. 1996, 281, 263

\bibitem[Metzger et al.(2015)]{Metzger2015} Metzger, B.~D., Margalit, B., Kasen, D., et al. 2015, MNRAS, 454, 3311

\bibitem[Milisavljevic et al.(2018)]{Milisavljevic2018} Milisavljevic, D., Patnaude, D. J., Chevalier, R. A., et al. 2018, ApJ,  864, 36 

\bibitem[Milisavljevic et al.(2013)]{Milisavljevic2013} Milisavljevic, D., Soderberg, A. M., Margutti, R., et al 2013, ApJ, 770, 38 

\bibitem[Modjaz et al.(2014)]{Modjaz2014} Modjaz, M., Blondin, S., Kirshner, R. P., et al. 2014, AJ, 147, 99

\bibitem[Modjaz et al.(2019)]{Modjaz2019} Modjaz M., Gutiérrez C. P., \& Arcavi I., 2019, Nature Astronomy, 3, 717

\bibitem[Mohan et al.(1999)]{Mohan1999} Mohan V., Uddin W., Sagar R., Gupta S. K., 1999, BASI, 27, 601

\bibitem[Morgan \& Edmunds(2003)]{Morgan2003} Morgan, H.~L. \& Edmunds, M.~G. 2003, MNRAS, 343, 427

\bibitem[Moriya et al.(2018)]{Moriya2018} Moriya T. J., Sorokina E. I., Chevalier R. A., 2018, Space Sci. Rev., 214, 59

\bibitem[Morozova et al.(2015)]{Morozova2015} Morozova, V., Piro, A.~L., Renzo, M., Ott, C. D. 2015, ApJ, 814, 63

\bibitem[Nadyozhin(1994)]{Nadyozhin1994} Nadyozhin, D.K. 1994, ApJS, 92, 527

\bibitem[Nicholl(2018a)]{Nicholl2018a} Nicholl, M. 2018a, Research Notes of the AAS, 2, 230

\bibitem[Nicholl et al.(2018b)]{Nicholl2018b} Nicholl, M., Blanchard, P.~K., Berger, E., et al. 2018b, ApJL, 866, L24

\bibitem[Nicholl et al.(2016)]{Nicholl2016} Nicholl, M., Berger, E., Smartt, S. J., et al. 2016, ApJ, 826, 39

\bibitem[Nicholl et al.(2013)]{Nicholl2013} Nicholl, M., Smartt, S. J., Jerkstrand, A., et al. 2013, Natur, 502, 346

\bibitem[Nicholl et al.(2015)]{Nicholl2015} Nicholl, M., Smartt, S. J., Jerkstrand, A., et al. 2015, MNRAS, 452, 3869

\bibitem[Nicholl et al.(2017)]{Nicholl2017} Nicholl, M., Guillochon, J., \& Berger, E. 2017, ApJ, 850, 55N 

\bibitem[Nicholl et al.(2019)]{Nicholl2019} Nicholl, M., Berger, E., Blanchard, P. K., et al. 2019, ApJ, 871, 102

\bibitem[Nomoto et al.(1995)]{Nomoto1995} Nomoto K., I., Iwamoto, K.,  Suzuki, T. 1995, Phys. Rep., 256, 173

\bibitem[Nomoto et al.(2006)]{Nomoto2006} Nomoto, K., Tominaga, N., Umeda, H., et al. 2006, Nuclear Physics A, 777, 424

\bibitem[Nugis \& Lamers(2000)]{Nugis2000} Nugis, T., \& Lamers, H.~J.~G.~L.~M. 2000, A\&A, 360, 227

\bibitem[Osterbrock(1989)]{Osterbrock1989} Osterbrock D. E., 1989, Sky \& Telesc., 78, 491

\bibitem[Ouchi \& Maeda(2017)]{Ouchi2017} Ouchi, R., \& Maeda, K.\ 2017, ApJ, 840, 90

\bibitem[Pandey et al.(2009)]{Pandey2009} Pandey, J.~C., Medhi, B.~J., Sagar, R., et al. 2009, MNRAS, 396, 1004

\bibitem[Pandey et al.(2003)]{Pandey2003} Pandey, S. B., Anupama, G. C., Sagar, R., et al. 2003, MNRAS, 340, 375P

\bibitem[Pandey et al.(2018)]{Pandey2018} Pandey, S.~B., Yadav, R.~K.~S., Nanjappa, N., et al. 2018, Bull. Soc. R. Sci. Liege, 87, 42

\bibitem[Pastorello et al.(2010)]{Pastorello2010} Pastorello, A., Smartt, S. J., Botticella, M. T., et al. 2010, ApJL, 724, L16

\bibitem[Patat et al.(2001)]{Patat2001} Patat, F., Cappellaro, E., Danziger, J., et al. 2001, ApJ, 555, 900

\bibitem[Paxton et al.(2011)]{Paxton2011} Paxton, B., Bildsten, L., Dotter, A., et al. 2011, ApJS, 192, 3

\bibitem[Paxton et al.(2013)]{Paxton2013} Paxton, B., Cantiello, M., Arras, P., et al. 2013, ApJS, 208, 4

\bibitem[Paxton et al.(2015)]{Paxton2015} Paxton, B., Marchant, P., Schwab, J., et al. 2015, ApJS, 220, 15

\bibitem[Paxton et al.(2018)]{Paxton2018} Paxton, B., Schwab, J., Bauer, E.~B., et al. 2018, ApJS, 234, 34

\bibitem[Pettini et al.(2004)]{Pettini2004} Pettini, M., \& Pagel, B. E. J. 2004, MNRAS, 348, L59

\bibitem[Podsiadlowski et al.(1992)]{Podsiadlowski1992} Podsiadlowski P., Joss P. C., Hsu J. J. L. 1992, ApJ, 391, 246

\bibitem[Prentice \& Mazzali(2017)]{Prentice2017} Prentice, S.~J. \& Mazzali, P.~A. 2017, MNRAS, 469, 2672

\bibitem[Prentice et al.(2019)]{Prentice2019} Prentice, S. J., Ashall, C., James, P. A., et al. 2019, MNRAS, 485, 1559

\bibitem[Quimby et al.(2011)]{Quimby2011} Quimby R. M. et al., 2011, Nature, 474, 487

\bibitem[Quimby et al.(2018)]{Quimby2018} Quimby, R. M., De Cia, A., Gal-Yam, A., et al. 2018, ApJ, 855, 2

\bibitem[Ramaprakash et al.(1998)]{Ramaprakash1998} Ramaprakash, A. N.,  Gupta, R., Sen, A. K., ei al. 1998, A\&AS, 128, 369  

\bibitem[Rautela et al.(2004)]{Rautela2004} Rautela, B. S., Joshi, G. C., \& Pandey, J. C., et al. 2004, BASI, 32, 159

\bibitem[Reilly et al.(2016)]{Reilly2016} Reilly, E., Maund, J. R., Baade, D., et al. 2016, MNRAS, 457, 288R

\bibitem[Rho et al.(2021)]{Rho2021} Rho, J., Evans, A., Geballe, T.~R., et al. 2021, ApJ, 908, 232

\bibitem[Sahu et al.(2011)]{Sahu2011} Sahu, D. K., Gurugubelli, U. K., Anupama, G. C., \& Nomoto, K., 2011, MNRAS, 413, 2583

\bibitem[Sahu et al.(2018)]{Sahu2018} Sahu, D. K., Anupama, G. C., Chakradhari, N. K., et al. 2018, MNRAS, 475, 2591S 

\bibitem[Sanders et al.(2012)]{Sanders2012} Sanders, N. E., Soderberg, A. M., Levesque, E. M., et al. 2012, ApJ, 758, 132

\bibitem[Sarangi \& Cherchneff(2013)]{Sarangi2013} Sarangi A., Cherchneff I., 2013, ApJ, 776, 107

\bibitem[Sarangi et al.(2018)]{Sarangi2018} Sarangi, A., Dwek, E., \& Arendt, R.~G. 2018, ApJ, 859, 66

\bibitem[Schlafly \& Finkbeiner(2011)]{Schlafly2011} Schlafly, E. F., \& Finkbeiner, D. P. 2011, ApJ, 737, 103

\bibitem[Schmidt et al.(1992)]{Schmidt1992} Schmidt, G.~D., Elston, R., \& Lupie, O.~L. 1992, AJ, 104, 1563

\bibitem[Serkowski et al.(1975)]{Serkowski1975} Serkowski K., Mathewson D. S., Ford V. L., 1975, ApJ, 196, 261

\bibitem[Shapiro \& Sutherland(1982)]{Shapiro1982} Shapiro P. R., Sutherland P. G., 1982, ApJ, 263, 902

\bibitem[Shivvers et al.(2017)]{Shivvers2017} Shivvers, I., Modjaz, M., Zheng, W., et al. 2017, PASP, 129, 054201

\bibitem[Shivvers et al.(2019)]{Shivvers2019} Shivvers, I, Filippenko, A. V., Silverman, J. M., et al. 2019, MNRAS, 482, 1545

\bibitem[Smartt(2009)]{Smartt2009} Smartt S. J. 2009, ARA\&A, 47, 63

\bibitem[Smith et al.(2011)]{Smith2011} Smith N., Gehrz R. D., Campbell R., et al. 2011, MNRAS, 418, 1959

\bibitem[Spyromilio et al.(2001)]{Spyromilio2001} Spyromilio, J., Leibundgut, B., \& Gilmozzi, R. 2001, A\&A, 376, 188

\bibitem[Srivastav et al.(2014)]{Srivastav2014} Srivastav S., Anupama G. C., Sahu D. K., 2014, MNRAS, 445, 1932

\bibitem[Srivastav et al.(2016)]{Srivastav2016} Srivastav, S., Ninan, J.~P., Kumar, B., et al. 2016, MNRAS, 457, 1000

\bibitem[Stalin et al.(2008)]{Stalin2008} Stalin, C.~S., Hegde, M., Sahu, D.~K., et al. 2008, BASI, 36, 111

\bibitem[Stevance et al.(2017)]{Stevance2017} Stevance, H. F., Maund, J. R., Baade, D., et al. 2017, MNRAS, 469, 1897S 
 
\bibitem[Stevance(2019)]{Stevance2019} Stevance, H.~F. 2019, arXiv:1906.07184
 
\bibitem[Stritzinger et al.(2009)]{Stritzinger2009} Stritzinger, M. et al., 2009, ApJ, 696, 713

\bibitem[Stritzinger et al.(2020)]{Stritzinger2020} Stritzinger, M.~D., Taddia, F., Holmbo, S., et al. 2020, A\&A, 634, A21

\bibitem[Taddia et al.(2015)]{Taddia2015} Taddia F., et al. 2015, A\&A, 574, 60

\bibitem[Taddia et al.(2018)]{Taddia2018} Taddia, F., Stritzinger, M. D., Bersten, M., et al. 2018, A\&A, 609, 136

\bibitem[Takaki et al.(2013)]{Takaki2013} Takaki, K., Kawabata, K. S., Yamanaka, M., et al. 2013, ApJ, 772, 17

\bibitem[Tanaka et al.(2008)]{Tanaka2008} Tanaka, M., Kawabata K. S., Maeda K., Hattori T., Nomoto K., 2008, ApJ, 689, 1191

\bibitem[Tanaka(2017)]{Tanaka2017} Tanaka, M.\ 2017, RSPTA, 375, 20160273. doi:10.1098/rsta.2016.0273

\bibitem[Taubenberger et al.(2009)]{Taubenberger2009} Taubenberger S., et al., 2009, MNRAS, 397, 677

\bibitem[Thielemann et al.(1996)]{Thielemann1996} Thielemann F.-K., Nomoto K., Hashimoto M.-A., 1996, ApJ, 460, 408

\bibitem[Thomas et al.(2011)]{Thomas2011} Thomas, R. C., Nugent, P. E., \& Meza, J. C. 2011, PASP, 123, 237

\bibitem[Valenti et al.(2008)]{Valenti2008} Valenti, S., Benetti, S., Cappellaro, E., et al. 2008, MNRAS, 383, 1485

\bibitem[Valenti et al.(2011)]{Valenti2011} Valenti, S., Fraser, M., Benetti, S., et al., 2011, MNRAS,  416, 3138

\bibitem[van Leeuwen(2007)]{Van2007} van Leeuwen, F. 2007, A\&A, 474, 653

\bibitem[Van Dyk(2017)]{Van2017} Van Dyk, S.~D.\ 2017, Philosophical Transactions of the Royal Society of London Series A, 375, 20160277

\bibitem[Van Dyk et al.(2018)]{Van2018} Van Dyk S. D., et al., 2018, ApJ, 860, 90

\bibitem[Vink et al.(2001)]{Vink2001} Vink, J.~S., de Koter, A., \& Lamers, H.~J.~G.~L.~M. 2001, A\&A, 369, 574

\bibitem[Uomoto(1986)]{Uomoto1986} Uomoto A., 1986, ApJ, 310, L35

\bibitem[Wang et al.(2003)]{Wang2003} Wang, L., Baade, D., H{\"o}flich, P., et al. 2003, ApJ, 592, 457

\bibitem[Wang \& Wheeler(2008)]{Wang2008} Wang L. \& Wheeler J. C. 2008, ARA\&A, 46, 433

\bibitem[Wang et al.(2019)]{Wang2019} Wang S.-Q., Wang L.-J., Dai Z.-G., 2019, Res. Astron. Astrophys., 19, 063

\bibitem[Wheeler et al.(2015)]{Wheeler2015} Wheeler, J. C., Johnson, V., \& Clocchiatti, A. 2015, MNRAS, 450,1295

\bibitem[Wheeler et al.(2017)]{Wheeler2017} Wheeler, J.~C., Chatzopoulos, E., Vink{\'o}, J., et al. 2017, ApJL, 851, L14

\bibitem[Wheeler et al.(1987)]{Wheeler1987} Wheeler, J. C., Harkness, R. P., Barker, E. S., Cochran, A. L., \& Wills, D. 1987, ApJ, 313, L69

\bibitem[Wheeler et al.(1998)]{Wheeler1998} Wheeler, J. C., Hoeflich, P., Harkness, R. P., \& Spyromilio, J. 1998, ApJ, 496, 908

\bibitem[Woosley(2010)]{Woosley2010} Woosley, S. E. 2010, ApJL, 719, L204

\bibitem[Xiang et al.(2019)]{Xiang2019} Xiang D., et al., 2019, ApJ, 871, 176

\bibitem[Yoon et al.(2010)]{Yoon2010} Yoon, S.-C., Woosley, S. E., \& Langer, N. 2010, ApJ, 725, 940

\bibitem[Yoon(2015)]{Yoon2015} Yoon, S.-C. 2015, PASA, 32, E015

\end{thebibliography}
\end{document}